\documentclass[usenatbib]{mn2e}
\usepackage{psfig}

\newcommand{\etal}{et~al.} 
\newcommand{\ionhy}{H{\sc ii} }

\newcommand{\UCHII}{UCH{\sc ii} }
\newcommand{\UCHIIns}{UCH{\sc ii}}
\newcommand{\degrees}{$^\circ$}
\newcommand{\kms}{$\mbox{km~s}^{-1 }$}
\newcommand{\kmsns}{$\mbox{km~s}^{-1}$}

\newcommand\HII{H\,{\sc ii} }

\newcommand{\vsfig}[2]           
{
  \begin{center}
    \begin{minipage}[t]{0.05\textwidth}
      {\footnotesize \raisebox{40mm}{(#2)}}
    \end{minipage}
    \begin{minipage}[t]{0.42\textwidth}
      \psfig{file=./#1.ps,height=0.95\textwidth,angle=270}
    \end{minipage}
    \hfill
  \end{center}
}

\newcommand{\specdfig}[2]        
{
   \begin{center}
     \begin{minipage}[t]{0.45\textwidth}
         \psfig{file=eps/#1.eps,height=0.5\textwidth,width=1\textwidth,angle=0}
     \end{minipage}
     \hfill
     \begin{minipage}[t]{0.45\textwidth}
         \psfig{file=eps/#2.eps,height=0.5\textwidth,width=1\textwidth,angle=0}
     \end{minipage}
   \end{center}
}

\newcommand{\specsfig}[1]        
{
   \begin{center}
     \begin{minipage}[t]{0.45\textwidth}
         \psfig{file=eps/#1.eps,height=0.55\textwidth,width=1\textwidth,angle=0}
     \end{minipage}
   \end{center}
}

\newcommand{\boxfig}[1]        
{
   \begin{center}
     \begin{minipage}[t]{0.46\textwidth}
         \psfig{file=eps/#1.eps,height=0.45\textwidth,angle=270}
     \end{minipage}
   \end{center}
}

\newcommand{\twofig}[2]        
{
   \begin{center}
     \begin{minipage}[t]{0.5\textwidth}
         \psfig{file=eps/#1.eps,height=0.95\textwidth}
     \end{minipage}
     \hfill
     \begin{minipage}[t]{0.5\textwidth}
         \psfig{file=eps/#2.eps,height=0.95\textwidth}
     \end{minipage}
   \end{center}
}

\begin{document}

\title[Water, OH and methanol masers in SFRs]{Water masers accompanying OH and methanol masers in star formation regions}
\author[S.\ L. Breen \etal]{S.\ L. Breen,$^{1,2}$\thanks{Email: Shari.Breen@utas.edu.au}  J.\ L. Caswell,$^2$ S.\ P. Ellingsen$^1$ \& C.\  J. Phillips$^2$\\ \\
 $^1$ School of Mathematics and Physics, University of Tasmania, Private Bag 37, Hobart, Tasmania 7001, Australia\\
  $^2$ Australia Telescope National Facility, CSIRO, PO Box 76, Epping, NSW 1710, Australia}

\date{Accepted 2010 April 2.  Received 2010 March 30; in original form 2010 February 17}

\pagerange{\pageref{firstpage}--\pageref{lastpage}}
\pubyear{2010}
 
 \maketitle
  
 \begin{abstract}
  The Australia Telescope Compact Array (ATCA) has been used to measure
positions with arcsecond accuracy for 379 masers at the 22-GHz
transition of water.  The principal observation targets were 202 OH masers
of the variety associated with star formation regions (SFR)s in the 
Southern Galactic plane.  At a second
epoch, most of these targets were observed again, and new targets of
methanol masers were added. Many of the water masers reported here are new
discoveries and others had been reported, with position uncertainties  
exceeding 10 arcsec, from Parkes telescope single dish observations 
many years ago.  

Variability in the masers is often acute, with very few features directly 
corresponding to those discovered two decades ago.  Within our 
current observations, less than a year apart, spectra are often 
dissimilar, but positions at the later epoch, even when measured for 
slightly different features, 
mostly correspond to the detected maser site measured earlier, to 
within the typical extent of the whole site, of a few arcseconds.

The precise water positions show that approximately 79 per cent (160 of 
202) of the OH maser sites show coincident water maser emission, the best 
estimate yet obtained for this statistic;  however, there are many 
instances where additional water sites are present 
offset from the OH target, and consequently less than half of the 
water masers coincide with a 1665-MHz ground-state OH maser counterpart.  
Our less uniform  sample of methanol 
targets is not suitable for a full investigation of their association with 
water masers, but we are able to explore 
differences between the velocities of peak emission from the 
three species, and quantify the typically larger deviations shown by water 
maser peaks from systemic velocities. 

Clusters of two or three distinct but nearby sites, each showing
one or several of the principal molecular masing transitions, are found 
to be common.  We also report the detection of ultracompact \HII regions 
towards some of the sites.  In combination with an investigation of  
correlations with IR sources from the GLIMPSE catalogue, these comparative 
studies allow further progress in 
the use of the maser properties to assign relative evolutionary 
stages in star formation to individual sites.  
  
  \end{abstract}


\section{Introduction}

Masers of OH (hydroxyl), water and methanol are key tools for investigating 
the formation of massive stars.  The masers reside in the dusty molecular 
envelope or torus of a massive star in its earliest stage of formation, 
and the masers are a sensitive probe for discovering stars in this 
embryonic state when the star is not visible because of obscuration from 
the dust.

Detailed studies of selected maser sites have been made comparing OH with 
water, and OH with methanol (Forster \& Caswell 1989, 1999; Caswell 1997; 
Caswell Vaile \& Forster 1995; 
hereafter FC89, FC99, C97, CVF95).  These suggest that the maser 
spots of all species are usually contained inside a region of diameter 
less than 30 mpc (= $0.93 ~ 10^{15}$ m = $0.93 ~  10^{17}$ cm  $\approx 
6200$ au $\approx 0.1$ light year), corresponding to an angular diameter 
of about 1 arcsec at a typical distance of 6 kpc.  
Positional precision of about 1 arcsec is therefore desirable to establish 
whether masers of different species originate from the same site.  

Within many star formation regions (SFRs), on a larger scale of 100 mpc or 
more, the FC89 and C97 studies revealed many 
instances where a number of maser sites are present in a small cluster, 
often with different combinations of maser species present.  There 
has been considerable speculation that the various combinations are 
indicative of massive young stars at a different evolutionary stage, or in 
a different mass range \citep[e.g.][]{Breen09,Ellingsen07}.  
Water maser emission is especially puzzling.  It has commonly been thought 
to be most prolific at an early stage 
of stellar evolution,  but isolated water masers are only partly accounted 
for by very young sites preceding OH maser excitation.  Other offset 
positions indicate the additional occurrence of water masers towards lower 
mass stars, or in high velocity fragments that have travelled far from the 
initial site.  In the latter case, maser spots from 
individual fragments are sometimes ephemeral, and disappear on a timescale 
as short as months, but are often replaced by 
generally similar emission in the same region, although at a slightly 
different position and velocity.  

To date, the most extensive unbiased survey for masers of the SFR variety
completed in the Galactic plane is at the 1665-MHz transition of OH,
presented as a catalogue reporting positions of arcsec accuracy for more
than 200 masers (Caswell 1998;  hereafter C98). Sensitive 
observations of 6.6-GHz methanol at
these positions have already been made, with a detection rate of 80 per
cent after high precision positions are compared \citep{C98,C09}.  Early
22-GHz water maser surveys with the Parkes radio telescope towards
southern SFRs have been reported by \citet{C+89} and
references therein, with positions measured to about 10 arcsec accuracy.  
Until recent years, follow up observations to arcsec accuracy were
restricted to northerly objects accessible to the VLA (FC89, FC99).  The
availability of the 22-GHz frequency band at the ATCA has allowed us to
search with high sensitivity and high positional precision for water
masers towards the full sample of OH masers, as well as $\sim$100 methanol
maser sites.

\section[]{Observations and data reduction}

Water maser observations were made with the ATCA on two separate
occasions.
During the first epoch of observations (2003 October 4 and 5), OH maser 
sites from C98 were targeted (see also Section 5.3) and during the second epoch (2004 
July 25, 26 and 30) many water maser detections towards the OH
masers were re-observed as well as a selected set of methanol masers 
\citep[chiefly from][]{C09}. 
OH maser sources that were targeted in the first epoch but showed no 
detectable water maser emission were not observed at the second epoch, and 
only a few of the water maser sources that had previously been observed
by FC89 with the VLA, and successfully confirmed during the first 
epoch, were reobserved.  

Our first observations were made with an EW array 
yielding 10 baselines between 30 and 352 m (project c1190).  
The correlator sampled a single linear polarization, processed to 
give a 512-channel spectrum across a 32-MHz bandwidth. The observing 
strategy was to observe approximately 100 targets over each 12-h session.  
After initial calibration, the first 10 targets were observed for 1.5 min 
each, followed by a calibrator, and similarly for the remaining 90 
targets. Then the cycle was repeated 2 more times, so as to 
provide for each source adequate uv-coverage, combined with a total 
integration time of 4.5 min.  Primary flux calibration is relative to PKS B1934-638 and in general is expected to be accurate to $\sim$20 \%. PKS B1921-293 was used for bandpass calibration.

The AIPS reduction package was used for  processing of the data
collected in this first epoch, following the general procedure described
in C97.  In the realignment of channels using the CVEL task, the adopted
rest frequency was 22235.08 MHz, and the velocity scale was with respect
to the local standard of rest (lsr).  The channel separation was 0.84
\kms\ which, with uniform weighting of the correlation function, yields a 
final velocity resolution of 1.0 \kmsns.  The quite coarse resolution was 
a compromise chosen in order to allow a large velocity coverage of more 
than 400 \kmsns.  With this coverage, it was possible to recognise any 
high velocity features indicative of close association with outflows (for 
which the water masers are renowned).

Total intensity maps were then produced of the channels with
maser emission apparent in the scalar averaged spectrum, or in a vector averaged spectrum shifted to the location of the target OH or methanol maser emission.  The rms noise in an individual channel image was typically
150~mJy.  The synthesised beam has a halfpower width of approximately 8 
arcsec in right ascension, but is larger in declination by a factor 
cosec(declination) as expected for an array aligned East-West with 
maximum baseline of 352 m.  

Our second series of observations (project c1330) were made in similar
fashion but with a different array configuration, H168. The correlator
configuration, and therefore the spectral resolution and velocity
coverage, was identical to that used in the 2003 observations. A few sample observations from this epoch were reduced, firstly with AIPS (as for the 2003 data), and secondly with  {\tt miriad} software package
\citep{Sault}. There was excellent agreement between data reduced in the respective data reduction packages. The full data set from
this epoch were reduced using {\tt miriad}, applying the standard techniques for ATCA spectral line and
continuum observations. Image cubes of the entire primary beam and
velocity ranges were produced for each source. The flux densities of
sources that were located away from the centre of the primary beam have
been corrected to account for beam attenuation. Spectra for each source
detected at this epoch were produced by integrating the emission in the
ATCA image cubes for each source. The typical resultant rms noise in each spectrum was 40 - 50 mJy. For the H168 array used in 2004, the synthesised beam was 
typically 13x9 arcsec.

The ATCA observations are most sensitive at the targeted positions, but provide useful measurements, albeit at lower sensitivity, of any other sources that happen to lie within the field of view of the primary beam; the full width to the first null is nearly 5  
arcmin, and the HPBW is 2.29 arcmin = 137 arcsec.

\begin{table*}
 \caption{22-GHz water masers detected towards sites of OH and methanol masers. Column 1 shows the source name in Galactic coordinates, column 2 and 3 give the right ascension and declination, column 4, 5 and 6 give the velocity of the water maser peak, velocity range and peak flux density in the 2003 observations, while columns 7, 8 and 9 give the velocity of the water maser peak, velocity range and peak flux density in the 2004 observations. A `--' in either column 6 or 9 indicates that no observations were made of the given source during the 2003 or the 2004 epoch, respectively, while the presence of a number preceded by a `$<$' indicates that there was no emission detected above the quoted threshold. For some complicated sources a `t' is present in either column 6 or 9 and this indicates that the exact nature of the detection is discussed in Section~\ref{sect:ind}. Associations are given in column 10, where the presence of an `o' denotes an OH maser, an `m' denotes a methanol maser, a `c' denotes the presence of a 22-GHz radio continuum source, a `g' the presence of a GLIMPSE point source and the presence of a `$\gamma$' indicates that the water maser source is outside the range of the GLIMPSE survey region. A `$^{\#}$' indicates that the proceeding associated source is strictly outside our association threshold but has been added through special circumstances. See Section 3 for a more extensive description.}
  \begin{tabular}{lccrcrrcrl} \hline
    \multicolumn{1}{c}{\bf Water maser} & {\bf RA}  & {\bf Dec} & {\bf Vpeak} &{\bf Vrange} & {\bf Speak} & {\bf Vpeak} &{\bf Vrange} & {\bf Speak} & {\bf Associations}\\
   \multicolumn{1}{c}{\bf ($l,b$)} & {\bf (J2000)}  & {\bf (J2000)}&  {\bf (\kmsns)}&  {\bf (\kmsns)} & {\bf (Jy)} &{\bf (\kmsns)}&  {\bf (\kmsns)} & {\bf (Jy)}     \\	
      \multicolumn{1}{c}{\bf (degrees)} & {\bf (h m s)}&{\bf ($^{o}$ $'$ $``$)}& {\bf 2003} & {\bf 2003} & {\bf 2003}& {\bf 2004} & {\bf 2004} & {\bf 2004} \\ \hline \\
G\,240.316+0.071 & 07 44 51.94 & --24 07 41.9 & 89 	& 58,100		& 10 		&		&			&	--	& 	o$\gamma$\\
G\,263.250+0.514 & 08 48 47.91 & --42 54 27.1 & 19 	& 17,21		& 3.0 		& 20		& 17,21		&	0.7	& 	om$\gamma$\\
G\,284.350--0.418& 10 24 10.60 & --57 52 33.0 &--4	 &--42,72    	& 60		& 7		&--78 ,71		&	102	&	o$\gamma$	\\
G\,285.260--0.067& 10 31 24.57 & --58 03 03.7 &--28  	&--95,20    	& 22		&      	--92	&--96,20		&	30	&	$\gamma$	\\
G\,285.263--0.050& 10 31 29.64 & --58 02 18.9 & 2    	&--59,50    	&1100	& 3		& --36,63		& 	1651	& 	o$\gamma$	\\
G\,287.371+0.644 & 10 48 04.25 & --58 27 00.7 & 1    	&--14,6     		&3.5		&  --1	& --13,1		&	4.9	&	om$\gamma$    \\
G\,290.374+1.661 & 11 12 17.98 & --58 46 21.6 & --12 	&--20,--10   	&3.5		& --47	&--47,--33		&	0.26	&	om$\gamma$ \\ 
G\,290.384+1.663 & 11 12 22.53 & --58 46 29.0 & --38 	&--50,--35  	& 6 		&--37	&--41,--35		&	4.5	&$\gamma$	 \\
G\,291.270--0.719& 11 11 49.67 & --61 18 53.8 &--102 	&--110,--23 	&65 		& --102	&--109,--17 	&	53	&   m$\gamma$\\
G\,291.274--0.709& 11 11 53.28 & --61 18 24.1 & --32 	&--51,14  	&60 		&--32	& --59,--14	&	64	&   	om$\gamma$	\\
G\,291.284--0.716& 11 11 56.58 & --61 19 01.1 &--133 	&--142,--120	&930	&--133	&--139,--120	&	701	&$\gamma$		      \\
G\,291.578--0.434& 11 15 04.91 & --61 09 50.7 &		&			&$<$0.2	&18		&17,	27		&	16 	& $\gamma$\\ 
G\,291.579--0.431& 11 15 05.67 & --61 09 41.0 & 12   	&--29,41     		&215	& 13		&--31,22		&	608	& om$\gamma$\\
G\,291.581--0.435& 11 15 06.17 & --61 09 56.1 & 26   	& 25,27     	&4.0		&		&			& $<$0.2	& m$\gamma$\\
G\,291.610--0.529& 11 15 02.58 & --61 15 48.8 & 13   	& --66,20   	&18 		&	12	& --66,28		&	39	& oc$\gamma$\\
G\,291.627--0.529& 11 15 10.18 & --61 16 12.7 & 22   	&8,23       		&12 		& 22		&	20,25	&	31  	& $\gamma$ \\
G\,291.629--0.541& 11 15 08.88 & --61 16 54.8 & 11 	&8,16       		&70 		&    	10	&--2,21		&	46	& $\gamma$ \\
G\,294.511--1.622& 11 35 32.04 &--63 14 44.3 & --12 	&--20,--6   	&250	& --12	&	--20,--4	&	112 & om$\gamma$\\
G\,294.976--1.733& 11 39 13.77 & --63 29 03.3 & 	  	&           		&   	--	&      1	& --17,5		& 	2.2		& $\gamma$\\
G\,294.989--1.719& 11 39 22.56 & --63 28 25.1 &		&           		&   	--	&     --17  	&  --18,--10	& 	0.6	&m$\gamma$\\
G\,297.660--0.974& 12 04 08.76 & --63 21 37.3 & 29	&--80,38    	&90 		&	26	& --79,82		&	75	& o\\
G\,299.012+0.125 & 12 17 24.05 & --62 29 13.6 &		&			& $<$0.2	&--26	&--27,--24		&	0.44&    \\           
G\,299.013+0.128 & 12 17 24.58 & --62 29 04.8 & 19 	&	 18,30     	&100	&	19	&--1,	50		&	83	& omcg\\
G\,300.491--0.190& 12 29 55.99 & --62 57 33.8 & 	24	&	23,25	&2.3		&25		&22,25		&	1.6	   &\\
G\,300.504--0.176& 12 30 03.42 & --62 56 50.2 & 11   	&--37,14    	&180	&11		&--26,14		&	94	& omg\\
G\,300.968+1.143 & 12 34 52.51 & --61 39 57.9 &--61  	&--86,--42  	&70 		&--58	&--86,--3		&	26	& $\gamma$\\
G\,300.971+1.143 & 12 34 54.34 & --61 39 57.1 &		&			&$<$3	&--43	&--46,--42		&	3.0 &$\gamma$\\ 
G\,301.136--0.225& 12 35 34.76 & --63 02 28.5 &		&		  	&t 		&	--47	&	--48,--41	&	17 \\
G\,301.136--0.226a& 12 35 34.93 & --63 02 34.5&--29	&--56,--18  	&80 		&	--29	&--31,--28		& 	78	& g\\
G\,301.136--0.226b& 12 35 34.84 & --63 02 31.1&		&		  	&t 		&	--45	&	--63,--43	&	92	& (omc)\\
G\,301.137--0.225& 12 35 35.21 & --63 02 30.6 &		&		  	&t 		&	--36	&	--40,--33	&	39	& omc\\
G\,305.191--0.006& 13 11 12.95 & --62 47 27.7&31    	&28,36      	&25  		& 	32	&29,35		&	4.8  	& g   \\
G\,305.198+0.007 & 13 11 16.38 & --62 46 37.4 &--39  	&--42,--27  	& 8 		&     --35	&--41,--24		&	3.5 	& \\
G\,305.208+0.207 & 13 11 13.37 & --62 34 40.0 &--39	&--43,--37  	&300	&     	--39	&--44,--38		& 	235	& om \\
G\,305.361+0.150 & 13 12 35.61 & --62 37 18.9 &--43  	&--45,--28  	&250	&	--36	& --44,--29		&	126  & omg   \\
G\,305.799--0.245& 13 16 42.92 & --62 58 31.7 &--26  	&--45,35	   	&400	&	--34	&	--119,37	&	266	& omg \\
G\,306.318--0.331& 13 21 20.87 & --63 00 22.7 &--19	&--24,--14  	&2.1  		& 	--18	&	--23,--15	&	0.7     \\
G\,307.805--0.456& 13 34 27.32 & --62 55 12.4 &--13  	&	--15,--7   	&1.4		&	--7	&	--11,--7	&	0.6	& og\\
G\,308.754+0.549 & 13 40 57.47 & --61 45 42.4 &--49  	&--50,--48  	& 3.4 		&	--49	&	--50,--48	&	0.7	& omg	\\
G\,308.918+0.124 & 13 43 01.64 & --62 08 48.9 &--56	&--66,--45  	&0.7		& --61	&	--62,--49	&	1.5	& om$^{\#}$\\
G\,309.384--0.135& 13 47 24.01 & --62 18 11.4 &--50	&--57,--41  	&	3.5	&--50	&	--51,--48	&	2.2	& omg\\
G\,310.144+0.760 & 13 51 58.53 & --61 15 40.6 &--58  	&--60,--55  	&2.5		&--58	& --59,--57     	&	1.6	& omg \\
G\,310.146+0.760 & 13 51 59.61 & --61 15 39.7 &--63  	&--64,--62  	&8  		&      		&			&$<$0.2 \\
G\,311.643--0.380& 14 06 38.77 & --61 58 22.7 &35    	&18,50      	&320	&36		&8,56		&	167	& omcg \\
G\,312.106+0.278 & 14 08 46.06 & --61 12 30.1 &		&			&	--	&--54	&--55,--53		& 0.6		& g	\\
G\,312.109+0.262 & 14 08 49.45 & --61 13 23.4 &		&			&	--	&--48	&--48,--47		&     0.43	& mg\\
G\,312.596+0.045 & 14 13 14.13 & --61 16 57.6 &	--61	& --62,--58	&0.6		&--59	&   --64,--56	&  1.5	& m\\
G\,312.599+0.046 & 14 13 15.19 & --61 16 51.9 &--75  	&--102,--58 	&2.7		& --79	& --96,--56		&	9	& om\\
G\,313.457+0.193 & 14 19 35.05 & --60 51 54.2 & 	--1	& --2,0		& 1.9		&45 		&   --1,46		&1.0        	& c \\
G\,313.470+0.191 & 14 19 40.97 & --60 51 46.2 & --5 	&--10,--2   	&6  		&--6		&--10,--1		& 4.5 	& omg\\

  \hline
\label{tab:masers}   
\end{tabular}
\end{table*}

\begin{table*}\addtocounter{table}{-1}
 \caption{-- {\emph {continued}}}
  \begin{tabular}{lccrcrrcrl} \hline
    \multicolumn{1}{c}{\bf Water maser} & {\bf RA}  & {\bf Dec} & {\bf Vpeak} &{\bf Vrange} & {\bf Speak} & {\bf Vpeak} &{\bf Vrange} & {\bf Speak} & {\bf Associations}\\
   \multicolumn{1}{c}{\bf ($l,b$)} & {\bf (J2000)}  & {\bf (J2000)}&  {\bf (\kmsns)}&  {\bf (\kmsns)} & {\bf (Jy)} &{\bf (\kmsns)}&  {\bf (\kmsns)} & {\bf (Jy)}     \\	
      \multicolumn{1}{c}{\bf (degrees)} & {\bf (h m s)}&{\bf ($^{o}$ $'$ $``$)}& {\bf 2003} & {\bf 2003} & {\bf 2003}& {\bf 2004} & {\bf 2004} & {\bf 2004} \\ \hline \\
G\,313.578+0.325 & 14 20 08.63 & --60 41 59.0 &--46	&--58,--32   	&55 		&--47	&--54,--32		&	46	& omg \\
G\,313.767--0.862& 14 25 01.71 & --61 44 57.2 & --54 	&--56,--48  	&160	&--54	&	--57,--36	&	162 	& omg\\
G\,314.320+0.112 & 14 26 26.37 & --60 38 29.4 & --44 	&--70,--40  	&34 		& --45	&--70,--40		&16		& om \\
G\,316.360--0.361& 14 43 11.07 & --60 17 10.3 & 3    	&--4,21     		& 30		&3		& --7,20		&16 \\ 
G\,316.361--0.363& 14 43 12.22 & --60 17 16.6 &--3	& --4,--1		& 6		&--3 		&--7,--1		&2.6		& \\ 
G\,316.412--0.308& 14 43 23.22 & --60 12 58.8 & --22 	&--30,6     		&6  		&--20	&--24,5		& 15		& omcg \\
G\,316.640--0.087& 14 44 18.39 & --59 55 10.7 & --19 	& --104,--2 	&12 		& --15	&--40,92		& 12		& omg\\
G\,316.763--0.011& 14 44 56.32 & --59 47 59.8 & --47 	&--51,--33  	&60 		& --48	&--49,--33		&4.1		& og\\
G\,316.812--0.057& 14 45 26.58 & --59 49 14.1 & --46 	&--47,--36  	&500	& --46	&--56,--11		&408 	& om\\
G\,317.429--0.561& 14 51 37.72 & --60 00 18.2 & 16 	& 12,25     	&0.5		&25		& 24,25		& 0.27	&  oc \\
G\,317.429--0.556& 14 51 36.75 & --60 00 03.4 &  26	&	25,36	&0.5		& 28	 	& 27,29		& 0.6		& 	\\
G\,318.044--1.404& 14 59 08.61 & --60 28 23.9 & 42   	& 31,44     	&3.5		& 42		& 32,42		& 3.5		& om$\gamma$ 	\\
G\,318.050+0.087 & 14 53 42.62 & --59 08 52.3 &--55  	&--61,--39  	&470	&--48	&--69,--38		& 50		& omg	\\	
G\,318.948--0.196a& 15 00 55.18 & --58 58 51.6 &--41	&--44,--28  	&5  		&	--36	&--44,--27		&	9	& g	\\
G\,318.948--0.196b& 15 00 55.33 & --58 58 53.6&		&			& 	t	&--38	& --39,--21       	&	7	& omg \\
G\,319.399--0.012& 15 03 17.50 & --58 36 11.4  &--4.5 	&--20,1     		&10 		& --5		& --7,2		&	3.8	& oc	\\
G\,319.836--0.196& 15 06 54.54 & --58 32 58.6  &--13	&--23,0     		&9  		& --11	& --19,0		&	2.8	& omg	\\
G\,320.120--0.440& 15 09 43.83 & --58 37 06.3  &--46  	&--70,--40  	&1.2		& --46	&--158,30		&	0.9	& o	\\
G\,320.221--0.281& 15 09 47.00 & --58 25 47.6 &		&			&$<$0.4	&--73	& --75,--72	&	0.7	&	\\
G\,320.232--0.284& 15 09 51.92 & --58 25 38.0 &--63  	&--70,--61  	&9  		& --67	&--81,--72		&	2	& om \\ 
G\,320.233--0.284& 15 09 52.50 & --58 25 35.7 &--60	& --61,--58	& 3.5		&--60	& --61,--54         &   	3.4	& c	\\
G\,320.255--0.305& 15 10 06.14 & --58 26 00.8 &		&			&	--	&--126 	&  --144,--111	&	45	& 	\\
G\,320.285--0.308& 15 10 18.88 & --58 25 16.5 &		&			&	--	&--69	&  --81,--56	&	15	&g	\\
G\,321.028--0.484& 15 15 50.90 & --58 11 19.7 & --60 	&--70,--55  	&2.5		& --58	& --70,--52	&	5	&	\\
G\,321.033--0.483 & 15 15 52.60 & --58 11 07.2 & 	--60	&--68,--58		&3.6		& --61	& --64,--49	&	0.25	& m	 \\
G\,321.148--0.529& 15 16 48.25 & --58 09 50.1 &--64  	&--66,--62  	&0.8		& --97	& --98,--61	&	1.6	& omg	\\
G\,322.158+0.636 & 15 18 34.52 & --56 38 24.7 &--73  	&--81,--61  	&5  		& --76	& --86,--65	&	2.7	& om \\
G\,322.165+0.625 & 15 18 39.74 & --56 38 46.7 & --40 	&--67,--36  	&9  		& --39	& --67,--36	&	2.8	&g	\\
G\,323.740--0.263& 15 31 45.48 & --56 30 49.6 & --50 	&--72,--46  	&140	& --50	& --88,--42	&	70	& omg \\
G\,324.201+0.122 & 15 32 52.76 & --55 56 04.9 & --87 	&--100,--47 	&50 		& --87	& --100,--48	& 	14	&o	\\
G\,324.716+0.342 & 15 34 57.41 & --55 27 22.3 & --55 	&--72,--47  	&10 		&--58	& --81,--30	& 26		& omg \\
G\,326.662+0.521 & 15 45 02.73 & --54 09 03.3 &--39  	&--50,--34  	&256	& 		&			&	--	&m	\\
G\,326.665+0.553 & 15 44 55.82 & --54 07 25.6 &		&			&	t	&--42 	&--127,--39	&	16	&g	\\ 
G\,326.670+0.554 & 15 44 57.03 & --54 07 10.6 &--42  	&--44,--40  	&26 		& --40	& --48,8		&	101	& o	\\
G\,326.780--0.241& 15 48 55.10 & --54 40 38.6 &--64  	&--66,--60  	&36 		&--66	&--92,--52		&	18	& og	\\
G\,326.859--0.676& 15 51 13.82 & --54 58 03.6 &		&			& --		&--103 	&  --104,--103	& 	0.42	& mg	\\
G\,327.119+0.511 & 15 47 32.56 & --53 52 39.3 &--87  	&--90,--80  	&25 		& --88	&--89,--57		&	19	& omg	\\
G\,327.291--0.578& 15 53 07.65 & --54 37 07.2 &--56	&--80,--39  	&400	& --63	&--84,--36		&	668	& omg	\\
G\,327.391+0.200 & 15 50 18.31 & --53 57 06.1 &		&			&	--	&--86	&--92,--86		& 0.40	& mg 	\\
G\,327.402+0.445 & 15 49 19.32 & --53 45 13.8 &--80  	&--83,--68  	&230	& --81	& --84,--69	& 195	& o$^{\#}$mcg \\
G\,327.581--0.077& 15 52 29.50 & --54 02 51.6 &		&			&	--	&--101  	& --102,--95	& 1.0	         &   	\\
G\,327.594--0.095& 15 52 38.19 & --54 03 11.5 &		&			&	--	& --99 	& --102,--91	& 0.8		&g	\\
G\,327.619--0.111& 15 52 50.31 & --54 03 00.0 &		&			&	--	&--85 	& --85,--84	& 0.20	& mg 	\\
G\,327.935--0.123& 15 54 33.09 & --53 51 29.1 &		&			&	--	&--98	& --100,--76	&2.1		&	\\
G\,328.236--0.548& 15 57 58.21 & --53 59 25.4 &--38  	&--39,--37 	&30  		&--38    	& --40,10		&20		&omcg	\\
G\,328.254--0.532& 15 57 59.69 & --53 58 00.7 &--50 	&--53,--48 	&200 	& --50	& --51,--48	& 155	& omg	\\
G\,328.306+0.432 & 15 54 05.91 & --53 11 37.4 &--96  	&--97,--87		&40  		& --93	& --96,--87	&	141	& c	\\
G\,328.808+0.633 & 15 55 48.23 & --52 43 05.2 &--46  	&--47,--44 	& 10 		& --46	& --48,--44	&	4.4	& omcg	\\
G\,329.021--0.186& 16 00 24.32 & --53 12 16.9 &--42  	&--43,--41 	&2.4   		&    --44	& --44,--43	&0.34	&g 	\\
G\,329.029--0.199& 16 00 30.22 & --53 12 34.3 &--38  	&--40,--37 	&1.6 		&		&			&$<$0.2	& og	\\
G\,329.030--0.205& 16 00 31.90 & --53 12 48.7 &--39  	&--54,--34 	&8   		&     --46 	& --52,--35	& 6		& om	\\
G\,329.031--0.198& 16 00 30.34 & --53 12 26.5 &--39  	&--54,33   		&5   		&  --52	& --65,33		& 1.2		& omg \\
G\,329.066--0.307& 16 01 09.89 & --53 16 01.5 &--48  	&--50,--47 	&1.4 		& --45	& --46,--45	& 0.6		& omg	\\	
G\,329.183--0.313& 16 01 46.90 & --53 11 41.7 &--51	&--66,--36 	&24  		& --50	&	--60,--39	& 34		& omg	\\
G\,329.342+0.130 & 16 00 38.87 & --52 45 22.9 &	--112	&	--115,--95		& 1.1		& 	--112 	& --114,--100	&	2.2	\\
G\,329.404--0.459& 16 03 31.81 & --53 09 30.8 &--113 	&--117,--106	&3.1 		&     --113 	& --116,--111	&   2.4	& 	\\
G\,329.405--0.459& 16 03 32.15 & --53 09 29.0 &--78  	&--80,--60 	&15  		&	--77	&	--79,--44	& 10		& om	\\
G\,329.407--0.459& 16 03 32.77 & --53 09 25.0 &--74  	&--76,--72 	& 80 		&      		&                      	&  $<$0.2	&  mg\\
G\,329.421--0.167& 16 02 19.85 & --52 55 41.8 &		&			&$<$0.8	& --77	&  --78,--75        & 	2.4	&	\\	
G\,329.424--0.164& 16 02 20.03 & --52 55 25.9 &--78  	&--83,--60 	&0.8 		&                 &      		&$<$0.2	&g	\\

 \hline

\end{tabular}
\end{table*}

\begin{table*}\addtocounter{table}{-1}
 \caption{-- {\emph {continued}}}
  \begin{tabular}{lccrcrrcrl} \hline
    \multicolumn{1}{c}{\bf Water maser} & {\bf RA}  & {\bf Dec} & {\bf Vpeak} &{\bf Vrange} & {\bf Speak} & {\bf Vpeak} &{\bf Vrange} & {\bf Speak} & {\bf Associations}\\
   \multicolumn{1}{c}{\bf ($l,b$)} & {\bf (J2000)}  & {\bf (J2000)}&  {\bf (\kmsns)}&  {\bf (\kmsns)} & {\bf (Jy)} &{\bf (\kmsns)}&  {\bf (\kmsns)} & {\bf (Jy)}     \\	
      \multicolumn{1}{c}{\bf (degrees)} & {\bf (h m s)}&{\bf ($^{o}$ $'$ $``$)}& {\bf 2003} & {\bf 2003} & {\bf 2003}& {\bf 2004} & {\bf 2004} & {\bf 2004} \\ \hline \\
G\,329.426--0.161& 16 02 19.71 & --52 55 12.8 &--73	&--78,--71 	&8   		&   --73	& --74,--72	&	10	&	\\
G\,329.457+0.503 & 15 59 36.93 & --52 23 53.6 &		&			&	--	&--66	& --68,--65	&	4.3	&	\\	
G\,329.622+0.138 & 16 02 00.28 & --52 33 57.7 &		&			&	--	&--82	& --110,--66	&	30	& mg	\\
G\,330.070+1.064 & 16 00 15.56 & --51 34 25.7 &		&			&	--	&--50	&  --75,--45        &	11	& m$\gamma$	\\	
G\,330.879--0.367& 16 10 20.04 & --52 06 06.8 &--64  	&--72,--28 	& 90 		& --60	& --72,--25	& 95		& omc \\
G\,330.954--0.182& 16 09 52.65 & --51 54 54.6 &--80  	&--150,70  	&240 	& --91	& --191,56	& 323	& ocg	\\
G\,331.132--0.244& 16 10 59.73 & --51 50 22.5 &--99	&--102,--73	&280 	& --99	& --118,--55	& 47		& omg	\\
G\,331.278--0.188& 16 11 26.51 & --51 41 55.8 &--86  	&--104,--79	&55  		& --90	& --118,--64	& 42		& omg	\\
G\,331.342--0.346& 16 12 26.49 & --51 46 14.9 &--60  	&--62,--59 	&1.8 		& --62	& --64,--60	& 11		& omg \\
G\,331.418+0.252 & 16 10 10.56 & --51 16 52.2 &		&			&	--	&--71	&--71,--70  	& 0.6		& g	\\
G\,331.442--0.187& 16 12 12.46 & --51 35 09.3 &--88	&--93,--72 	&70  		& --88	&--113,--80	& 212	&	omcg \\
G\,331.512--0.103& 16 12 10.01 & --51 28 36.7 &--89  	&--162,--33	&700 	& --90	& --159,--32	& 534	&	ocg \\
G\,331.555--0.122& 16 12 27.20 & --51 27 42.6 &--99  	&--170,--96	&20  		& --99	& --166,--86	& 9		&	\\
G\,332.094--0.421& 16 16 16.68 & --51 18 26.2 &		&			&	--	& --59	&   --59,--58	&   0.29	&	m \\
G\,332.296--0.094& 16 15 45.84 & --50 55 52.7 &		&			&	--	&--50	& --71,--43	& 6		&	m\\	
G\,332.349--0.433& 16 17 30.03 & --51 08 16.6 &		&			&	--	& --67	& --68,--67	& 2.9		&	\\
G\,332.352--0.117& 16 16 07.10 & --50 54 31.0 &--45  	&--48,--41 	&2.0   		& --60	&--61,--60		& 0.5		&	om \\
G\,332.604--0.167& 16 17 29.45 & --50 46 11.7 &		&			&	--	&--46	& --48,--45	& 2.4		&	mg \\
G\,332.725--0.621& 16 20 02.91 & --51 00 33.1	&--42  	&--43,--41 	&0.6 		& --58	& --59,--56	& 5		&	omg \\
G\,332.826--0.549& 16 20 11.17 & --50 53 14.6 &--56	&--72,--30 	&45  		& --59	& --71,--35	& 70		&	mc \\
G\,332.964--0.679& 16 21 23.03 & --50 52 57.3 &		&			&	--	&--52	&  --52,--50	& 2.9		&	mg \\
G\,333.030--0.063& 16 18 56.86 & --50 23 53.6 &		&			&	--	&--40	& --153,--40	& 3.4		&	mc\\
G\,333.055--0.436& 16 20 42.47 & --50 38 46.4 &		&			&	--	&--47	& --57,--48	&0.39	&	\\
G\,333.114--0.439& 16 20 59.30 & --50 36 21.9 &	--62	&	--63,--60	& 4.6  	&--62	& --62,--53	& 2.0		&	\\
G\,333.121--0.434& 16 20 59.70 & --50 35 50.8 &--57  	&--59,12   		&38 		&  --47	& --91,--33	& 21		&	m\\
G\,333.126--0.440& 16 21 02.69 & --50 35 54.1 &--50  	&--70,--48 	&19  		&   --52	& --72,--47	& 12		&	m \\
G\,333.128--0.440& 16 21 03.18 & --50 35 51.8 &		&			&$<$0.2  	&--124	& --125,--124	& 0.79	&	m\\
G\,333.130--0.425& 16 20 59.75 & --50 35 05.1 & --39 	&--67,--38 	&23  		& --64	& --65,--31	&1.9		&	\\
G\,333.132--0.560& 16 21 36.46 & --50 40 45.4 &		&			&	--	&--53	&   --67,--46	& 1.9		&	\\
G\,333.219--0.062& 16 19 47.40 & --50 15 53.8 &	 	&          		&  $<$0.3	& --13	& --14,84		&   0.5	& g	\\
G\,333.234--0.060& 16 19 50.85 & --50 15 09.7 &--88  	&--102,--83	&140 	&--88	& --102,82	& 117	&	og \\
G\,333.315+0.106 & 16 19 28.75 & --50 04 39.7 &--48	&--68,--41 	&2.2 		& --48	& --60,--48	& 6		&	omg \\
G\,333.387+0.032 & 16 20 07.52 & --50 04 47.4 &--61	&--63,--60 	&0.4 		&  --61	& --61,--60	& 0.14	&	omg \\
G\,333.467--0.164& 16 21 20.20 & --50 09 46.1 &--44  	&--46,--40 	&3.5 		& --42	& --47,--40	& 3.2		&	om\\
G\,333.608--0.215& 16 22 11.08 & --50 05 56.3 &--51	&--76,--45 	&50  		& --49	& --83,--41	& 24		&	o\\
G\,333.646+0.058 & 16 21 09.12 & --49 52 45.1 &		&			&	--	&--89	&   --90,--84	& 3.3		&	m \\
G\,333.682--0.436& 16 23 29.67 & --50 12 07.4 &		&			& 	--	&--3		& --3,--2		& 0.24	&	mg \\
G\,333.930--0.134& 16 23 14.68 & --49 48 48.8 &		&			&	--	&--46	& --50,--45	&  0.18	&	m \\
G\,334.635--0.015& 16 25 45.83 & --49 13 37.0 &		&			&	--	& --26	& --29,--15	& 49		&	mg \\
G\,334.935--0.098& 16 27 24.22 & --49 04 11.0 &		&			&	--	&--17	& --18,--14	&   1.0	&	mg\\
G\,334.951--0.092& 16 27 26.96 & --49 03 14.7 &		&			&	--	&--21	& --25,--20	&  1.3	&	\\
G\,335.059--0.428& 16 29 23.20& --49 12 31.3 & 		&			&   t		& --38	& --39,--37	& 1.7		& 	\\ 
G\,335.060--0.428& 16 29 23.24 & --49 12 28.0 &--46  	&--50,--37 	&12  		&--37	& --44,15		& 3.0		&	omg\\
G\,335.070--0.423& 16 29 24.72 & --49 11 47.7 &--88	& --109,--84	&  1.0	&--90	& --105,--84	& 5		& g	\\
G\,335.585--0.285& 16 30 57.34 & --48 43 39.4 &--45  	&--50,--40 	&30  		& --42	& --49,--32	&25		&	omg \\
G\,335.586--0.290& 16 30 58.73 & --48 43 51.2 &--48  	&--61,--33 	& 5  		& --56     	& --57,--42	& 20		&	omg \\
G\,335.588--0.264& 16 30 52.52 & --48 42 39.5 &--51  	&--56,--48 	& 16  	&      		&                      	&  $<$0.2	&g	\\   
G\,335.727+0.191 & 16 29 27.52 & --48 17 51.9 &		&			&	--	&--51	& --51,--42	& 13		&	m\\
G\,335.787+0.177 & 16 29 46.18 & --48 15 49.1 &--55  	&--56,--48 	& 3 .0	&   --49   	& --59,--45	&	10	&	\\
G\,335.789+0.174 & 16 29 47.33 & --48 15 50.8 &--46	&--51,--45 	& 3.0  	&		&			& $<$0.2	&	omg \\
G\,335.789+0.183 & 16 29 45.10 & --48 15 30.4 &--91  	&--112,--89	&4.2 		&		&			& $<$0.2	&	\\
G\,336.018--0.827& 16 35 09.35 & --48 46 47.7 &--54  	&--59,--36 	&120 	&  --54	& --59,--36	&	82	&	omc\\
G\,336.352--0.149& 16 33 30.73 & --48 04 27.6 &	--79	& --81,--78	& 0.4		& --79	& --81,--78	&    0.43	&	\\
G\,336.359--0.137& 16 33 29.37 & --48 03 41.5 &--67	&--67,--66 	&0.5 		&		&			&$<$0.2	&	omc\\
G\,336.433--0.262& 16 34 20.31 & --48 05 30.5 &		&			&	--	&--89	& --90,--88	& 0.6		&	mg \\
G\,336.496--0.258& 16 34 34.52 & --48 02 34.3 &		&			&	--	&--25	& --38,--14	& 8		&	\\
G\,336.830--0.375& 16 36 26.19 & --47 52 29.5 &		&			&	--	&--20	& --45,--19	& 0.28	&	mg \\
G\,336.864+0.005 & 16 34 54.50 & --47 35 37.7 &--78	&--80,--64 	&3.0   	& --66	& --79,--65	& 2.2		&	om\\
G\,336.864--0.002& 16 34 56.02 & --47 35 55.3 &--73  	&--76,--71 	&4.5   	& --73      	&    --77,--71    	& 1.5		&	\\        
G\,336.870--0.003& 16 34 57.91 & --47 35 42.7 &	--77	& --78,--55	& 2.0		&--77	&   --78,--71	& 3.7		& 	\\
 \hline

\end{tabular}
\end{table*}

\begin{table*}\addtocounter{table}{-1}
 \caption{-- {\emph {continued}}}
  \begin{tabular}{lccrcrrcrl} \hline
    \multicolumn{1}{c}{\bf Water maser} & {\bf RA}  & {\bf Dec} & {\bf Vpeak} &{\bf Vrange} & {\bf Speak} & {\bf Vpeak} &{\bf Vrange} & {\bf Speak} & {\bf Associations}\\
   \multicolumn{1}{c}{\bf ($l,b$)} & {\bf (J2000)}  & {\bf (J2000)}&  {\bf (\kmsns)}&  {\bf (\kmsns)} & {\bf (Jy)} &{\bf (\kmsns)}&  {\bf (\kmsns)} & {\bf (Jy)}     \\	
      \multicolumn{1}{c}{\bf (degrees)} & {\bf (h m s)}&{\bf ($^{o}$ $'$ $``$)}& {\bf 2003} & {\bf 2003} & {\bf 2003}& {\bf 2004} & {\bf 2004} & {\bf 2004} \\ \hline \\
G\,336.983--0.183& 16 36 12.38 & --47 37 59.1 &	--76 	&	--77,--74 		& 0.4		& 45  	& --78,45		&  0.18	& o$^{\#}$mc	\\ 
G\,336.991--0.024& 16 35 32.53 & --47 31 12.4 & --48 	&--61,--44 	& 4  		& --49	& --54,--44	& 1.0		&	c\\
G\,336.994--0.027& 16 35 34.01 & --47 31 12.2 &--121 	&--137,--80	&160 	& --120	& --177,--47	& 158	&	omg \\ 
G\,336.995--0.024& 16 35 33.53 & --47 31 00.6 &		&			&$<$0.2	& --54	& --54,--48	& 1.1          &  g	\\
G\,337.258--0.101& 16 36 56.36 & --47 22 27.5 &--69  	&--71,--52 	&1.6 		& --69	& --69,--68	& 0.33	&	omg \\

G\,337.404--0.402& 16 38 50.57 & --47 28 00.8 &--40	&--53,--37 	&140 	&--40	& --49,--35	&	137	&	omcg \\
G\,337.612--0.060& 16 38 09.46 & --47 05 00.3 &--52  	&--101,--47	&38  		& --51	& --99,--46	& 17		&	omg \\
G\,337.687+0.137 & 16 37 35.60 & --46 53 46.3 &		&			&	--	&--74	& --151,--73	&      0.6	&	mg \\
G\,337.705--0.053& 16 38 29.72 & --47 00 35.7 &--48  	&--147,5   		&54  		& --49	& --159,--31	&	39	&	omcg \\
G\,337.916--0.477& 16 41 10.49 & --47 08 02.9 &--46	&--80,--28 	&400 	& --33	& --65,--27	& 321	&	o\\ 
G\,337.920--0.456& 16 41 06.14 & --47 07 02.3 &--40	&--42,--39 	&50  		& --40	& --69,--27	& 22		&	om \\
G\,337.994+0.133 & 16 38 48.63 & --46 40 15.7 &--114 	&--117,--110	&4.5		& --113	& --116,--111	& 4.5		&	\\
G\,337.998+0.137 & 16 38 48.53 & --46 39 56.4 &--39  	&--50,--32 	&30  		& --38	& --47,--30	& 27		&	omg \\
G\,338.069+0.011 & 16 39 37.98 & --46 41 44.9 &--37  	&--40,--21 	&1.3		&  --28	& --46,--22	& 3.9		&	g \\
G\,338.075+0.012 & 16 39 38.88 & --46 41 26.8 &--50  	&--51,--50 	&0.5		&		&			& $<$0.2	&	omc\\
G\,338.075+0.010 & 16 39 39.78 & --46 41 31.8 &--132 	&--139,--21	&1.2		& --48	& --51,--28	& 1.2		&	m\\
G\,338.077+0.019 & 16 39 37.69 & --46 41 05.6 &	--40	&	--41,--38	& 1.0 	& --40	&   --40,--39	& 1.3		&	g \\
G\,338.281+0.542 & 16 38 09.16 & --46 11 02.6 &--61 	&--68,--59  	&12  		& --64	& --67,--58	& 4.6		&	omg \\
G\,338.427+0.051 & 16 40 50.25 & --46 24 05.5 &		&			& --		&--30	&  --46,--29	&   0.48	&		\\
G\,338.430+0.053 & 16 40 50.50 & --46 23 53.0 &		&			&	--	&--44	&   --51,--29	& 3.2		&	g \\
G\,338.433+0.057 & 16 40 50.08 & --46 23 33.8 &		&			&	--	&--29	& --30,--29	&  0.19	&	m\\
G\,338.435+0.055 & 16 40 51.23 & --46 23 34.5 &		&			&	--	&--31	& --31,--30	&   0.58	&	\\
G\,338.436+0.057 & 16 40 51.05 & --46 23 26.8 &		&			&	--	&--35	&  --56,--34	&    0.30	&	\\
G\,338.440+0.064 & 16 40 49.87 & --46 22 59.5 &		&			&	--	&--81	& --81,--80	& 0.6		&	g \\
G\,338.461--0.245& 16 42 15.57 & --46 34 18.6 &--53 	&--119,--51 	&9   		& --52	& --114,--49	& 8		& 	omg \\
G\,338.462--0.259& 16 42 19.49 & --46 34 48.3 &		&			& $<$0.2	&--54	&--55,--53		& 	0.35	&	\\
G\,338.472+0.289 & 16 39 58.99 & --46 12 36.2 &--54 	&--62,--22  	&25  		& --29	& --62,--25	& 10		&	omg \\
G\,338.562+0.217 & 16 40 38.12 & --46 11 24.8 &		&			&	--	&--39	& --40,--39	&  0.22	&	mg \\
G\,338.567+0.110 & 16 41 07.16 & --46 15 28.1 &		&			&	--	&--76	& --92,--74	& 3.2		&	m \\
G\,338.682--0.084& 16 42 24.12 & --46 17 59.1 &--6  	&--18,--5   	&2.2 		& --16	& --18,--6		& 0.7		&	ocg \\
G\,338.920+0.550 & 16 40 34.02 & --45 42 07.9& 	--110&	--125,--65	& 17		& --68	& --127,--64	& 5.3		&	mg \\
G\,338.925+0.556 & 16 40 33.63 & --45 41 37.9 &--63 	&--85,--61  	&160 	& --62	&--86,--6		& 116	&	om\\
G\,338.924--0.060& 16 43 13.25 & --46 06 04.2 &		&			&	--	&--64	& --69,--63	& 2.3		&	\\
G\,339.582--0.127& 16 45 58.88 & --45 38 47.4 &		&			&	--	&--28	& --32,--26	& 0.27	&	mg \\
G\,339.584--0.128& 16 45 59.63 & --45 38 44.3 &		&			&	--	&--40	& --41,--26	& 1.8		&	g \\
G\,339.585--0.126& 16 45 59.18 & --45 38 36.7 &		&			&	--	&--41	& --75,--39	&  0.40	&	g \\
G\,339.586--0.128& 16 45 59.92 & --45 38 39.7 &		&			&	--	&--106	& --112,--105	&  0.5	&	g \\
G\,339.609--0.115& 16 46 01.57 & --45 37 07.3 &		&			& $<$0.2	&--71	&  --82,--67	& 3.0		&	\\
G\,339.622--0.121& 16 46 06.03 & --45 36 44.5 &--34 	&--38,--32  	&90  		&--33	& --36,--32	& 43		&	om \\
G\,339.762+0.055 & 16 45 51.56 & --45 23 31.0 &		&			&	--	&--57	&  --58,--57	&   0.25	&	mg \\
G\,339.884--1.259& 16 52 04.71 & --46 08 33.6 &--32 	&--50,--28  	&50  		& --51	& --52,--24	& 4.3		&	om$\gamma$\\
G\,340.054--0.243& 16 48 13.82 & --45 21 43.9 &--54 	&--58,--44  	&35  		&      		&			&	--	&	omg \\
G\,340.785--0.096& 16 50 14.84 & --44 42 24.7 &--120	&--121,--111	&1.0   		&		&			& $<$0.2	&	omg \\
G\,341.218--0.212& 16 52 17.92 & --44 26 51.6 &--43 	&--47,--38  	&120 	& --39	& --49,--23	& 33		&	omg \\
G\,341.276+0.062 & 16 51 19.50 & --44 13 44.0 &--77 	&--80,--58  	&5   		& --64	& --83,--62	& 10		&	omg \\
G\,342.484+0.183 & 16 55 02.39 & --43 12 59.3 &		&          		&	--	&--43	& --44,--36	& 0.6		&	mg \\
G\,343.126--0.065& 16 58 17.54 & --42 52 15.8&	--16	& --17,--15	& 2.5		&--21	& --23,--20	& 8		&	\\
G\,343.127--0.063& 16 58 17.29 & --42 52 06.6 &--35 	&--45,--20  	&250         & --30	& --46,--16	& 208	&	o\\	
G\,344.226--0.576& 17 04 09.36 & --42 18 58.0 &		&			&$<$0.2	&--19	&  --19,--18	&   0.33	&	\\
G\,344.228--0.569& 17 04 07.85 & --42 18 38.9 & --24	&--28,31    	&8   		& --25	& --52,--9		&	1.4	&	omg \\
G\,344.421+0.046 & 17 02 08.53 & --41 46 56.4 &--27 	&--32,--18  	&1.8 		& --26	& --31,--24	& 0.9		&	\\
G\,344.582--0.024& 17 02 57.94 & --41 41 54.1 &--4  	&--20,5     		&250 	& --4		& --52,6		&	127	&	omcg \\
G\,345.004--0.224& 17 05 11.12 & --41 29 04.1 &--23	&--46,16   		&3.0   		& 15		& --89,15		&	3.7	&	om$^{\#}$cg \\
G\,345.010+1.793 & 16 56 47.51 & --40 14 23.9 &--17	& --19,10    	&2.0   		&		&			& $<$0.2	&	omc$\gamma$\\
G\,345.010+1.802 & 16 56 45.41 & --40 14 04.2 &--28	&--33,--26   	&7   		& --25	& --31,--18	& 23		&	$\gamma$\\
G\,345.012+1.797 & 16 56 47.01 & --40 14 08.5 &--12	&--13,7      	& 50 		& --12	& --12,7		& 29		&	m$\gamma$\\
G\,345.397--0.950& 17 09 33.08 & --41 36 20.4 &	--21	& --22,--20	&3.1		&--21	&  --27,--20	& 6		&	\\
G\,345.402--0.948& 17 09 33.62 & --41 36 02.9 &--26	&--32,--21   	&8   		& --23	& --32,--21	& 2.9		&	\\
G\,345.405--0.947& 17 09 33.83 & --41 35 50.8 &		&			& $<$0.5	&--28	&  --28,--23	&   0.5	&	\\
G\,345.406--0.942& 17 09 32.71 & --41 35 37.6 &--15	&--19,--12   	&3.0 		&  --18	& --20,--17	& 1.9		&	\\

\hline
\end{tabular}
\end{table*}

\begin{table*}\addtocounter{table}{-1}
 \caption{-- {\emph {continued}}}
  \begin{tabular}{lccrcrrcrl} \hline
    \multicolumn{1}{c}{\bf Water maser} & {\bf RA}  & {\bf Dec} & {\bf Vpeak} &{\bf Vrange} & {\bf Speak} & {\bf Vpeak} &{\bf Vrange} & {\bf Speak} & {\bf Associations}\\
   \multicolumn{1}{c}{\bf ($l,b$)} & {\bf (J2000)}  & {\bf (J2000)}&  {\bf (\kmsns)}&  {\bf (\kmsns)} & {\bf (Jy)} &{\bf (\kmsns)}&  {\bf (\kmsns)} & {\bf (Jy)}     \\	
      \multicolumn{1}{c}{\bf (degrees)} & {\bf (h m s)}&{\bf ($^{o}$ $'$ $``$)}& {\bf 2003} & {\bf 2003} & {\bf 2003}& {\bf 2004} & {\bf 2004} & {\bf 2004} \\ \hline\\
G\,345.408--0.953& 17 09 35.85 & --41 35 56.5 &		&			&$<$0.5	&--15	& --16,36		&    0.5	&	o$^{\#}$m$^{\#}$c\\
G\,345.412--0.955& 17 09 37.08 & --41 35 48.9 &		&			&$<$0.2	&--55	& --55,--54	& 1.6		&	\\
G\,345.425--0.951& 17 09 38.64 & --41 35 03.3 & --13	&	--15,--12	&1.5		&--13	& --16,--13	& 1.1		&	m \\
G\,345.438--0.074& 17 05 56.75 & --41 02 54.9 &--11	&--32,--9    	& 40 		& --12	& --37,--8		& 15		&	o\\
G\,345.482+0.309 & 17 04 28.41 & --40 46 52.1 &--52	&--80,--24   	&15  		& --55	& --82,--51	& 1.8		&	 \\

G\,345.487+0.314 & 17 04 28.19 & --40 46 28.6 &--16	&--24,2      	&6   		& --13	& --39,--12	& 0.7		&	m \\
G\,345.493+1.469 & 16 59 41.47 & --40 03 46.2 &5   		&3,6         		&6   		&		&			& $<$0.2	&	o$\gamma$\\
G\,345.494+1.470 & 16 59 41.15 & --40 03 39.9 &	1	&	--16,2	&  1.0	& 0	 	& --18,--4		& 1.8		&	$\gamma$\\
G\,345.495+1.473 & 16 59 40.78 & --40 03 28.9 &--60  	&--62,--16 	&4.0   		& --9		& --10,--8		& 1.5		&	$\gamma$\\
G\,345.505+0.348 & 17 04 23.02 & --40 44 23.5 &--3   	&--43,1    		&22  		&--4		& --42,4		& 4.5		&	om \\ 
G\,345.505+0.343 & 17 04 24.35 & --40 44 32.1 &		&			& $<$0.2	& --68	& --70,--67	& 0.8		&	\\
G\,345.699--0.090& 17 06 50.72 & --40 50 58.9 &--10	&--87,102    	&200 	& --5		& --92,141	& 216	&	og \\
G\,346.480+0.132 & 17 08 22.67 & --40 05 26.9 &--10	&--12,--8      	&1.2 		&		&			& $<$0.2	&	omg \\

G\,346.522+0.085 & 17 08 42.35 & --40 05 08.1 & 		&			&	--	&4		& --2,14		& 2.9		&	m \\
G\,346.529+0.106 & 17 08 38.12 & --40 04 03.9 &		&			&	--	& 4		& --1,5		& 1.6		&	\\
G\,347.588+0.213 & 17 11 27.61 & --39 09 08.1 &		&			&	--	&--93	& --94,--93	& 0.7		&	\\
G\,347.623+0.148 & 17 11 50.09 & --39 09 45.8 &--118	&--120,--117	&0.5 		& --118	& --122,--116	& 1.0		&	\\
G\,347.628+0.149 & 17 11 50.85 & --39 09 29.6 &--125	&--133,--122 	&0.5 		&		&			& $<$0.2	&	omg \\
G\,347.632+0.210 & 17 11 36.30 & --39 07 06.5 &		&			&	--	& --88	& --95,--24	&	6	&	mc\\
G\,348.533--0.974& 17 19 16.20 & --39 04 30.6 &--56 	&--59,--10  	&3.2 		& --31	& --93,24		& 	1.8	&	\\
G\,348.534--0.983& 17 19 18.64 & --39 04 47.3 &--21 	&--30,--11  	&2.2 		& --14	& --109,--12	& 0.9		&	\\
G\,348.551--0.979& 17 19 20.61 & --39 03 49.2 &--30 	&--32,--15  	&1.4 		& --18	& --18,--17	& 	0.28	&	mg \\
G\,348.726--1.038& 17 20 06.42 & --38 57 13.2 &--11 	&--64,42    	&115 	& --10	& --77,61		&	162	&	$\gamma$\\
G\,348.885+0.096 & 17 15 50.25 & --38 10 12.3 &--81 	&--84,--77  	&8   		& --80	& --87,--77	&	8	&	omg \\
G\,348.892--0.180& 17 17 00.26 & --38 19 27.9 & 7   	&6,12       		&2.0   		&		&			&	--	&	omg \\
G\,349.052+0.002 & 17 16 43.25 & --38 05 18.2 &15   	&14,17      	&3.0 		&      		&                    	&    $<$0.2&	 \\
G\,349.067--0.018& 17 16 50.82 & --38 05 13.8 & 5   	&3,7        		&1.6 		& 13		& --14,19		&	1.0	&	omg \\
G\,349.074--0.015& 17 16 51.23 & --38 04 49.3 &--27 	&--32,--20  	&1.6		& --26	& --32,--24	& 1.2		&	g \\
G\,349.068+0.110 & 17 16 19.24 & --38 00 48.3 &--25 	&--32,--22  	&3.0 		& --21	& --22,--20	&	1.4	&	\\
G\,349.092+0.105 & 17 16 24.66 & --37 59 45.4 &--80 	&--84,--74  	&43  		& --80	&   --84,--72	& 154	&	omg \\
G\,350.015+0.433 & 17 17 45.43 & --37 03 12.9 &--35 	&--50,--26   	&4.7 		&		&			&	--	&	omg \\
G\,350.098+0.099 & 17 19 21.82 & --37 10 41.2 &--44 	&--45,--42  	&1.8 		&      		&                      	&  $<$0.2	&	 \\
G\,350.098+0.080 & 17 19 26.67 & --37 11 20.8 &--66 	&--68,--65  	&6 		& --67	& --69,--67	& 2.6		&	g \\
G\,350.100+0.081 & 17 19 26.74 & --37 11 09.8 &	--68	&	--69,--67	&4.0		& --68	& --69,--63	&	5	&	\\
G\,350.105+0.084 & 17 19 26.78 & --37 10 51.5 &--71 	&--102,--38 	&6 		& --71	& --74,--29	& 14		&	m\\
G\,350.110+0.087 & 17 19 26.96 & --37 10 29.0 &--72 	&--74,--71  	&8   		& --72	& --73,--70	& 14		&	g \\
G\,350.112+0.089 & 17 19 26.96 & --37 10 19.9 &--172	&--175,--110	&2.0		& --128	& --164,--106	& 2.9		&	\\
G\,350.113+0.095 & 17 19 25.69 & --37 10 04.8 &--66 	&--70,--64  	&2.5 		& --64	& --79,--63	&	1.4	&	og \\
G\,350.274+0.120 & 17 19 47.07 & --37 01 16.6 &		&			&	--	&--63	&  --63,--61	& 2.9		&	 \\
G\,350.299+0.122 & 17 19 50.89 & --37 00 00.6 &		&			& 	--	&--68	& --68,--67	& 0.17	&	m\\
G\,350.330+0.100 & 17 20 01.79 & --36 59 14.3 &--68 	&--69,--60  	&1.0		& --62	& --68,--48	& 1.9		&	o\\
G\,350.341+0.140 & 17 19 53.60 & --36 57 19.0&		&			& --		&--105	& --105,--58	&	3.2	&	g \\
G\,350.686--0.491& 17 23 28.71 & --37 01 47.9 &--23  	&--22,--24  	&0.6		& --14	& --14,--13	& 1.0		&	omg \\
G\,350.690--0.490& 17 23 29.19 & --37 01 35.6 &		&			& $<$0.6	& --22  	&   --24,--21	& 3.4		&	 \\
G\,351.160+0.696 & 17 19 57.50 & --35 57 54.1 &--3   	&--11.5,--0.5	&9  		&   --3	& --10,2		& 18		&    	omc\\
G\,351.163+0.696 & 17 19 58.13 & --35 57 48.9 &		&			& t		&--10	& --13,--9		& 10		&	\\ 
G\,351.240+0.668 & 17 20 17.98 & --35 54 57.6&		&			&--	& --24  	& --38,31           & 21		&	\\
G\,351.243+0.671 & 17 20 17.76 & --35 54 42.8 &		&		 	&	--	& 	--77	& --108,84	& 453	& m	\\	
G\,351.246+0.668 & 17 20 19.01 &--35 54 38.2 &		&			& 	--	&21 		& 19,22		& 1.3		&c	\\
G\,351.417+0.646 & 17 20 53.29 & --35 46 58.3 &--10  	&--58,50    	&1400	&		&			&	--	&om	\\
G\,351.582--0.353& 17 25 25.35 & --36 12 44.0 &--89  	&--120,--87 	&1600	&		&			&	--	&omg 	\\
G\,351.775--0.536& 17 26 42.50 & --36 09 15.9 &--2   	&--32,21     		&85 		&		&			&	--	&om	\\
G\,352.098+0.160 & 17 24 45.25 &--35 29 48.8 &		&			&	--	&--75  	& --75,--73	&	2.5	& g	\\
G\,352.111+0.176 & 17 24 43.79 &--35 28 37.7 &		&			&	--	&--60  	& --61,--60	& 	1.3	&mg	\\
G\,352.133--0.944& 17 29 22.46 &--36 04 59.9 &		&			&	--	& --11	&  --17,--6		& 	27	&mg	\\
G\,352.162+0.199 & 17 24 46.36 & --35 25 19.9 &--45 	&--46,--44   	&1  		& --45	& --119,--43	& 0.28	&og	\\
G\,352.517--0.155& 17 27 11.34 & --35 19 32.0 &--49 	&--53,--46   	&12 		&      		&			&	--	&om	\\
G\,352.525--0.158	& 17 27 13.44	&--35 19 15.7	& --51	&--53,--49	&	3.6	&	&	&	--	&	mg	\\ 
G\,352.623--1.076& 17 31 14.93 & --35 44 47.5 &2	 	&0,2         		&28 		& 2		& --7,3		&	8	&	$\gamma$\\
G\,352.630--1.067& 17 31 13.94 & --35 44 08.8 &--2	&--10,20     	&35 		&  0		& --13,18		&	700	&om$\gamma$\\

\hline

\end{tabular}
\end{table*}

\begin{table*}\addtocounter{table}{-1}
 \caption{-- {\emph {continued}}}
  \begin{tabular}{lccrcrrcrl} \hline
    \multicolumn{1}{c}{\bf Water maser} & {\bf RA}  & {\bf Dec} & {\bf Vpeak} &{\bf Vrange} & {\bf Speak} & {\bf Vpeak} &{\bf Vrange} & {\bf Speak} & {\bf Associations}\\
   \multicolumn{1}{c}{\bf ($l,b$)} & {\bf (J2000)}  & {\bf (J2000)}&  {\bf (\kmsns)}&  {\bf (\kmsns)} & {\bf (Jy)} &{\bf (\kmsns)}&  {\bf (\kmsns)} & {\bf (Jy)}     \\	
      \multicolumn{1}{c}{\bf (degrees)} & {\bf (h m s)}&{\bf ($^{o}$ $'$ $``$)}& {\bf 2003} & {\bf 2003} & {\bf 2003}& {\bf 2004} & {\bf 2004} & {\bf 2004} \\ \hline \\
G\,353.273+0.641 & 17 26 01.57 & --34 15 14.7&		&			&	--	&--49 	& --110,--5	&	366	&m	\\
G\,353.408--0.350& 17 30 23.38 &--34 41 29.8 &		&			& $<$0.2	&--14	& --15,--13	& 0.6		& g	\\
G\,353.411--0.356& 17 30 25.31 & --34 41 35.0&		&			& $<$0.2	&--20 	& --21,--17	& 6.4		&	\\
G\,353.411--0.362& 17 30 26.78 & --34 41 46.6&		&			&$<$0.2	&--7		& --7,--5		& 1.8		& c	\\
G\,353.413--0.367& 17 30 28.47 & --34 41 49.1 &--17 	&--29,--8    	&2.0		& --20	& --26,--9		&	7	& g	\\
G\,353.413--0.365& 17 30 27.84 & --34 41 44.6 &--20 	&--24,--18    	&4.5		& --19	& --25,--18	&	2.2	& 	\\
G\,353.414--0.363& 17 30 27.56 & --34 41 38.9 &--10 	&--11,--9    	&2  		&    1		& --22,7		&	1.7	&	\\
G\,353.463+0.563 & 17 26 51.25 & --34 08 25.3 &--47 	&--52,--46   	& 1 		&		&			&	--	& g	\\
G\,353.464+0.562 & 17 26 51.61 & --34 08 24.1 & --60	& --61,--59         & 0.7         &                &                        &     --         & omg\\ 
G\,354.594+0.469 & 17 30 14.47 & --33 15 03.2 &--22 	&--25,--20   	&9		& 		&			&$<$0.2	&	\\
G\,354.615+0.472 & 17 30 17.22 & --33 13 54.4 &--38 	&--40,--12   	&1.6		&		&			&$<$0.2	& om	\\
G\,354.703+0.297 & 17 31 12.96 & --33 15 17.6&		&			& $<$0.4	&105	& 104,112		&	1.6	&	\\
G\,354.712+0.293 & 17 31 15.38 & --33 14 57.8 &96   	&95,105       	&1.1		&      		&			&$<$0.2	& g	\\
G\,354.722+0.302& 17 31 14.87 & --33 14 08.3 &		&			&$<$0.2	&99		& 96,101		&	0.38	& 	\\
G\,355.343+0.147 & 17 33 29.00 & --32 48 00.1 & 17  	&9,35        		&34 		&		&			&--		& om	\\
G\,355.345+0.149 & 17 33 28.83 & --32 47 49.9 & 72	& 8,109              & 2.2         &                &                          &  --           & m \\
G\,357.965--0.164& 17 41 20.12 & --30 45 15.9 & --4	& --5,--3		& 53		&--19 	& --20,--19	& 0.9		&mg	\\
G\,357.967--0.163& 17 41 20.30 & --30 45 07.0 &0    	&--80,100    	&40 		& --65	&--81,87		& 57		&om	\\
G\,358.371--0.468& 17 43 32.01 & --30 34 11.4 &10   	&--5,12     		&2.7   		& 2		&2,31		& 1.4		&mg	\\
G\,358.386--0.483& 17 43 37.72 & --30 33 50.6 &--1  	&--4,1      		&4.5 		& 0		& --3,1		& 3.6		&omcg	\\
G\,359.137+0.032 & 17 43 25.61 & --29 39 18.6 &--1  	&--117,26    	&300 	&		&			&	--	&omg	\\
G\,359.419--0.104& 17 44 38.27 & --29 29 12.0&		&			&$<$0.2	&--26 	& --59,2		& 1.0		&	\\
G\,359.436--0.102& 17 44 40.23 & --29 28 12.2 &--59 	&--61,--50  	&15  		& --59	& --61,--54	& 9		&mg	\\
G\,359.436--0.104& 17 44 40.66 & --29 28 16.1&--56	& --57,--55	& 1.5		& --47	& --48,--47	& 0.37	&om	\\
G\,359.441--0.111& 17 44 42.87 & --29 28 15.4 &--65 	&--66,--64  	&1.8 		&      		&                      	&$<$0.2	&	\\
G\,359.442--0.106& 17 44 42.00 & --29 28 00.6 &--53 	&--54,--52  	&9   		& --53	& --56,--52	& 0.928	& g 	\\
G\,359.442--0.104& 17 44 41.47 & --29 27 58.8&		&			& t		&--60 	& --62,--58	& 0.8		&	\\
G\,359.443--0.104& 17 44 41.82 & --29 27 56.5&		&			& t		& --49	& --52,--48 	& 0.6		& g	\\ 
G\,359.615--0.243& 17 45 39.12 & --29 23 29.6 &22   	&--16,73    	&7   		& 64		& --15,68		& 14		&omg	\\
G\,359.969--0.457& 17 47 20.08 & --29 11 59.0 &11   	&10,16      	&27  		&		&			& 	--	&om	\\
G\,0.209--0.002  & 17 46 07.44 &--28 45 32.1 &		&			&	--	&39		& 18,42		&	2.2	&c	\\
G\,0.212--0.002  & 17 46 07.86 &--28 45 23.0 &		&			&	--	&56		& 55,65		&	0.9	&m	\\
G\,0.216--0.023  & 17 46 13.20 &--28 45 49.8 &		&			&	--	&--11	& --16,8		&	1.7	& g	\\
G\,0.308--0.177  & 17 47 02.38 &--28 45 55.2 &		&			&	--	&--26 	&  --27,--26	&	1.8	& g	\\
G\,0.316--0.201  & 17 47 09.29 &--28 46 15.5 &		&			&	--	&23		&   14,31		&	22	&mg	\\
G\,0.376+0.040   & 17 46 21.29 & --28 35 39.7 &40   		&5,58       		&55  		&		&			&	--	&omg	\\
G\,0.497+0.188   & 17 46 03.96 & --28 24 50.7 &--8  		&--72,25    	&7   		& 26		& --50,34		&	2.5	&omg	\\
G\,0.547--0.851  & 17 50 14.52 & --28 54 28.8 &20   		&--60,110   	&200 	&		&			&	--	&omg	\\
G\,0.655--0.045 & 17 47 20.88 & --28 23 59.9  &  --52      & --55,--50           &      4.0   &       &      &    --&\\ 
G\,0.657--0.042 & 17 47 20.44 & --28 23 46.4  &   62        & 60,64                &    70      &        &       &   --& og\\
G\,0.665--0.032 & 17 47 19.07 & --28 23 03.5  & 4            &       0,7                  & 4.3          &               &         & -- &g\\ 
G\,0.668--0.035  & 17 47 20.26 & --28 23 02.6 &59   		&50,90      	&408 	&		&			&	--	& og	\\
G\,0.677--0.028 & 17 47 19.64 & --28 22 18.9  & 36           & 14,132              & 330       &       &       &   --    & g\\ 
G\,2.143+0.009   & 17 50 36.13 & --27 05 46.4 &37   		&35,70      	&0.6 		&		&			&	--	&omg	\\
G\,2.536+0.198   & 17 50 46.66 &--26 39 44.9 &		&			&	--	&25		& --1,63		&	30	&m	\\
G\,5.886--0.392  & 18 00 30.52 & --24 03 58.2 &11   		&--10,23    	&40  		&		&			&	--	&	o$^{\#}$m$^{\#}$\\
G\,5.897--0.445  & 18 00 43.90 & --24 04 58.1&		&			&	--	&19		& 19,20	&	0.9	&	\\
G\,5.901--0.430  & 18 00 40.95 &--24 04 19.6 &		&			&	--	&14		&   --40,30	&	3.9	&m	\\
G\,5.913--0.388  & 18 00 33.12 & --24 02 28.4 &--50 		&--65,--47  	&9.2 		&      		&                      	&	--	& g	\\
G\,6.049--1.447  & 18 04 53.19 & --24 26 40.3 &20   		&18,22      	&1.0   		&		&			&	--	&o$\gamma$	\\
G\,6.534--0.105  & 18 00 49.44 &--23 21 40.2 &		&			&	--	&22		&  6,23		&	0.5	& g	\\   
G\,6.611--0.082  & 18 00 54.15 &--23 17 00.8 & 		&			&	--	&6		& --34,11		&	7	&mg	\\
G\,6.796--0.257  & 18 01 57.71 & --23 12 32.5 &14   		&10,20      	&10  		& 1		& --1,24		&	5	&omg	\\
G\,8.139+0.226   & 18 03 00.83 &--21 48 09.5 &		&			&	--	&16		& 1,24		&	18	&mg	\\
G\,8.670--0.356  & 18 06 19.17 &--21 37 31.0 &34		&30,44      	&10  		& 36		& --9,43		&	3.4	&omc	\\
G\,8.673--0.354  & 18 06 18.98 &--21 37 17.9 &	--4	&	--23,--3	&0.5		&--8		&--20,4		&	0.36	&	\\
G\,9.620+0.194   & 18 06 14.97 & --20 31 37.5 &5    		&--30,50    	&25  		& 6		& --19,11		&	12	&o	\\
G\,9.622+0.195   & 18 06 14.88 &--20 31 31.0 &	21	&	0,22		& 1.4		& 22		& 0,23		&   1.3	&om$^{\#}$	\\
G\,9.986--0.028  & 18 07 50.20 &--20 18 55.9 &		&			&	--	&49		& 44,60		&	4.0	&mg	\\
G\,10.288--0.125 & 18 08 49.44 &--20 05 57.4 &		&			&	--	&9 		& 9,11		&	0.5	&m	\\
 \hline

\end{tabular}
\end{table*}

\begin{table*}\addtocounter{table}{-1}
   \caption{-- {\emph {continued}}}
  \begin{tabular}{lccrcrrcrl} \hline
    \multicolumn{1}{c}{\bf Water maser} & {\bf RA}  & {\bf Dec} & {\bf Vpeak} &{\bf Vrange} & {\bf Speak} & {\bf Vpeak} &{\bf Vrange} & {\bf Speak} & {\bf Associations}\\
   \multicolumn{1}{c}{\bf ($l,b$)} & {\bf (J2000)}  & {\bf (J2000)}&  {\bf (\kmsns)}&  {\bf (\kmsns)} & {\bf (Jy)} &{\bf (\kmsns)}&  {\bf (\kmsns)} & {\bf (Jy)}     \\	
      \multicolumn{1}{c}{\bf (degrees)} & {\bf (h m s)}&{\bf ($^{o}$ $'$ $``$)}& {\bf 2003} & {\bf 2003} & {\bf 2003}& {\bf 2004} & {\bf 2004} & {\bf 2004} \\ \hline \\
G\,10.307--0.270 & 18 09 24.29 & --20 09 08.8&		&			&	--	&33		& 32,33		&	0.8	&	\\
G\,10.323--0.160 & 18 09 01.57 &--20 05 07.6 &		&			&	--	&--3		& --6,62		&	3.2	&m	\\
G\,10.331--0.159 & 18 09 02.30 &--20 04 40.2 &		&			&	--	&11		& 7,22		& 	1.0	&	\\
G\,10.342--0.143 & 18 09 00.11 & --20 03 35.8&		&			&	--	&8		& --40,61		&	4.8	&mg	\\
G\,10.445--0.018 & 18 08 45.01 & --19 54 35.1 &71  		&58,85       	&6   		& 70		& 69,81		&	2.7	&	\\ 
G\,10.473+0.027  & 18 08 38.30 & --19 51 48.8 &60  		&30,93       	&45  		& 62		& 28,129		& 169	&omcg	\\
G\,10.480+0.034  & 18 08 37.69 & --19 51 12.4 &64  		&63,65       	&12  		&		&			& $<$0.2	&om\\
G\,10.623--0.383 & 18 10 28.57 & --19 55 49.4 &2   		&--11,5      	&350 	&		&			&	--	&og	\\
G\,10.625--0.335 & 18 10 18.07 & --19 54 20.5&		&			&	--	&	--5 	& --6,--5		&	0.40	&	\\
G\,10.959+0.022  & 18 09 39.38 & --19 26 27.0&		&			&	--	&25		& 7,26		&	2.9	&	mcg \\

G\,11.034+0.062  & 18 09 39.75 & --19 21 21.1 &18  		&16,19       	&	0.6 	&		&			&	--	&	om\\
G\,11.498--1.486 & 18 16 22.32 & --19 41 26.1&		&			&	--	&17		& --3,21		&	112	&	m$\gamma$\\
G\,11.903--0.142 & 18 12 11.41 & --18 41 33.0 &36  		&35,38       	&0.3 		&      		&			& $<$0.2	&	om \\
G\,12.203--0.107 & 18 12 40.25 & --18 24 47.4&	33	&	32,35		& 6&35		& 32,37		& 8		&	m\\
G\,12.209--0.102 & 18 12 39.88 & --18 24 17.1 &21  		&--12,42     	&141  		&   22	&   --15,50		&	51	&	mcg \\
G\,12.216--0.119 & 18 12 44.55 & --18 24 25.2 &26  		&25,42       	&6   		& 26		& 0,38		& 10		&	og \\
G\,12.681--0.183 & 18 13 54.82 & --18 01 47.0 &61		&55,63       	&1200 		& 61		& 17,62		& 445	&	omg \\ 
G\,12.884+0.502  & 18 11 47.92 &--17 31 21.3 &	42	&	41,43	& 1.3		&40		& 34,41		& 1.1		&	\\
G\,12.889+0.489  & 18 11 51.54 & --17 31 27.9 &30  		&28,32       	&45  		& 30		& 28,39		& 59		&	omg \\
G\,12.901--0.242 & 18 14 34.50 & --17 51 51.6 &35  		&23,37       	&8		& 36		& 23,37		& 12		& g	\\
G\,12.908--0.260 & 18 14 39.45 & --17 52 01.4 &37  		&    21,39      		& 0.8		&		&			&$<$0.2	&	om\\
G\,15.016--0.679 & 18 20 23.11 & --16 12 38.8&		&			& t		&19		& 14,21		&11		&	\\ 
G\,15.025--0.658 & 18 20 19.50 & --16 11 31.9&		&            		& t		& 25		&  14,31		& 4.2		&	\\
G\,15.026--0.654 & 18 20 18.72 & --16 11 23.3 &27  		&27,31       	&29   		&  28		& 26,31		& 16		&	\\
G\,15.028--0.673 & 18 20 23.04 & --16 11 48.4 &19  		&17,28       	&197 	& 20		& 14,31		& 67		&	 \\
G\,15.032--0.667 & 18 20 22.35 & --16 11 27.2 &	45	&	44,46	& 1.8		&40		& --8,41		& 1.1		& g	\\
G\,15.032--0.670 & 18 20 23.01 & --16 11 30.1&		&			&t	 	&26		& 14,28            	& 12		&	\\
G\,15.034--0.667 & 18 20 22.58 & --16 11 18.4&		&			& t		&17		& 17,21		& 3.8		&	\\
G\,17.638+0.156  & 18 22 26.45 & --13 30 12.4 &27  		&15,29       	&230 	& 27		& 17,35		& 245	&	om\\ \hline
\end{tabular}
\end{table*}

\section[]{Results}

The search for 22-GHz water masers carried out with the ATCA in 2003
October and 2004 July towards 202 OH maser sites and 104 methanol maser
sites (with no reported OH maser emission) resulted in the detection of 379 distinct water maser sites
(Table~\ref{tab:masers}). Spectra of all detected sources are shown in
Fig.~\ref{fig:spectra}. For the majority of sources, the spectra are taken 
from the 2004 data, except where sources were either not observed
or not detected at this epoch. For these latter cases we show the 2003 
spectra and distinguish them from the 2004 spectra with a `2003' in the 
top left hand corner of each spectrum.  The 2003 spectra were obtained 
directly from the uv data with a phase shift to the source position, and 
amplitude correction for offset from the field centre.  For eight
sources we show a spectrum from 
each epoch to either highlight that a weak source is a genuine detection
(333.387+0.032 and 336.983-0.183) or to give an indication of the
level of variability seen over the 10 month time-scale (284.350--0.418,
321.148--0.529, 327.291--0.578, 345.004--0.224, 15.026--0.654 and 15.028--0.673). A velocity range of 200 
\kms\ is
shown for the majority of sources, but there are several instances
where we either decreased this value to clearly show individual features in spectra that are complex (or include multiple nearby sources)  
or increased it in order to display extremely high velocity features. A decreased velocity range of 100
\kms\ is shown for the following sources; 301.136--0.225,
301.136--0.226a, 301.136--0.226b,
301.137--0.225, 335.060--0.428/335.059--0.428, 336.991--0.024, 336.995--0.024, 359.441--0.111,
359.442--0.106, 359.442--0.104, 359.443--0.104. An increased range of 300 
\kms\ was used 
for 320.120--0.440, 330.954--0.182, 333.219--0.062,
333.234--0.060, 345.699--0.090, 357.965--0.164,
357.967--0.163, 0.547--0.851, 0.668--0.035, 0.665--0.032,
0.655--0.045, 0.657--0.042 and 0.677--0.028.

A number of the sources that we detect have been observed previously
and have been presented in the literature \citep[e.g.][and references
therein]{John72,C74,K76,GD77,B80,BS82,BE83,C+89,HC96} but the
majority of these earlier observations (performed up to 20 years ago) were made with 
relatively poor positional accuracy.  Due to the intrinsically variable
nature of water masers, many sources exhibit levels of variability 
so extreme that they display no common spectral features at 
epochs separated by many years. This, combined with the tendency of water
masers to form in clusters, and the previously poor positional 
information, mean that it
is almost impossible to accurately match up sources from the literature
with our present data. We have therefore limited our references (in
Section ~\ref{sect:ind}) to previous detections of sources that were 
observed with high positional
precision \citep[e.g.][]{FC89,Breen,CP08}, or where there was little doubt
that the sources were the same.

The majority of OH maser targets were observed in both 2003 and 2004,
whereas the methanol maser targets were observed in 2004 only. Where
appropriate data were available for both epochs, reported positions are
the average of the two since, in general, it provides the most accurate
positions for the sources. For sources north of declination --20\degrees,
we have used a weighting of 2:1 for the declinations in favour of the 2004
data to account for the three times more elongated beam of the
2003 observations (a consequence of the different array configurations). 
Sources that
were observed at both epochs allowed a direct comparison of the positions
for each of the sources and therefore afford verification of the
positional uncertainties. 

Additional to direct comparison of 2003 with 2004 data, an overall assessment 
of data quality and reliability was made in several other 
ways.  FC89 and FC99 used their VLA observations of a sample of more than 70 SFR targets 
with OH and water masers, to show that more than half were a 
simple association of 
water and OH masers coincident to within their combined relative errors 
(of typically 1 arcsec).  Subsequent observation of the more southerly 
OH masers in that sample with the ATCA (C98) showed that the most 
southerly ones (observed by the VLA inevitably at low elevation) had 
significantly larger position uncertainties, and corrections to the 
positions resulted in an increased number of close OH/water maser associations.  
Thus we may expect the majority of our sample to show a water maser 
position coincident with OH, and thus the OH position is an indirect check 
on the accuracy of the newly derived water positions.

A further assessment was made using the 35
masers north of declination --47 degrees which are present in the
FC89 target list. The FC89 absolute positions for
the more southerly targets, although of variable quality for the OH masers
(where ionospheric effects at low elevation can be significant), appear to
remain excellent for the water masers.  Thus we can directly compare our
positions to those of FC89, to assess the errors in our current data.  
Furthermore, in some fields there is a strong ultracompact \ionhy 
(\UCHIIns)
region that has been measured to subarcsecond accuracy, such as in the
6-GHz observations of C97 and C2001; where these are detectable in the
current 22-GHz observations, they allow a further check on the positions,
without the need for any assumptions concerning the true relative
positions of the masers.

From these many comparisons, we are able to estimate our rms 
positional uncertainty as 2 arcsec.  The target
OH and methanol masers have rms position uncertainties of 0.4 arcsec 
(C98, Caswell 2009). 
An additional positional uncertainty in characterising any water maser 
site by a single position arises because a single site sometimes consists 
of many separate spots with angular separations as
extensive as 4 arcsec \citep[e.g.][]{R88}, explicable by  
an outflow \citep[e.g.][]{CP08}. 
We therefore regard our water maser sources to be associated with OH or 
methanol masers when they are separated by less than 3 arcsec, a 
threshold which captures most associations without diluting them with too 
many false, chance, coincidences;  see also further discussion in Section 
5.2.  Where 
the water maser positions are derived from a single epoch, we relax 
this threshold to 4.5 arcsec. As the positions of
the 22-GHz radio continuum have also been determined from a single epoch,
a threshold of 4.5 arcsec is similarly adopted for continuum associations. 
Most of our proposed associations correspond to a much better
accuracy (see Table 3) than our thresholds. There are, however, some more complex cases
that have been judged on individual merit as discussed in detailed 
considerations summarized in Section~\ref{sect:ind}.  For example, the 
required precision of agreement was relaxed for sources believed
to be nearby, at a distance of less than 2 kpc.   
Comparison of our 379 water maser positions with the
positions of OH and methanol masers shows that 128 are coincident with both species, 33
are coincident with OH only and 70 are coincident with the location of
methanol masers (see Section 5.3 for more extensive discussion). Surprisingly, 148 sources have no association with other maser species and we describe these as `solitary'.

Details for the 379 sources that we detect are presented in 
Table~\ref{tab:masers}
and, following the usual practice, the Galactic longitude and latitude of
each source, listed in the first column, is used as an identifying source
name for each water maser.  These Galactic coordinates are derived 
from the more precise measurements of equatorial coordinates given in 
columns 2 and 3. The peak velocity and velocity range (w.r.t. lsr), 
followed by the peak flux density, are given in columns 4, 5 and 6 for 
the 2003 epoch and in columns
7, 8 and 9 for the 2004 epoch. The presence of a `--' in either column 6 or
9 indicates that no observations were made for that source during the 2003
or 2004 observations respectively and a `t' in either column indicates
that there is a comment in the text of Section~\ref{sect:ind} explaining 
the nature of
the detection status at the indicated epoch. The presence of a number
preceded by a `$<$' in either column 6 or 9 indicates that no emission
above the quoted flux density was detected at that epoch. Column 10 gives
a list of associations for each water maser source; here, the presence of
an `o' denotes the presence of an associated OH maser, `m' the presence of
an associated methanol maser, `c' the presence of associated 22-GHz
continuum emission (in our observations) and `g' the presence of an 
associated
GLIMPSE point source. A `$\gamma$' in this column indicates that the source is outside the GLIMPSE survey region. A `$^{\#}$' following an `o' indicates that the OH
maser is strictly outside our association threshold but is associated with
the methanol maser that falls within our threshold for a given source,
meaning that either all three sources are coincident or the water maser is
offset;  similarly for the case where a `$^{\#}$' follows an `m'. In some cases we have the situation that both the OH and the methanol masers are strictly located outside the association threshold but we regard them as associated through special circumstances, in which case we have used the `$\#$' after both the `o' and the `m'.
In the case of 301.136-0.226 a second water maser site lies within the association threshold for the same methanol and OH masers and the association is shown in parentheses.

OH masers that were searched and resulted in no water maser
detection are listed in Table~\ref{tab:nowater}. The first column gives
the name of the OH maser followed by its right ascension and declination. 
Column 4 gives the angular separation between the OH maser
and the nearest methanol maser \citep{C09} within 2.5 arcsec, and when a
`--' is present, this signifies that there is no methanol maser within 2.5 
arcsec of the OH maser.

An extensive list of the OH and methanol masers associated with our water
maser sources, as well as water maser associations with 22-GHz continuum
sources is given in Table~\ref{tab:ass}.  All OH and methanol masers as
well as 22-GHz radio continuum that fall within 5 arcsec of the water
masers are presented. Column 1 in Table~\ref{tab:ass} gives the water
maser source name, and the name of the nearby OH and methanol masers are
given in columns 2 and 4 respectively.  The source names, based on the 
precise positions of individual species, inevitably differ slightly in a 
few cases due to different small position errors. 
The angular separations between the water masers and OH masers are given 
in column 3, and between water and methanol masers in column 5. Columns 6, 
7 and 8 give the peak velocity of the water (2004 values are given unless 
not available, in which case the 2003 value is used) and the coincident 
OH and methanol maser
sources. Column 9 gives the Galactic coordinates of 22-GHz continuum
sources that we detect, followed by the angular separation between the
continuum source and the water maser in column 10. The discussion of individual sources in Section~\ref{sect:ind} includes some comparisons between the positions of water maser sources and
other masers, continuum and GLIMPSE sources.

A complete list of the continuum sources detected towards water maser sources in the 2004
observations is given in Table~\ref{tab:cont}. Associations between the
29 water maser sources observed only during the 2003 observations and
possibly associated 22-GHz continuum sources (see Section~\ref{sect:cont}) have not 
been determined.

\begin{table}

   \caption{OH masers with no associated water maser emission. Listed in column 1 is the OH maser source name followed in columns 2 and 3 by the right ascension and declination. Column 4 shows the angular separation between the listed OH maser and a nearby methanol maser; a -- in this column indicates that there is no known methanol maser sources within 2.5 arcsec.}
  \begin{tabular}{lccrcrrcrl} \hline
    \multicolumn{1}{c}{\bf OH maser}& {\bf RA}  & {\bf Dec} & {\bf Methanol}\\
   \multicolumn{1}{c} {\bf ($l,b$)} & {\bf (J2000)}  & {\bf (J2000)} & {\bf maser}  \\	
      \multicolumn{1}{c}{\bf degrees}& {\bf (h m s)}&{\bf ($^{o}$ $'$ $``$)}& {\bf sep. (arcsec)} \\ \hline \\
G\,232.621+0.996		&07 32 09.82	 &--16 58 13.0	& 0.7\\
G\,300.969+1.147		&12 34 53.24	 &--61 39 40.3	&  0.5\\
G\,305.200+0.019		&13 11 16.90	 &--62 45 54.7	&  0.5\\
G\,305.202+0.208		&13 11 10.61	 &--62 34 37.8	&  1.3\\
G\,306.322--0.334		&13 21 23.00	 &--63 00 30.4	&  0.9\\
G\,309.921+0.479		&13 50 41.73	 &--61 35 09.8	&0.9 \\
G\,313.705--0.190		&14 22 34.72	 &--61 08 27.4	& 0.8\\
G\,316.359--0.362		&14 43 11.00	 &--60 17 15.3	& 2.5\\
G\,321.030--0.485		&15 15 51.67	 &--58 11 18.0	& 0.9\\
G\,323.459--0.079		&15 29 19.36	 &--56 31 21.4	& 1.4\\
G\,328.307+0.430		&15 54 06.48	 &--53 11 40.3	& --\\
G\,329.339+0.148		&16 00 33.15	 &--52 44 39.8	& 0.2 \\
G\,331.542--0.066		&16 12 09.05	 &--51 25 47.2	&  0.5 \\
G\,331.543--0.066		&16 12 09.16	 &--51 25 45.3	&0.2\\
G\,331.556--0.121		&16 12 27.19	 &--51 27 38.1	&0.2\\
G\,332.295+2.280		&16 05 41.72	 &--49 11 30.5	& 0.2 \\
G\,332.824--0.548		&16 20 10.23	 &--50 53 18.1	& --\\
G\,333.135--0.431		&16 21 02.97	 &--50 35 10.1	& 2.4\\
G\,335.556--0.307		&16 30 56.00	 &--48 45 51.0	& 0.8\\
G\,336.822+0.028		&16 34 38.26	 &--47 36 33.0	&  0.8 \\
G\,336.941--0.156		&16 35 55.22	 &--47 38 45.7	& 0.4\\
G\,338.875--0.084		&16 43 08.23	 &--46 09 12.8	&  0.2\\
G\,339.053--0.315		&16 44 49.16	 &--46 10 14.4	& 2.2\\
G\,339.282+0.136		&16 43 43.12	 &--45 42 08.4	& 0.4\\
G\,339.682--1.207		&16 51 06.21	 &--46 15 57.8	& 0.4\\
G\,343.930+0.125		&17 00 10.92	 &--42 07 18.7	& 0.6 \\
G\,344.419+0.044		&17 02 08.67	 &--41 47 08.6	& 1.8 \\
G\,345.498+1.467		&16 59 42.81	 &--40 03 36.2	&  0.4\\
G\,347.870+0.014		&17 13 08.80	 &--39 02 29.5	& --\\
G\,348.550--0.979		&17 19 20.39	 &--39 03 51.8	&0.3 \\
G\,348.579--0.920		&17 19 10.56	 &--39 00 24.5	&  0.6\\
G\,348.698--1.027		&17 19 58.91	 &--38 58 14.1	& --\\
G\,348.703--1.043		&17 20 03.96	 &--38 58 31.3	&1.2\\
G\,348.727--1.037		&17 20 06.55	 &--38 57 08.2	&  0.9\\
G\,350.011--1.342		&17 25 06.50	 &--38 04 00.7	& 0.5\\
G\,353.410--0.360		&17 30 26.20	 &--34 41 45.5		&0.3\\
G\,354.724+0.300		&17 31 15.52	 &--33 14 05.3	&0.5\\
G\,356.662--0.264		&17 38 29.22	 &--31 54 40.6	& 2.0 \\
G\,3.910+0.001		&17 54 38.77	 &--25 34 45.2	&0.5\\
G\,8.683--0.368		&18 06 23.46	 &--21 37 10.2	& 0.4\\
G\,12.025-0.031		&18 12 01.88	 &--18 31 55.6	&0.3\\
G\,15.034--0.677		&18 20 24.75	 &--16 11 34.9	& 0.6\\ \hline
\end{tabular}
\label{tab:nowater}
\end{table}

\begin{figure*}
	\psfig{figure=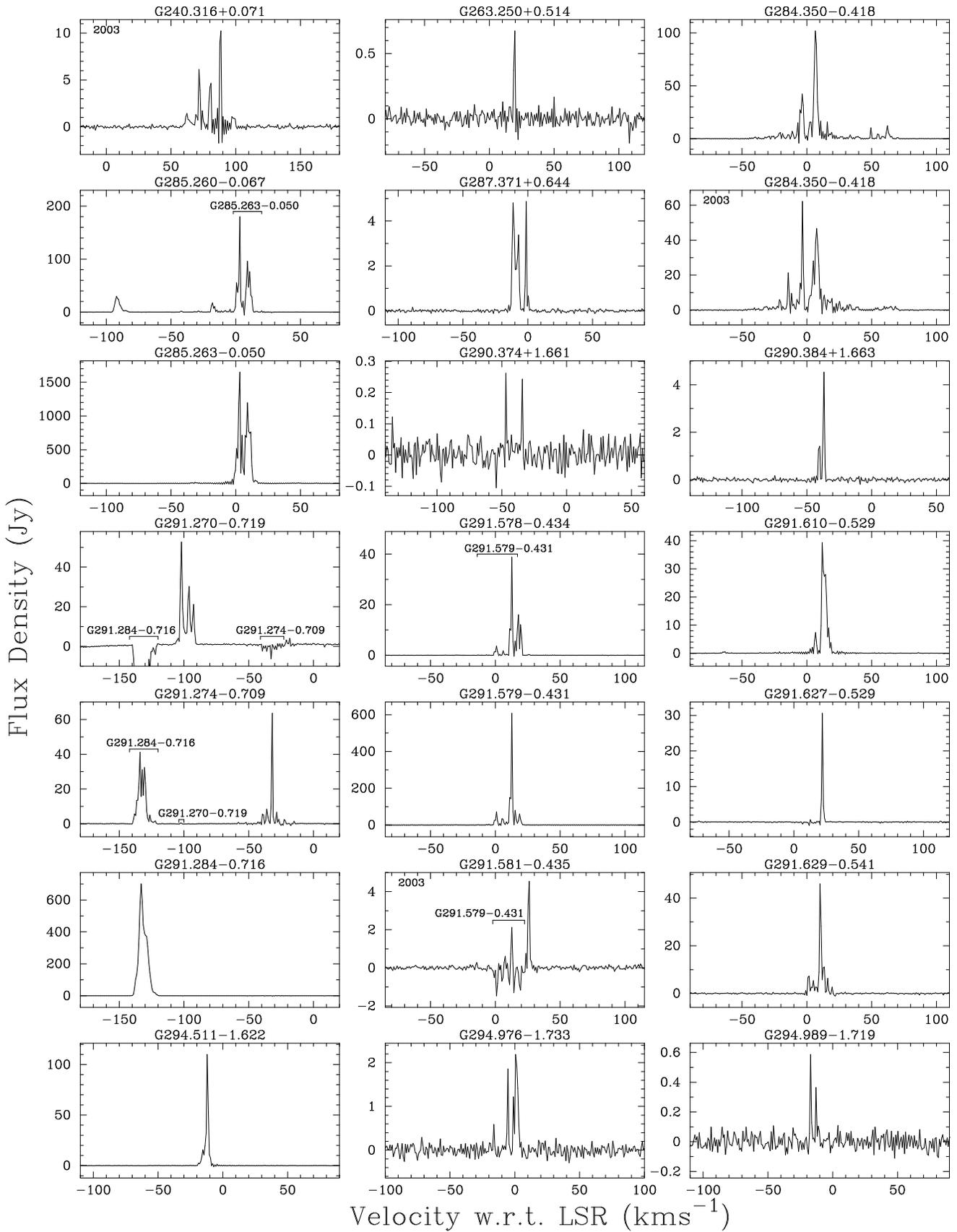}
\caption{Spectra of the 22-GHz water masers detected in 2004 towards sites of OH and methanol masers.}
\label{fig:spectra}
\end{figure*}

\begin{figure*}\addtocounter{figure}{-1}
	\psfig{figure=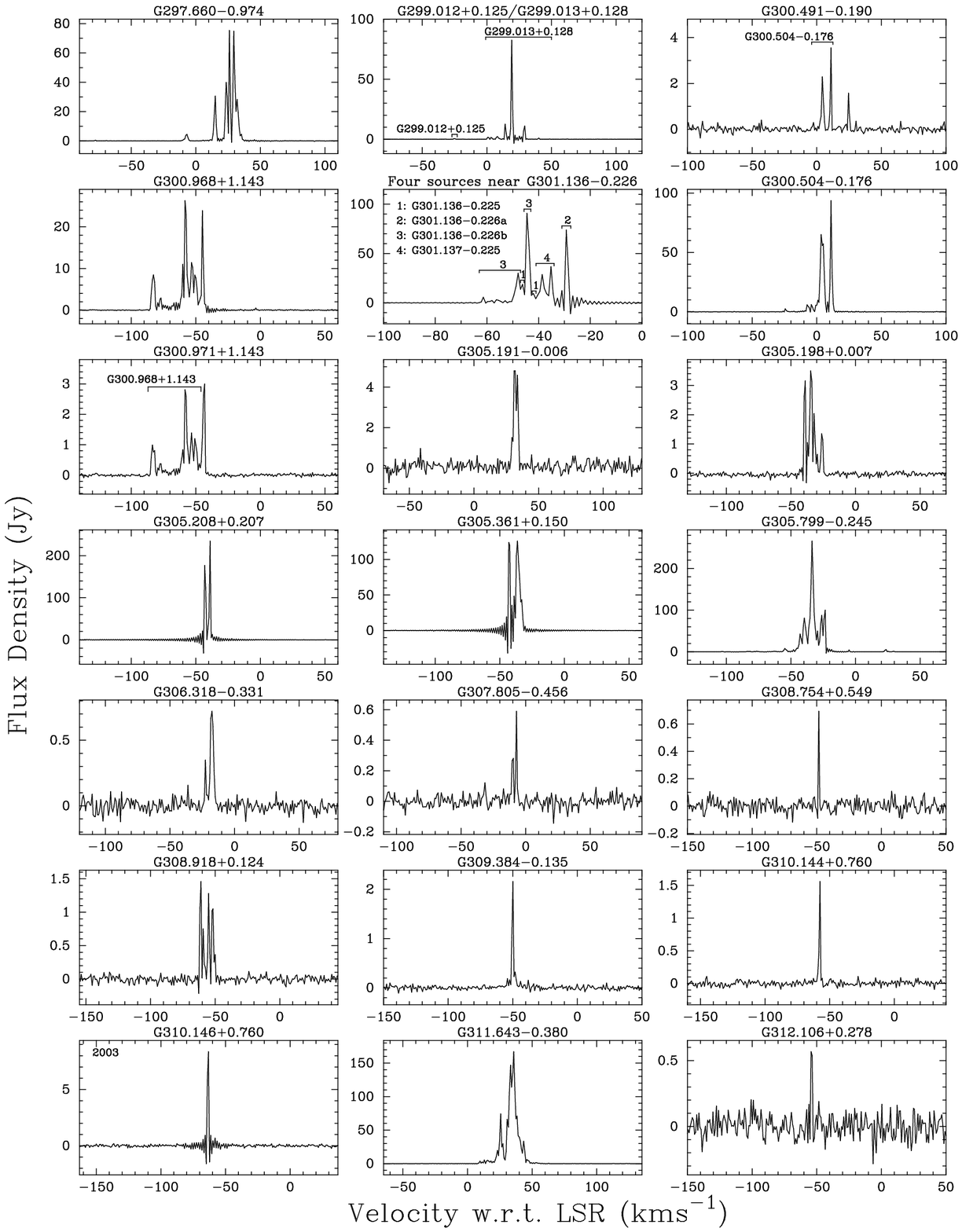}
\caption{--{\emph {continuued}}}
\end{figure*}

\begin{figure*}\addtocounter{figure}{-1}
	\psfig{figure=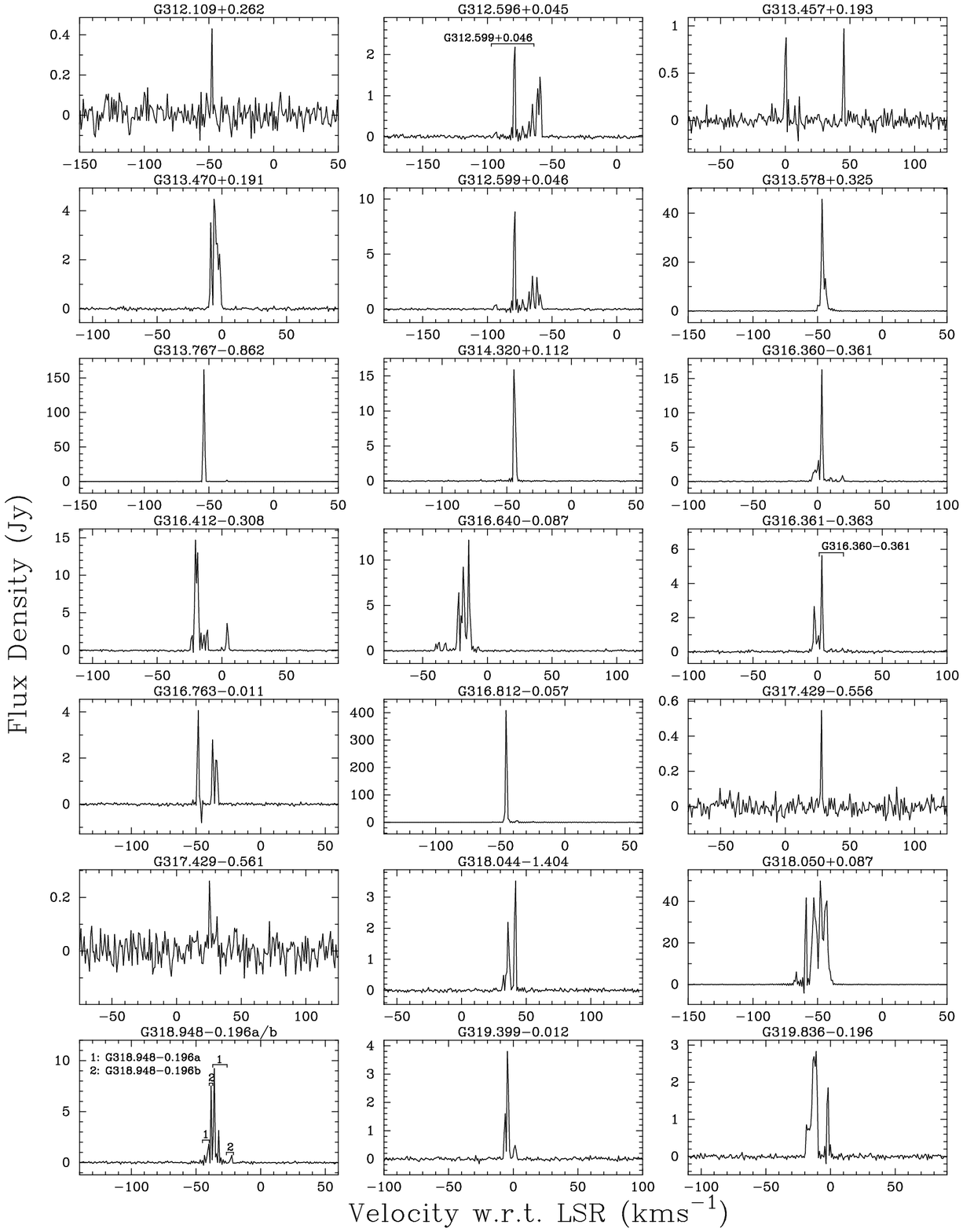}
\caption{--{\emph {continuued}}}
\end{figure*}

\begin{figure*}\addtocounter{figure}{-1}
	\psfig{figure=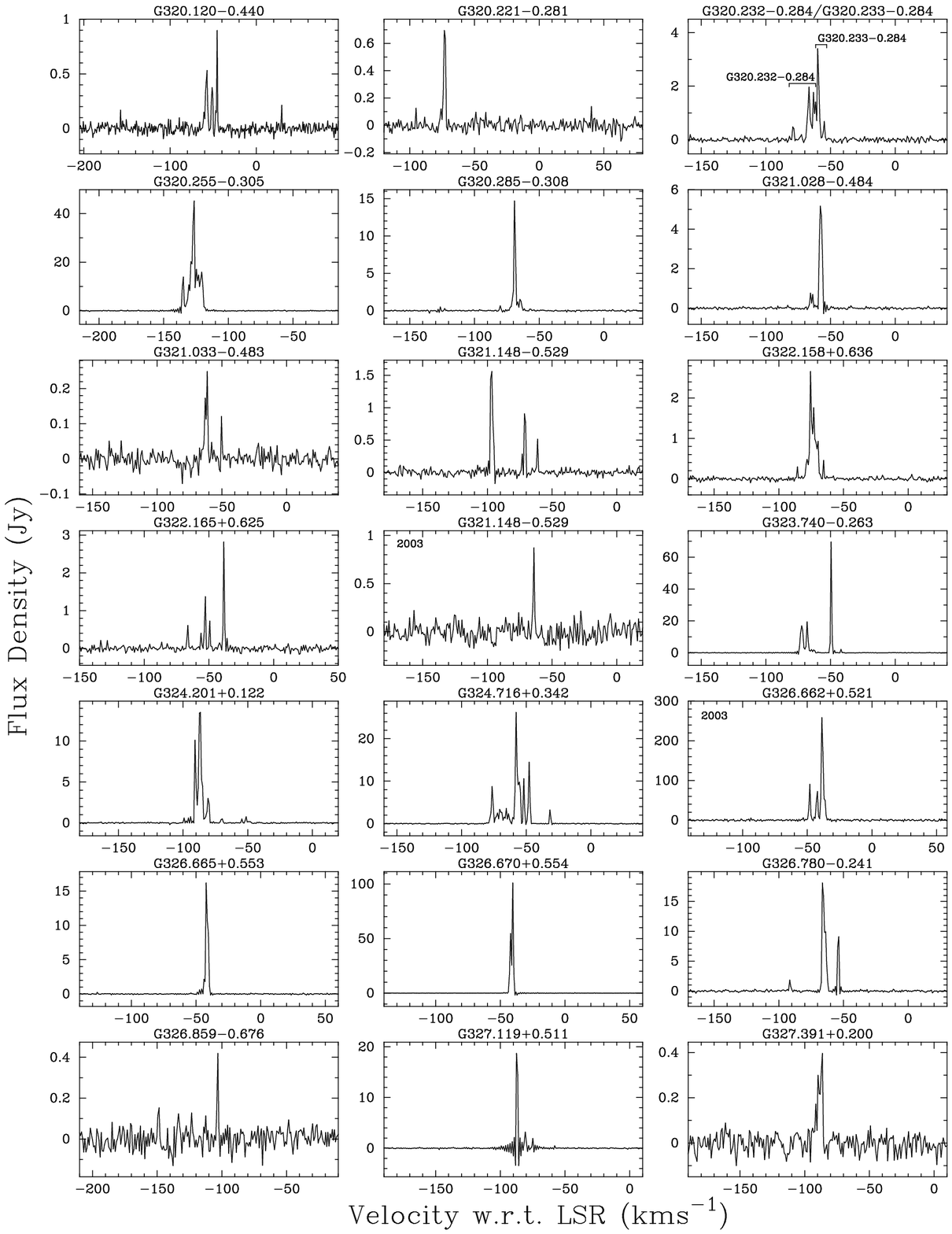}
\caption{--{\emph {continuued}}}
\end{figure*}

\begin{figure*}\addtocounter{figure}{-1}
	\psfig{figure=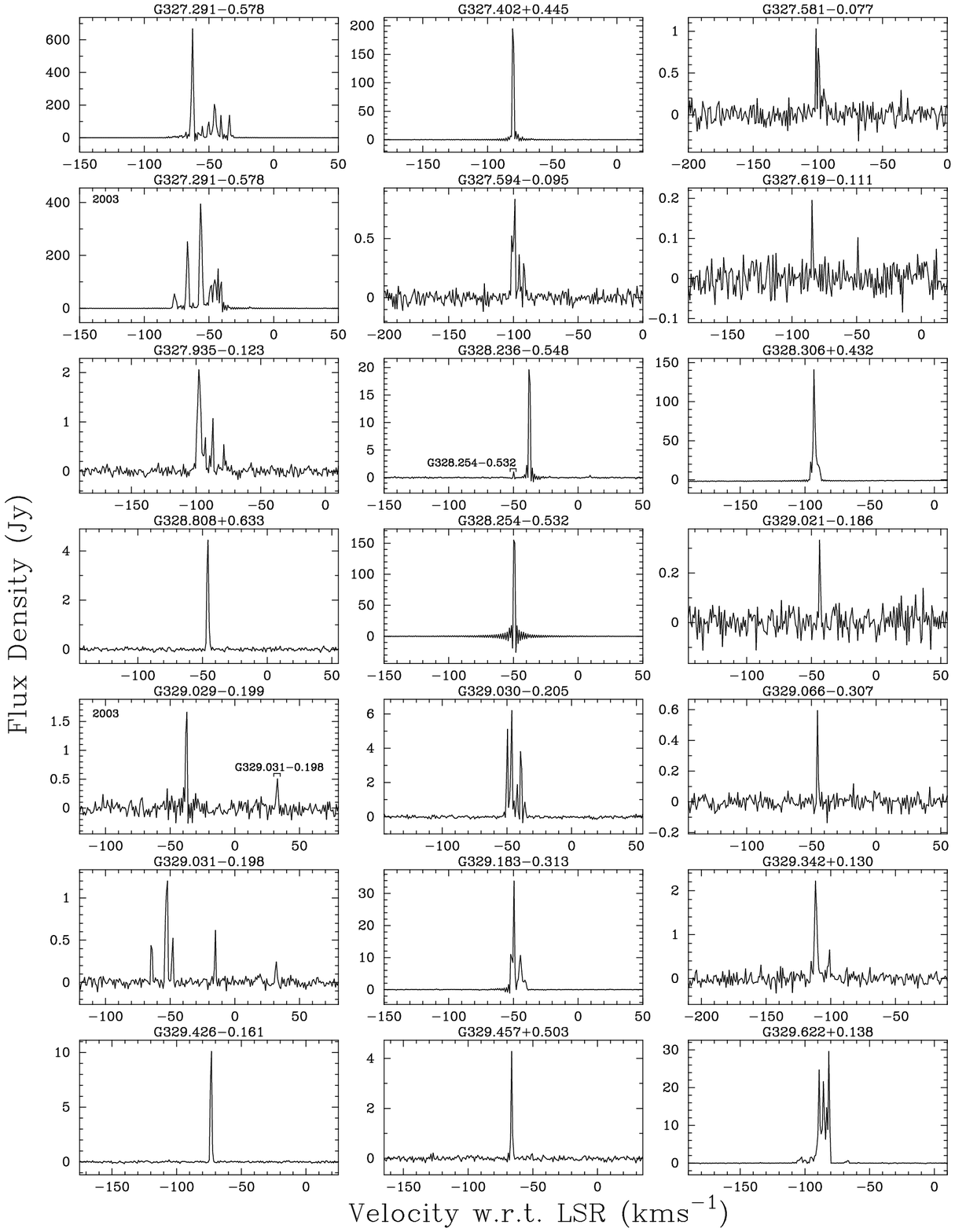}
\caption{--{\emph {continuued}}}
\end{figure*}

\begin{figure*}\addtocounter{figure}{-1}
	\psfig{figure=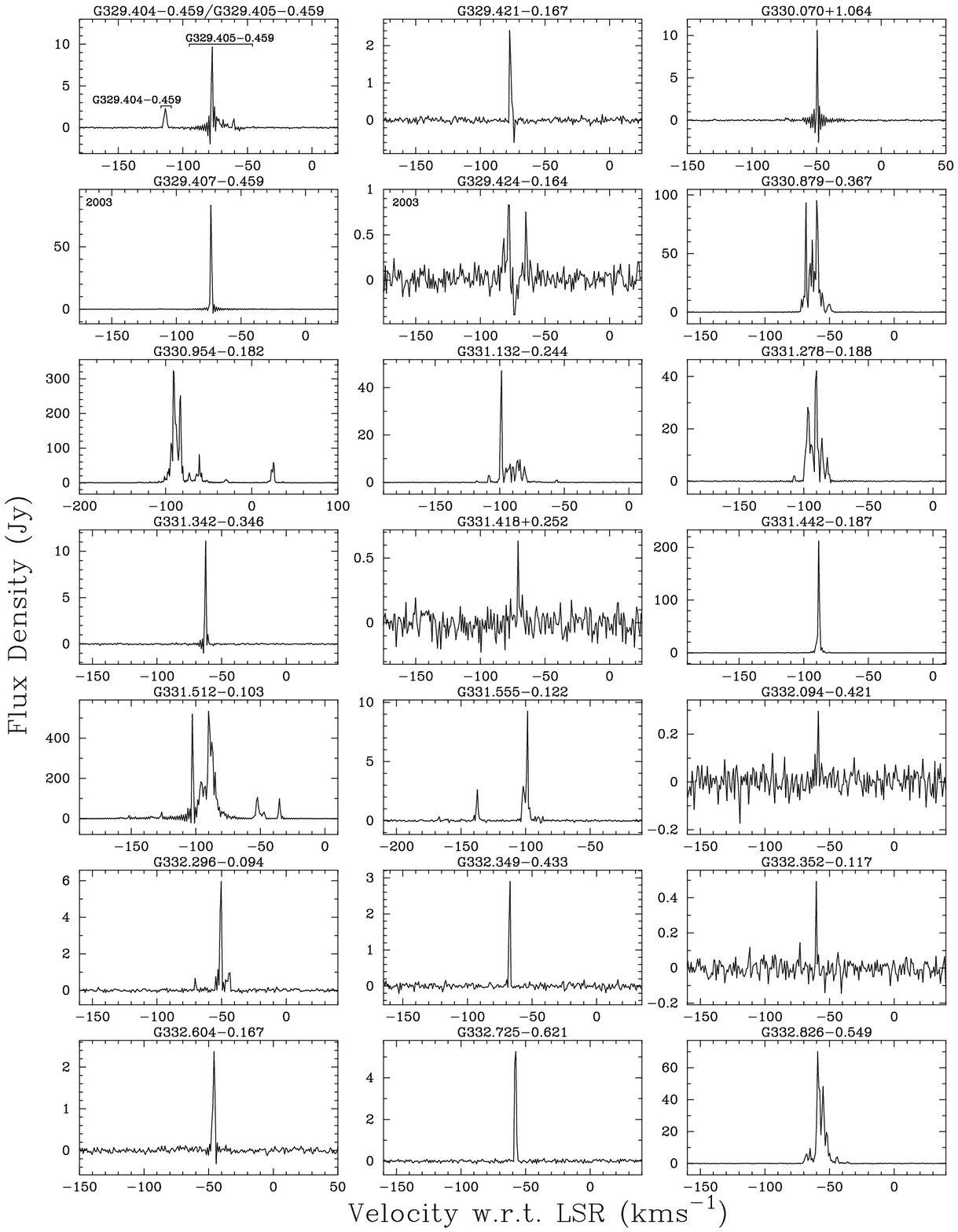}
\caption{--{\emph {continuued}}}
\end{figure*}

\begin{figure*}\addtocounter{figure}{-1}
	\psfig{figure=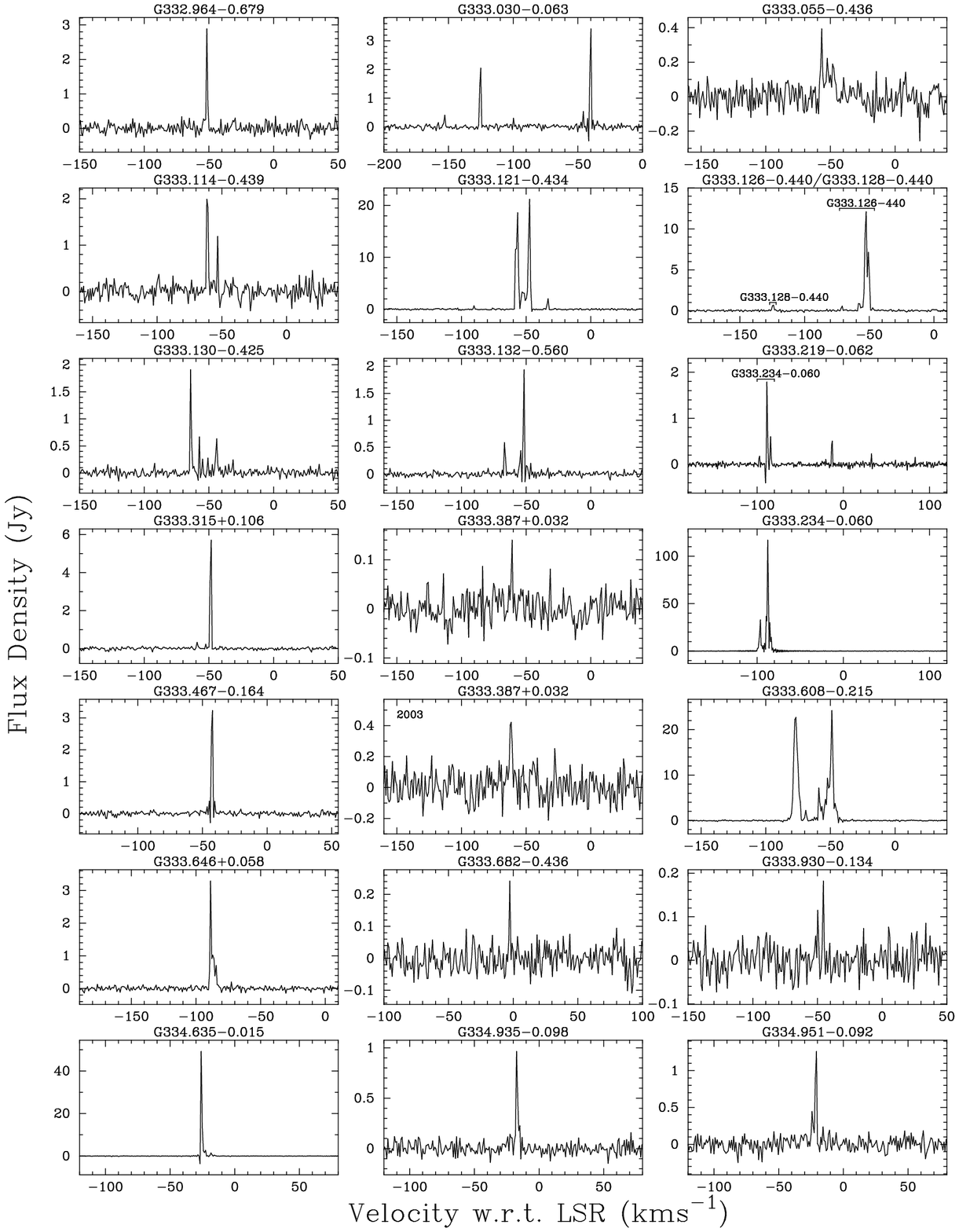}
\caption{--{\emph {continuued}}}
\end{figure*}

\begin{figure*}\addtocounter{figure}{-1}
	\psfig{figure=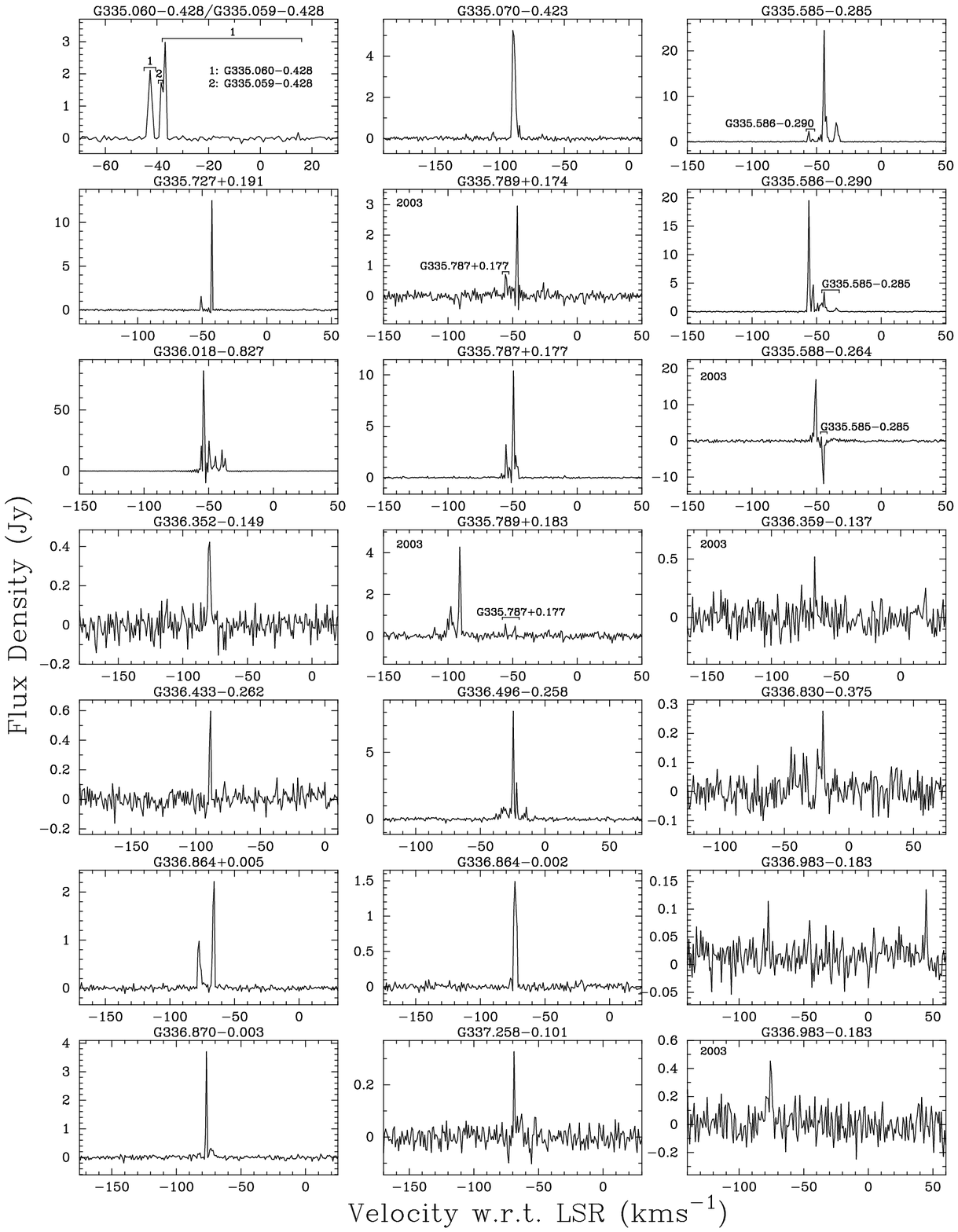}
\caption{--{\emph {continuued}}}
\end{figure*}

\begin{figure*}\addtocounter{figure}{-1}
	\psfig{figure=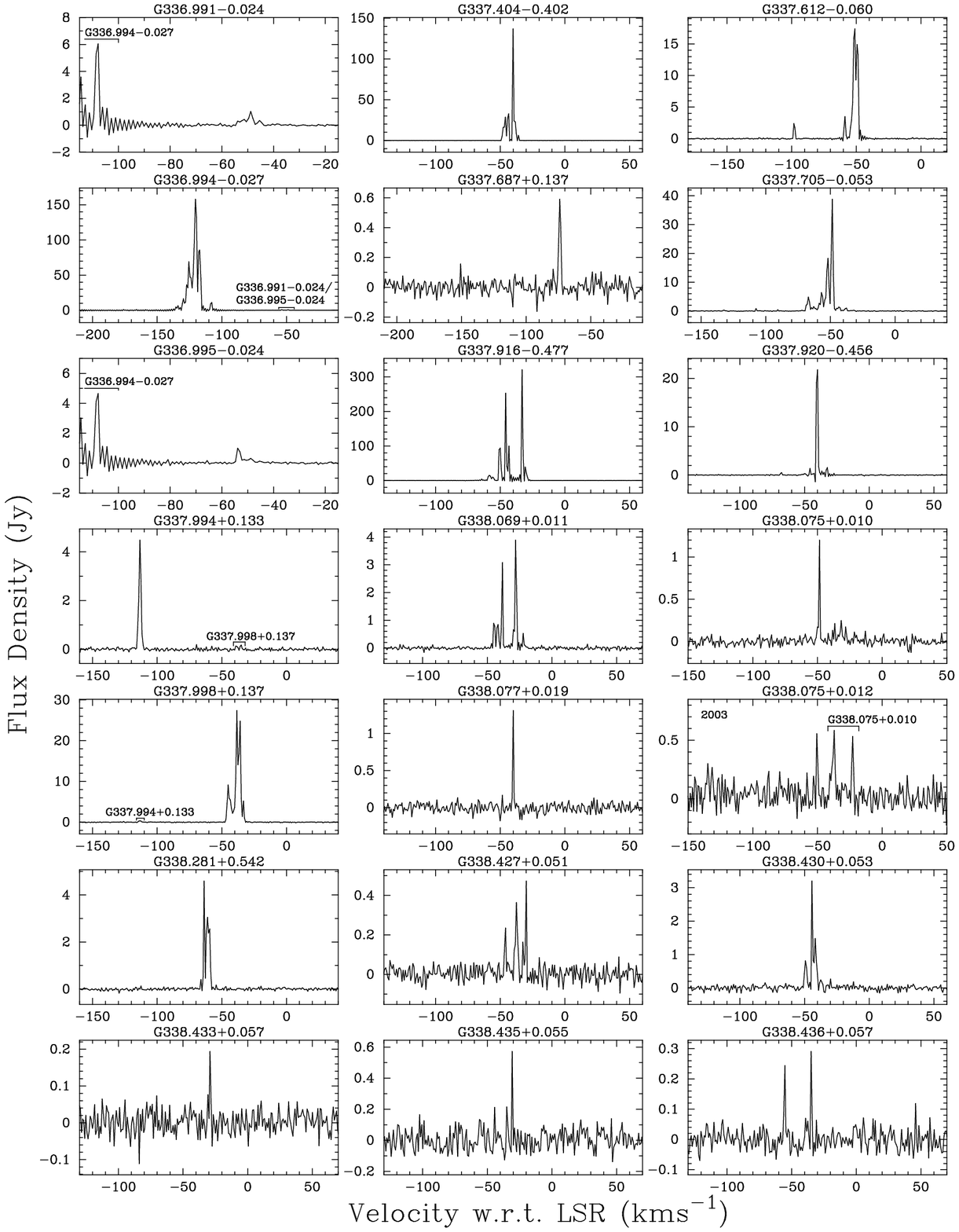}
\caption{--{\emph {continuued}}}
\end{figure*}

\begin{figure*}\addtocounter{figure}{-1}
	\psfig{figure=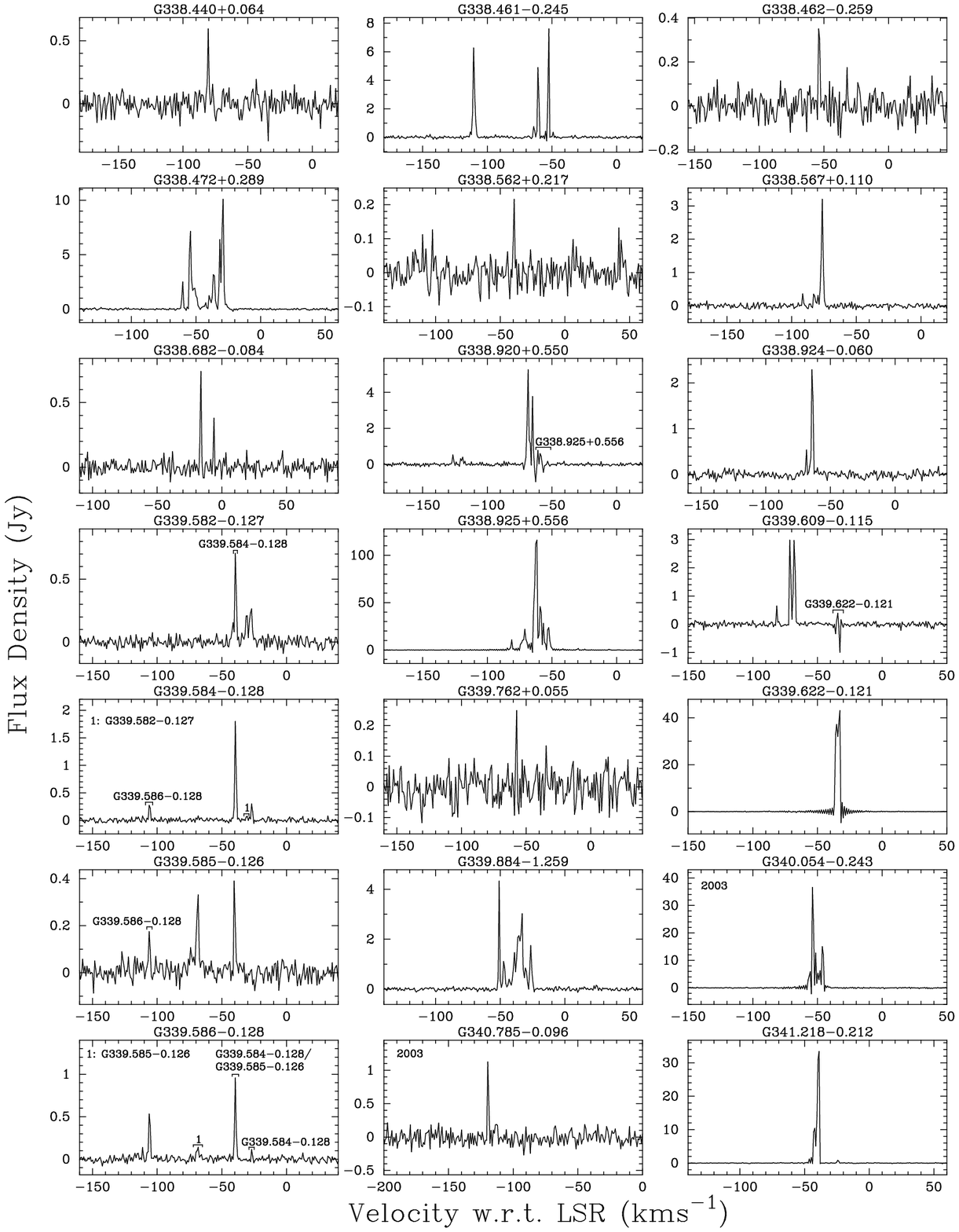}
\caption{--{\emph {continuued}}}
\end{figure*}

\begin{figure*}\addtocounter{figure}{-1}
	\psfig{figure=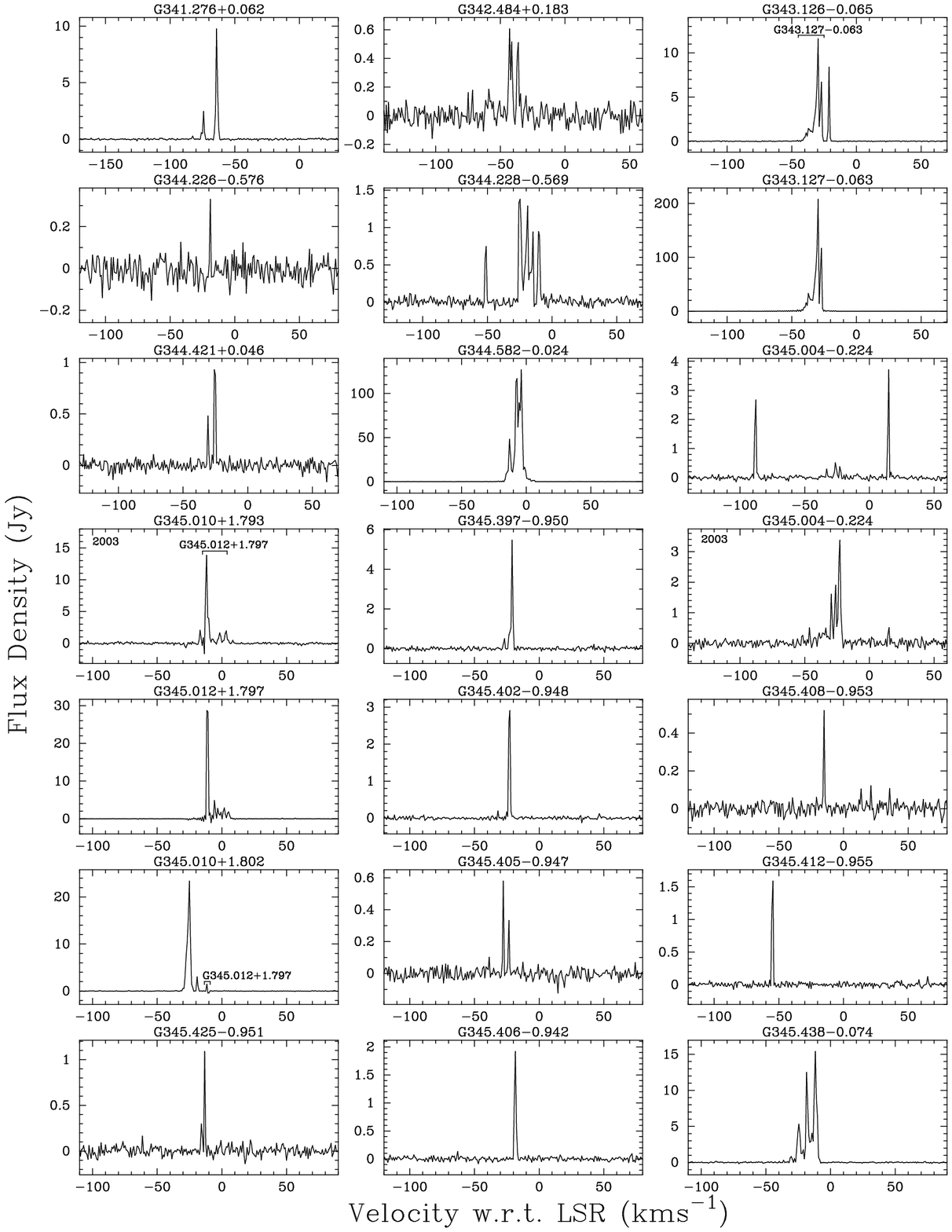}
\caption{--{\emph {continuued}}}
\end{figure*}

\begin{figure*}\addtocounter{figure}{-1}
	\psfig{figure=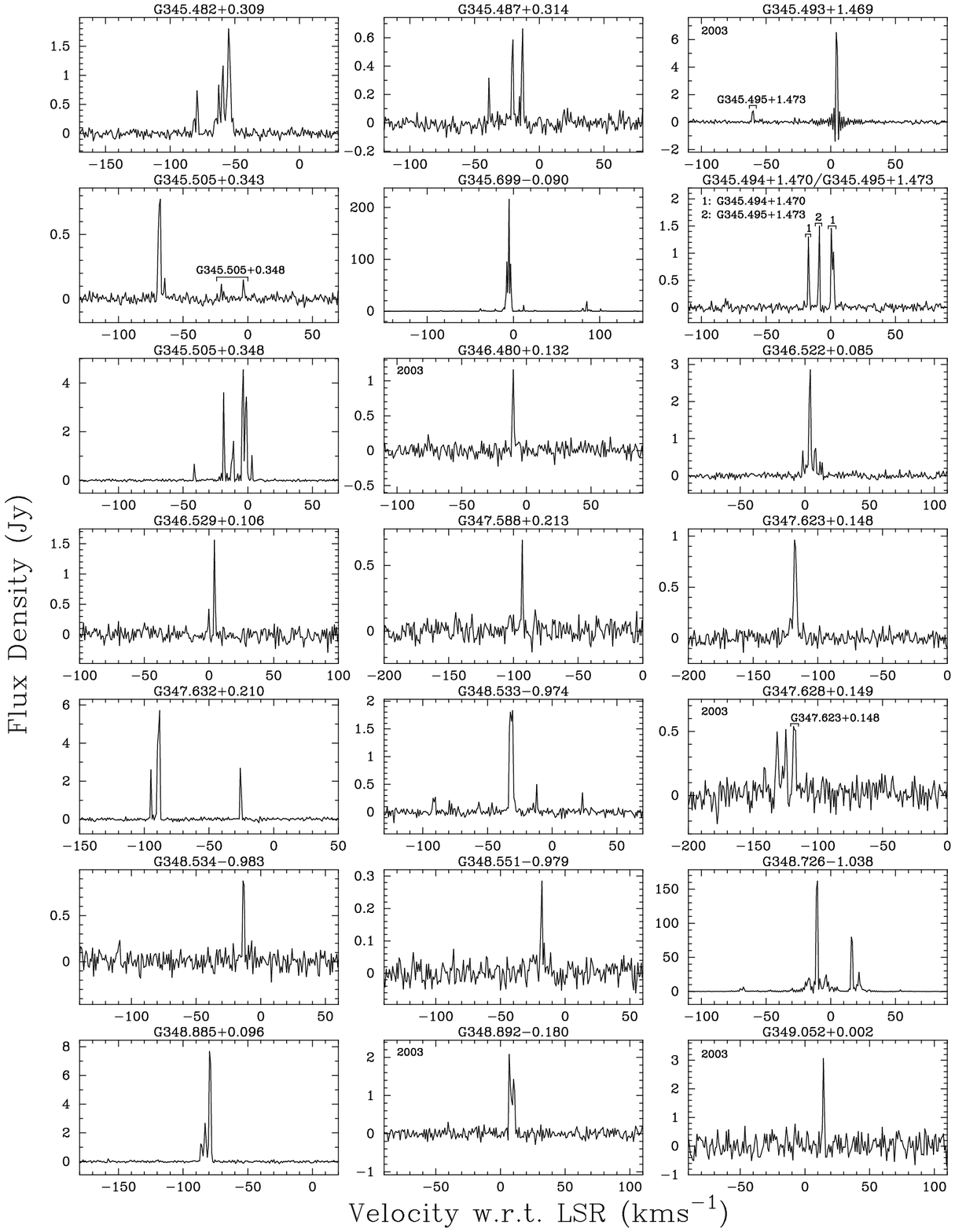}
\caption{--{\emph {continuued}}}
\end{figure*}

\begin{figure*}\addtocounter{figure}{-1}
	\psfig{figure=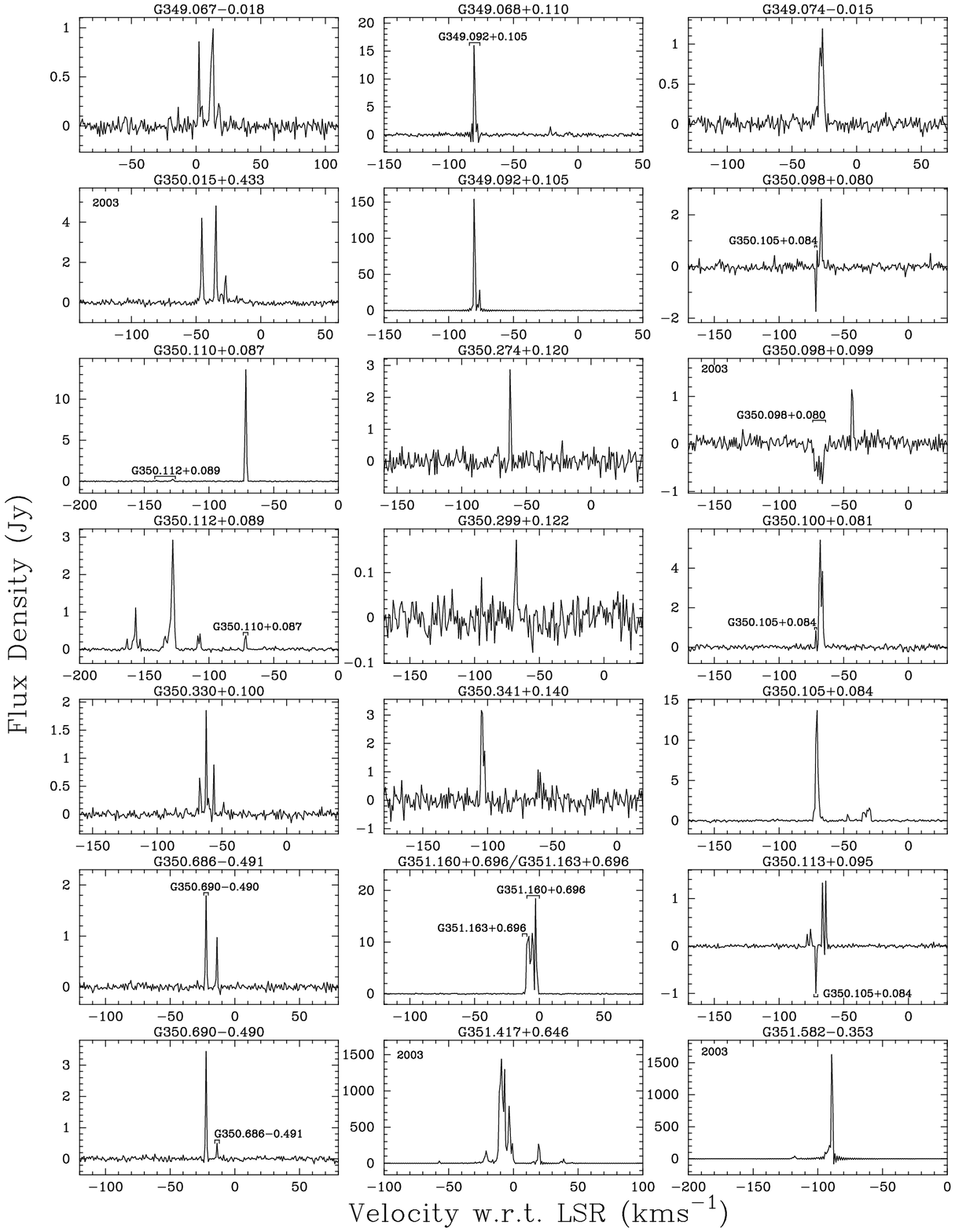}
\caption{--{\emph {continuued}}}
\end{figure*}

\begin{figure*}\addtocounter{figure}{-1}
	\psfig{figure=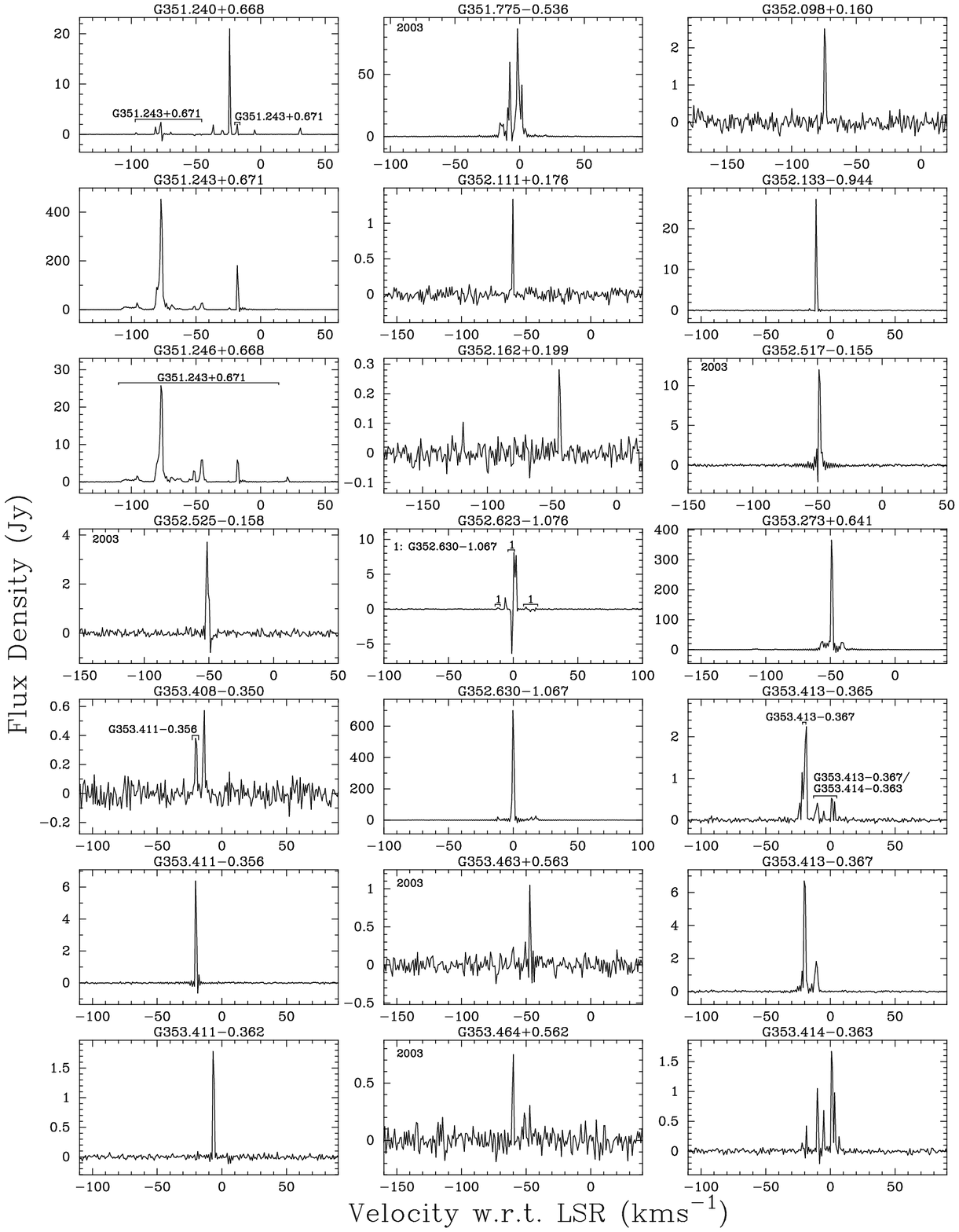}
\caption{--{\emph {continuued}}}
\end{figure*}

\begin{figure*}\addtocounter{figure}{-1}
	\psfig{figure=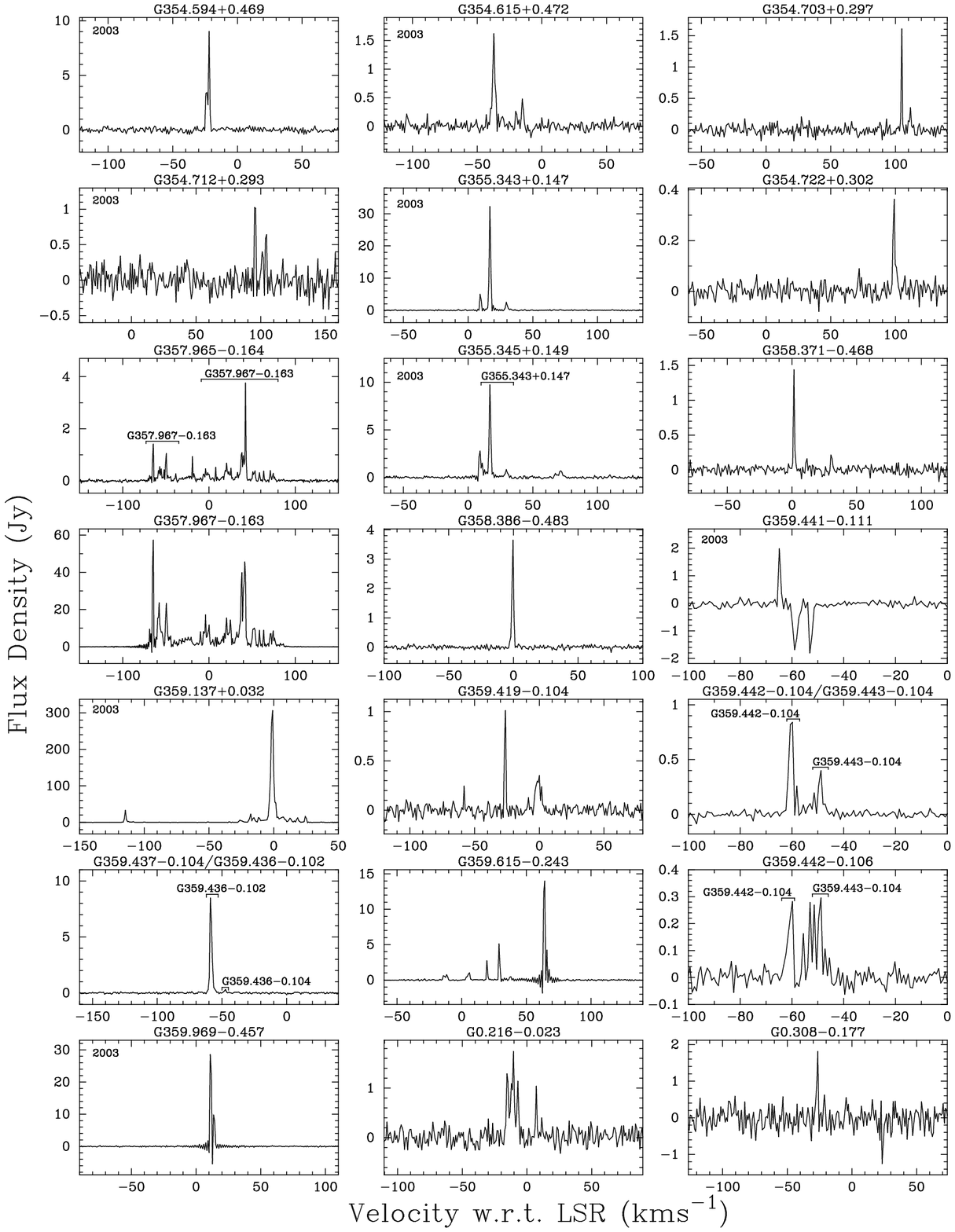}
\caption{--{\emph {continuued}}}
\end{figure*}

\begin{figure*}\addtocounter{figure}{-1}
	\psfig{figure=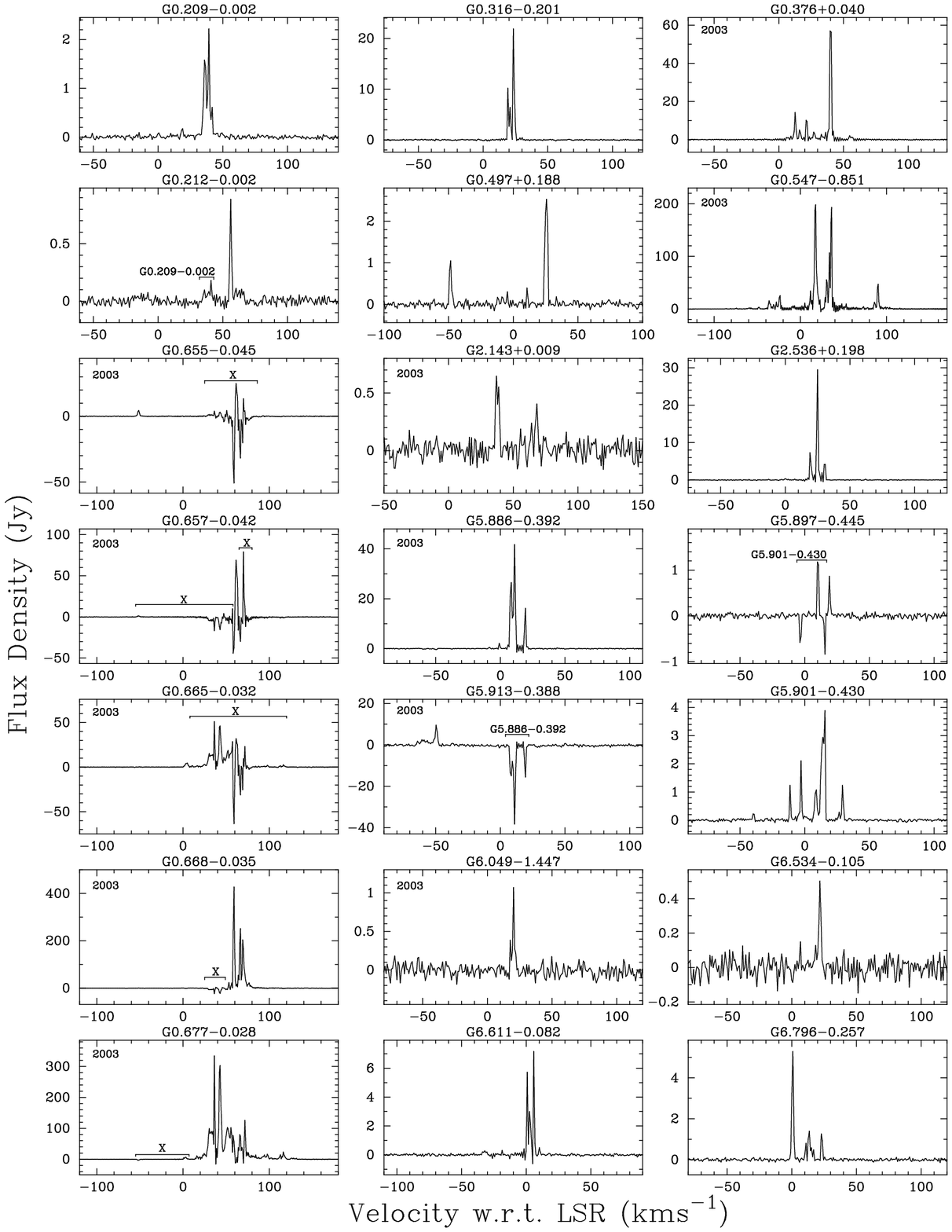}
\caption{--{\emph {continuued}}}
\end{figure*}

\begin{figure*}\addtocounter{figure}{-1}
	\psfig{figure=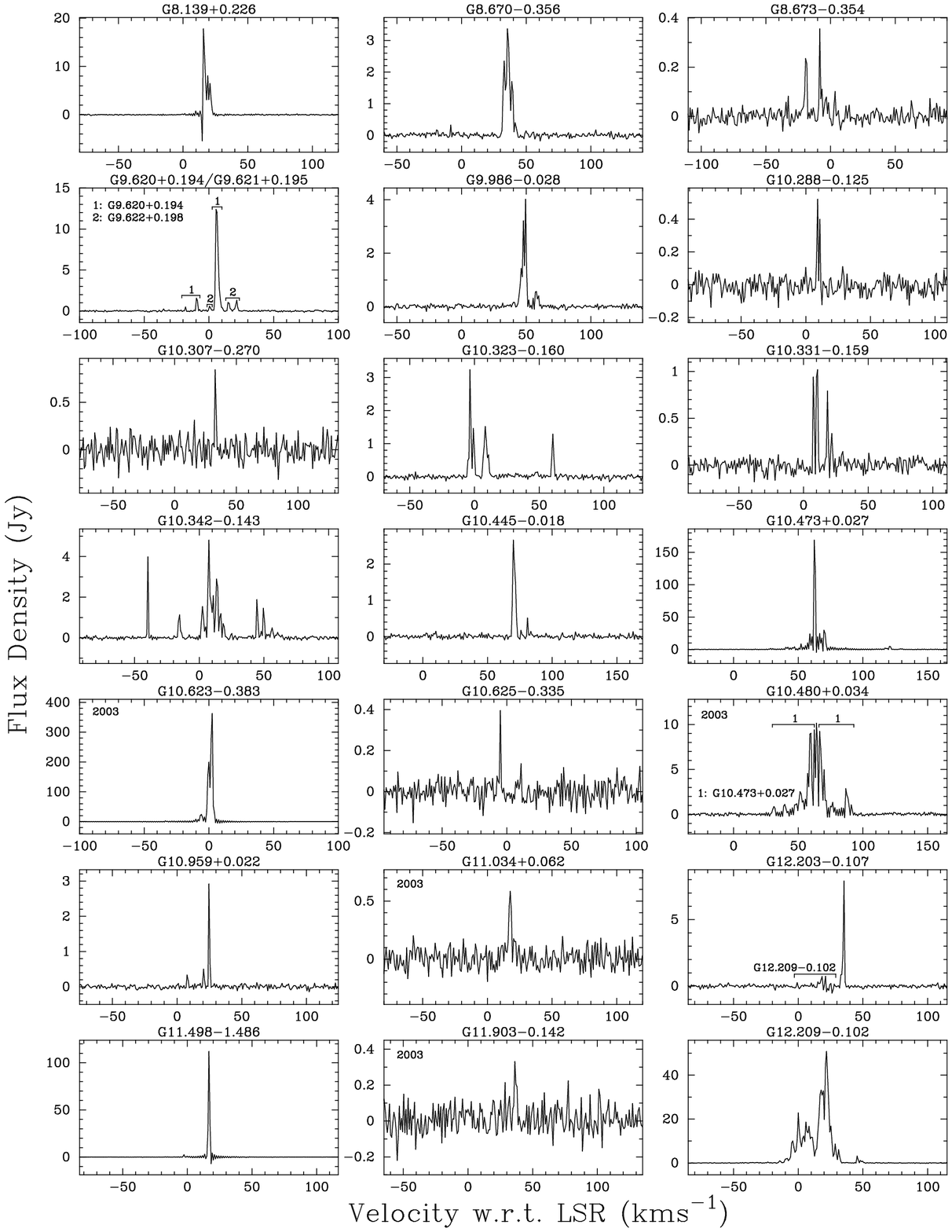}
\caption{--{\emph {continuued}}}
\end{figure*}

\clearpage

\begin{figure*}\addtocounter{figure}{-1}
	\psfig{figure=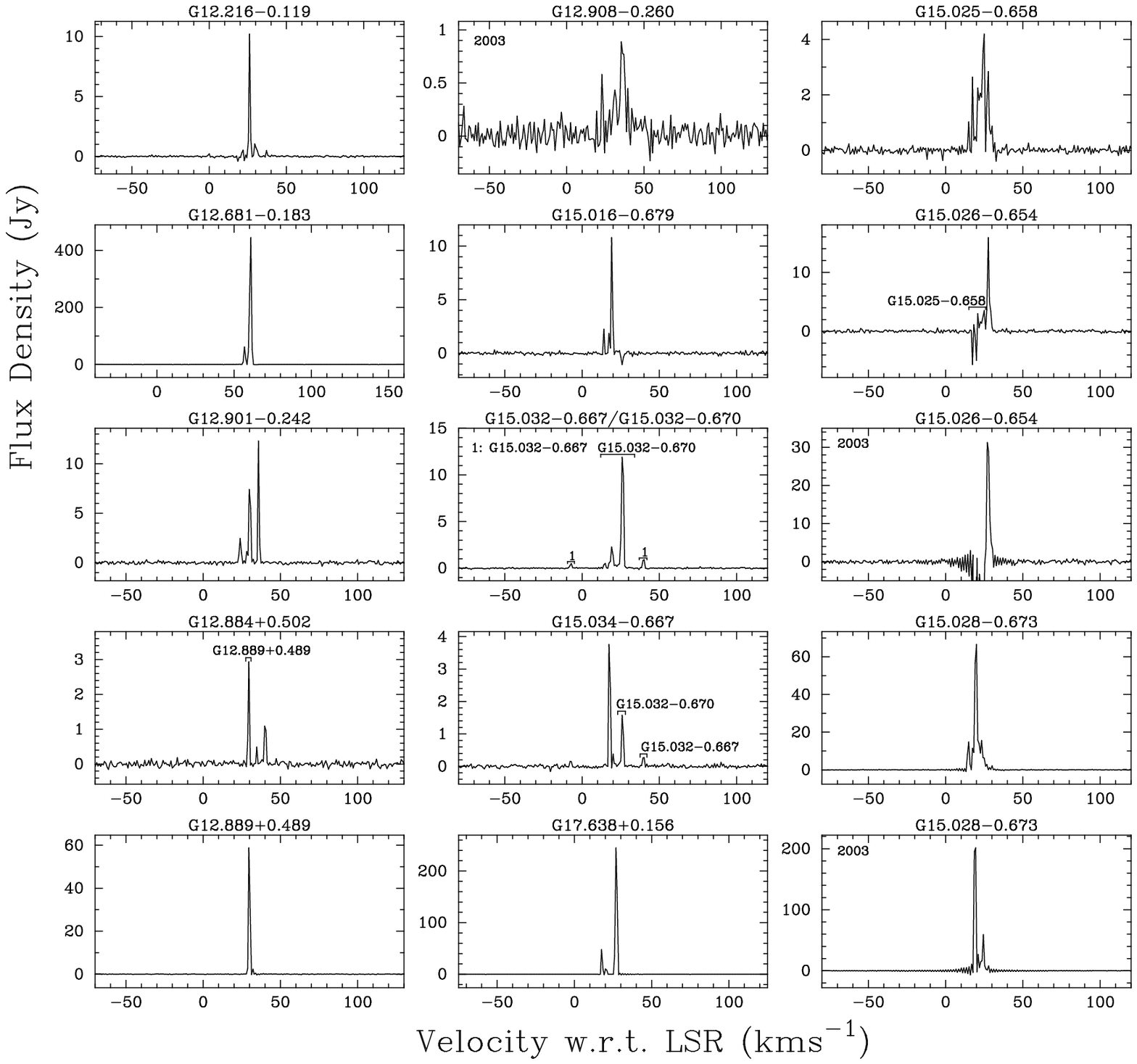}
\caption{--{\emph {continuued}}}
\end{figure*}

\begin{table*}
  \caption{Water maser sources with associated OH and methanol masers as well as 22-GHz continuum emission. Column 1 shows the water maser source name; column 2 gives the source name of the nearest OH maser within 5 arcsec (-- if none) of the detected water maser; column 3 gives the angular separation between the water and the OH masers; column 4 gives the source name of the nearest methanol maser within 5 arcsec (-- if none) of the detected water maser; column 5 gives the angular separation between the water maser and the methanol maser; columns 6, 7 and 8 give the water maser, OH maser and methanol maser peak velocities; column 9 gives detected \UCHII regions with 5 arcsec of the detected water masers (-- if none); and column 10 gives the angular separation between the \UCHII region and the detected water maser.}
  \begin{tabular}{llclcllrlcl} \hline
    \multicolumn{1}{c}{\bf Water} & {\bf OH} & {\bf Sep.} & {\bf Methanol}  & {\bf Sep.} & {\bf Water}& {\bf OH} & {\bf Methanol} & {\bf Continuum} & {\bf Sep.} \\
     \multicolumn{1}{c}{\bf ($l,b$)}  &{\bf ($l,b$)} & &{\bf ($l,b$)} & & {\bf Vpeak} &  {\bf Vpeak} &  {\bf Vpeak}&{\bf ($l,b$)} \\	
      \multicolumn{1}{c}{\bf (degrees)} & {\bf (degrees)} & {\bf (arcsec)}&{\bf (degrees)} &{\bf (arcsec)}& {\bf (\kms)} & {\bf (\kms)} &{\bf (\kms)} & {\bf (degrees)} &{\bf (arcsec)} \\ \hline \\
G\,240.316+0.071		&	G\,240.316+0.071	&	0.7	&	--				&            	&   	89	&	63	&		&		--				&		&	\\ 
G\,263.250+0.514	 	&	G\,263.250+0.514	&	2.1	&	G\,263.250+0.514	&	1.5   	& 	20	&	15.3	&	12.3	&		--				&		&	\\
G\,284.350--0.418	 	&	G\,284.351--0.418	&	1.1	&	--				&      		&	7	&	6	&		&		--				&		&	\\
G\,285.263--0.050		&	G\,285.263--0.050 	&	2.0	&	--				&		&	3	&	6	&		&	--					&		&\\
G\,287.371+0.644	 	&	G\,287.371+0.644 	&	1.6	&	G\,287.371+0.644	&	1.5    &	--11	&	--4	&	--1.8	&		--				&		&\\
G\,290.374+1.661	 	&	G\,290.374+1.661 	&	1.6	&	G\,290.374+1.661 	&	1.0    &	--12	& --23.3	& --24.2	&		--				&		&\\
G\,291.270--0.719	 	&	--				&	 	&	G\,291.270--0.719	&      2.5    &	--102	&		& --26.5	&		--				&		&\\
G\,291.274--0.709	 	&	G\,291.274--0.709	&	1.3	&	G\,291.274--0.709	&	0.7   	&	--32	& --24.5	& --29.6	&		--				&		&\\
G\,291.579--0.431	 	&	G\,291.579--0.431	&	0.6	&   	G\,291.579--0.431  	&	0.7   	&	13	&	13	&	14.5	&		--				&		&\\
G\,291.581--0.435	 	&	--				&		&	G\,291.582--0.435	&      3.8    &	26	&		&	10.5	&		--				&		&	\\	
G\,291.610--0.529	 	&	G\,291.610--0.529 	&	0.7	&	--				&            	&	12	&	18	&		&	G\,291.611--0.529		&	2.6 	&\\
G\,291.627--0.529	 	&	--				&		&	--				&      		&		&		&		&	G\,291.626--0.531 		&	4.8	&	\\
G\,294.511--1.622	 	&	G\,294.511--1.621 	&	2.1	&	G\,294.511--1.621	&	1.8   	&	--12	&--12.7	&--12.3	&		--				&		&\\
G\,294.989--1.719	 	&	--				&		&	G\,294.990--1.719	&      2.5    & 	--17	&		&	--12.3	&		--				&		&\\
G\,297.660--0.974		&	G\,297.660--0.973	&	2.0	&	--				&            	&	26	&	27.6	&		&	--					&		&	\\
G\,299.013+0.128	 	&	G\,299.013+0.128	&	1.2	&	G\,299.013+0.128	&	1.1    &	19	&	20.3	&	18.4	&	G\,299.012+0.128		&	3.3	&	\\
G\,300.504--0.176	 	&	G\,300.504--0.176 	&	0.6	&	G\,300.504--0.176	&	1.8    &	11	&	22.4	&	7.5	&	--			     		&		&\\
G\,301.136--0.226b	 	&	G\,301.136--0.226	&	2.0	&	G\,301.136--0.226	&	 2.5  	&	--44	&  --40.2	&	--39.8&	G\,301.136--0.226     		&	1.0	&\\		
G\,301.137--0.225	 	&	G\,301.136--0.226	&	2.0	&	G\,301.136--0.226	&      2.0    &	--35	& --40.2	&--39.8	&	--					&		&\\
G\,305.208+0.207		&      G\,305.208+0.206	&	3.0	&	G\,305.208+0.206  	&	2.7	&	--42	&	--38	&	--38.3	&	--					&		&\\
G\,305.361+0.150	 	&	G\,305.362+0.150 	&	2.1	&	G\,305.362+0.150  	&	2.0	&	--36	&	--39.5	&	--36.5	&	--				    	&		&\\
G\,305.799--0.245	 	&	G\,305.799--0.245	&	3.0	&	G\,305.799--0.245  	&    	2.5    & 	--34	&	--36.7	&	--39.5	&	--			&\\	
G\,307.805--0.456	 	&	G\,307.805--0.456	&	1.5	&	--				&		& 	--7	&	--14.5	&		&		--			&\\
G\,308.754+0.549	 	&	G\,308.754+0.549	&	0.8	&	G\,308.754+0.549	&	1.4    &	--48	&	--43.5	&	--51.0	&	--			&\\
G\,308.918+0.124	 	&	G\,308.918+0.123 	&	3.0	&	G\,308.918+0.123	&	3.6    &	--61	&	--54	&	--54.7	&		--			&\\
G\,309.384--0.135	 	&	G\,309.384--0.135	&	1.3	&	G\,309.384--0.135	&	0.6    &	--50	&	--52	&	--49.6	&		--			&\\
G\,310.144+0.760	 	&	G\,310.144+0.760	&	2.2	&	G\,310.144+0.760   	&	1.0	&	--63	&	--57	&	--55.6	&		--			&\\
G\,311.643--0.380	 	&	G\,311.643--0.380 	&	1.3	&	G\,311.643--0.380   	&	0.4	&	36	&	38	&	32.5	&	G\,311.643--0.380		&	1.3	&\\
G\,312.109+0.262	 	&	--				&		&	G\,312.108+0.262	&     1.9     &	--48	&		& 	--50.0	&		--			&\\
G\,312.596+0.045	 	&	--				&		&	G\,312.597+0.045	&     1.6     &	--59	&		&	--60.0	&   		--			&\\	
G\,312.599+0.046	 	&	G\,312.598+0.045	&	2.1	&	G\,312.598+0.045 	&	2.1   	&	--79	&	--65.2	&	--67.9	&	--			&\\	
G\,313.457+0.193	 	&	--				&		&	--			        	&            	&    		&		&		&	G\,313.458+0.193		&     2.1	&\\	
G\,313.470+0.191	 	&	G\,313.469+0.190	&	0.9	&	G\,313.469+0.190	&	1.1    &	--15	&	--10	&	--9.4	&--\\	
G\,313.578+0.325	 	&	G\,313.577+0.325	&      1.0	&    	G\,313.577+0.325	&	1.9    &	--47	&	--47	&	--47.9	&	--	\\
G\,313.767--0.862	 	&	G\,313.767--0.863	&      1.1	&	G\,313.767--0.863	&	0.9    &	--54	&	--53.5	&	--54.6	&--\\	
G\,314.320+0.112	 	&	G\,314.320+0.112 	&	2.2  	&	G\,314.320+0.112	&	2.3    &	--45	&	--45	&	--43.7	&--\\
G\,316.361--0.363	 	&	--				&		&	G\,316.359--0.362	&     3.2     	&	--3	&		&	3.5	&--\\
G\,316.412--0.308	 	&	G\,316.412--0.308  	&	1.5	&	G\,316.412--0.308	&	2.3   	&	--20	&	--2	&	--5.7	&	G\,316.412--0.308  		&	0.7	&\\
G\,316.640--0.087	 	&	G\,316.640--0.087  	&	0.7	&	G\,316.640--0.087 	&	0.9   	&	--15	&	--22	&	--19.8	&	--\\
G\,316.763--0.011	 	&	G\,316.763--0.012 	&	1.0	&	--				&            	&	--48	&	--40	&		&--\\			 	
G\,316.812--0.057	 	&	G\,316.811--0.057	&	2.2	&	G\,316.811--0.057	&	2.5   	&	--46	&	--43.5	&	--46.3	&--\\
G\,317.429--0.561	 	&	G\,317.429--0.561 	&	2.1	&	--				&            	&	25	&	25.5	&		&	G\,317.430--0.561		&	2.6	&\\
G\,318.044--1.404	 	&	G\,318.044--1.405	&	2.0	&	G\,318.043--1.404	&	1.6   	&	42	&	45	&	46.2	&	--\\	
G\,318.050+0.087	 	&	G\,318.050+0.087	&	0.6	&	G\,318.050+0.087	&	0.4    &	--48	&	--53	&	--46.5	&	--\\
G\,318.948--0.196b	 	&	G\,318.948--0.196	&	0.8	&	G\,318.948--0.196	&	0.9    &	--38	&	--35.5	&	--34.7	&	--\\
G\,319.399--0.012	 	&	G\,319.398--0.012	&	1.1	&	--				&            	&	--5	&	--1	&		&	G\,319.399--0.012		&	0.8	&\\		 
G\,319.836--0.196	 	&	G\,319.836--0.196 	&	1.5	&	G\,319.836--0.197	&	1.6    &	--11	&	--10.5	&	--9.1	&	--\\
G\,320.120--0.440	 	&	G\,320.120--0.440 	&	0.5	&	--				&            	&	--46	&	--55.5	&		&	--\\		
G\,320.232--0.284		&	G\,320.232--0.284 	&	0.4	&	G\,320.231--0.284	&	0.6  	&	--67	&	--64	&	--66.5	&	--\\
G\,320.233--0.284	 	&	--				&		&	--				&      		&		&		&		&	G\,320.234--0.283		&	3.5	&\\
G\,321.033--0.483	 	&	--				&		& 	G\,321.033--0.483	&	0.5    &	--61	&		&	--61.6	&	--\\
G\,321.148--0.529	 	&	G\,321.148--0.529	&	1.1	&	G\,321.148--0.529	&	1.1	&	--97	&	--63	&	--66.1	&	--\\
G\,322.158+0.636	 	& 	G\,322.158+0.636	&	1.2	&	G\,322.158+0.636 	&	1.2	&	--76	&	--61	&	--63.3	&		--\\			
G\,323.740--0.263	 	&	G\,323.740--0.263	&	1.1	&	G\,323.740--0.263   	&	0.6	&	--50	&	--39	&	--51.1	&	--	&\\
\hline

\end{tabular}
\label{tab:ass}
\end{table*}

\begin{table*}\addtocounter{table}{-1}
  \caption{-- {\emph {continued}}}
   \begin{tabular}{llclcllrlcl} \hline
     \multicolumn{1}{c}{\bf Water} & {\bf OH} & {\bf Sep.} & {\bf Methanol}  & {\bf Sep.} & {\bf Water}& {\bf OH} & {\bf Methanol} & {\bf Continuum} & {\bf Sep.} \\
     \multicolumn{1}{c}{\bf ($l,b$)}  &{\bf ($l,b$)} & &{\bf ($l,b$)} & & {\bf Vpeak} &  {\bf Vpeak} &  {\bf Vpeak}&{\bf ($l,b$)} \\	
      \multicolumn{1}{c}{\bf (degrees)} & {\bf (degrees)} & {\bf (arcsec)}&{\bf (degrees)} &{\bf (arcsec)}& {\bf (\kms)} & {\bf (\kms)} &{\bf (\kms)} & {\bf (degrees)} &{\bf (arcsec)} \\ \hline \\
G\,324.201+0.122	 	&	G\,324.200+0.121	&	2.9	&	--		   		&            	&	--87	&	--91.5	&		&	--	&	\\		
G\,324.716+0.342	 	&	G\,324.716+0.342	&	1.7	&	G\,324.716+0.342 	&	1.5   	&	--58	&	--50	&	--46	&	--	&\\
G\,326.662+0.521	 	&	--				&		&	G\,326.662+0.521	&      2.0    &	 --39	&		&	--38.6	& 	--\\
G\,326.670+0.554	 	&	G\,326.670+0.554	&	2.6	&	--				&           	&	--40	&	--40.8	&		&	--\\
G\,326.780--0.241		&	G\,326.780--0.241 	&	0.9	&	--				&	     	&      	--66	&	--65	&		&	--\\		
 G\,326.859--0.676		&	--				&		& 	G\,326.859--0.677	&	3.4	&	--103	&		&	--58.0	&	--\\     
G\,327.119+0.511		&	G\,327.120+0.511	&	2.3	&	G\,327.120+0.511 	&	1.8	&	--88	&	--80.5	&	--87.0	&	--\\
G\,327.291--0.578	 	&	G\,327.291--0.578  	&	1.2	&	G\,327.291--0.578	&	0.8	&	--63	&	--50.5	&	--36.8	&	--\\
G\,327.391+0.200	 	& 	--				&		&    	G\,327.392+0.199	&	1.5    &	--86	&		&	--84.6	&	--\\
G\,327.402+0.445		&	G\,327.402+0.444	&	3.2	&	G\,327.402+0.444 	&	1.6	&	--81	&	--77	&	--82.6	&	G\,327.402+0.445		&	0.6	&\\
G\,327.619--0.111	 	&	--				&		&     G\,327.618--0.111	&	0.9	&	--85	&		&	--97.6	&	--\\
G\,328.236--0.548	 	&  	G\,328.237--0.547 	&	2.9	&     G\,328.237--0.547 	&	2.6	&	--38	&	--41	&	--44.5	&	G\,328.236--0.547		&	2.1	&\\
G\,328.254--0.532	 	&	G\,328.254--0.532	&	1.5	&     G\,328.254--0.532 	&	0.8	&	--50	&	--37	&	--37.5	&	--\\
G\,328.306+0.432	 	&	--				&		&     --				&            	&		&		&		&	G\,328.307+0.431		&	3.3	&\\
G\,328.808+0.633	 	&	G\,328.809+0.633	&	2.9	&     G\,328.808+0.633	&	2.4	&	--46	&	--43.5	&	--43.8	&	G\,328.808+0.633 		&	2.0	&	\\	
G\,329.029--0.199		&	G\,329.029--0.200	&	1.9	&     --	      			&           	&	--38	&	--38.5	&		&	--	\\		   
G\,329.030--0.205	 	&	G\,329.029--0.205   	& 	1.5	&	G\,329.029--0.205	&	1.3	&	--46	&	--38.5	&	--37.4	&	--\\
G\,329.031--0.198	 	&	G\,329.031--0.198   	&	1.3	&	G\,329.031--0.198	&	0.8	&	--52	&	--45.5	&	--45.5	&	--\\
G\,329.066--0.307	 	&	G\,329.066--0.308   	&	1.5	&	G\,329.066--0.308	&	1.2	&	--45	&	--43.5	&	--43.8	&	--	\\
G\,329.183--0.313	 	&	G\,329.183--0.314   	&	2.3	&	G\,329.183--0.314	&	1.9	&	--50	&	--53	&	--55.7	&	--\\
G\,329.405--0.459	 	&	G\,329.405--0.459	&	2.0	&	G\,329.405--0.459	&	1.5	&	--77	&	--69.5	&	--70.5	&--		\\
G\,329.407--0.459	 	&	--				&		&	G\,329.407--0.459	&	2.1   	& 	--74	&		&	--66.7	&	--\\
G\,329.622+0.138	 	&	--				&		&	G\,329.622+0.138	&	1.7    &	--82	&		&	--84.8	&	--\\
G\,330.070+1.064	 	&	--				&		&	G\,330.070+1.064	&	1.2    &	--50	&		&	--38.8	&	--\\
G\,330.879--0.367		&	G\,330.878--0.367a	&	0.9	&	G\,330.878--0.367	&	2.6	&	--60	&	--61.8	&	--59.3	&	G\,330.879--0.367		&	1.7	&\\
					&	G\,330.878--0.367b	&	1.2	&	--				&           	&		&	--65.6	&		&	\\
G\,330.954--0.182	 	&	G\,330.954--0.182	&	1.3	&	G\,330.953--0.182	&	3.9	&	--91	&	--85.5	&	--87.6	&	G\,330.954--0.182		&	1.6	&	\\	
G\,331.132--0.244	 	&	G\,331.132--0.244	&	0.2	&	G\,331.132--0.244 	&	0.3	&	--99	&	--88.5	&	--84.3	&	--\\
G\,331.278--0.188	 	&	G\,331.278--0.188   	&	0.9	&	G\,331.278--0.188	&	1.1	&	--90	&	--89.5	&	--78.2	&	--	\\
G\,331.342--0.346	 	&	G\,331.342--0.346   	&	1.4	&	G\,331.342--0.346	&	1.6	&	--62	&	--67	&	--67.4	&	--\\	
G\,331.442--0.187	 	&	G\,331.442--0.186	&	0.5	&	G\,331.442--0.187	&	0.9	&	--88	&	--83	&	--88.4	&	G\,331.443--0.187		&	3.5	&\\
G\,331.512--0.103	 	&	G\,331.512--0.103	&	1.4	&	--				&	      	&	--90	&	--88.2	&		&	G\,331.512--0.103		&	1.0	&\\	
G\,331.555--0.122	 	&	G\,331.556--0.121 	&	4.5	&	G\,331.556--0.121	&	4.4	&	--99	&	--100	& --103.4	&	--\\		 
G\,332.094--0.421	 	&	--				&		&	G\,332.094--0.421	&	2.2    &	--59	&		&	--58.6	&	--\\
G\,332.296--0.094	 	&	--				&		&	G\,332.295--0.094	&	4.4    &	--50	&		&	--47.0	&	--\\	 
G\,332.352--0.117	 	&	G\,332.352--0.117	&	0.8	&	G\,332.352--0.117	&	0.2	&	--60	&	--44	&	--41.8	&	--\\
G\,332.604--0.167	 	&	--				&		&	G\,332.604--0.167	&	1.6  	&	--46	&		&	--50.9	&	--\\	
G\,332.725--0.621	 	&	G\,332.726--0.621	&	1.4	&	G\,332.726--0.621	&	1.0	&	--58	&	--48	&	--49.6	&	--\\	
G\,332.826--0.549	 	&	--				&		&	G\,332.826--0.549	&	3.0    &	--59	&		&	--61.7	&	G\,332.826--0.549		&	1.1	&\\
G\,332.964--0.679	 	&	--				&		&	G\,332.963--0.679	&	1.6    &	--52	&		&	--45.8	&	--\\
G\,333.030--0.063	 	&	--				&		&	G\,333.029--0.063	&	1.3    &	--40	&		&	--55.2	&	G\,333.030--0.063		&	0.7	&\\ 
G\,333.121--0.434	 	&	--				&		&   	G\,333.121--0.434	&	1.2    &	--47	&		&	--49.3	&	--\\
G\,333.126--0.440	 	&	--				&		&	G\,333.126--0.440	&	1.0    &	--52	&		&	--43.9	&	--\\
G\,333.128--0.440	 	&	--				&		&  	G\,333.128--0.440	&	2.5    &	--124	&		&	--44.6	&	--\\
G\,333.234--0.060	 	&	G\,333.234--0.060	&	0.6	&	--				&	      	&	--88	&	--84	&		&	--\\	
G\,333.315+0.106	 	&    	G\,333.315+0.105 	&	2.8	&	G\,333.315+0.105	&	3.0	&	--48	&	--47	&	--45	&	--\\
G\,333.387+0.032	 	&    	G\,333.387+0.032 	&	0.6	&	G\,333.387+0.032	&	1.1	&	--61	&	--74	&	--73.9	&	--\\
G\,333.467--0.164	 	&    	G\,333.466--0.164	&	2.1	&	G\,333.466--0.164	&	2.5	&	--42	&	--43.5	&	--42.5	&	G\,333.466--0.163		&	4.7	&\\
G\,333.608--0.215	 	&    	G\,333.608--0.215  	&	0.2	&	--				&	      	&	--49	&	--51	&		&--\\
G\,333.646+0.058	 	&	--				&		&  	G\,333.646+0.058 	&	0.9	&	--89	&		&	--87.3	&		--\\
G\,333.682--0.436	 	&	--				&		&    	G\,333.683--0.437	&	1.6	&	--3	&		&	--5.3	&	--\\	
G\,333.930--0.134	 	&	--				&		&    	G\,333.931--0.135	&	1.5	&	--46	&		&	--36.7	&	--\\
G\,334.635--0.015	 	&	--				&		&    	G\,334.635--0.015	&	1.0	&	--26	&		&	--30	&	--\\
G\,334.935--0.098	 	&	--				&		&    	G\,334.935--0.098	&	0.4	&	--17	&		&	--19.5	&	--\\
G\,335.060--0.428	 	&	G\,335.060--0.427 	&	1.5	&	G\,335.060--0.427	&	1.3	&	--37	&	--36	&	--47.0	&	--\\
G\,335.585--0.285	 	&    	G\,335.585--0.285  	&	0.5	&	G\,335.585--0.285	&	0.7	&	--42	&	--48	&	--49.3	&	--\\
G\,335.586--0.290	 	&	G\,335.585--0.289	&	1.1	&	G\,335.585--0.289	&	0.8	&	--56	&	--53.5	&	--51.4	&	--\\
					&					&		&	G\,335.585--0.290	&      2.3    &		&		&	--47.3	&\\		
G\,335.727+0.191	 	&	--				&		& 	G\,335.726+0.191	&	2.0    &	--51	&		&	--44.4	&--	\\
G\,335.789+0.174	 	&	G\,335.789+0.174	&	0.5	&	G\,335.789+0.174	&	0.9    &	--46	&	--51.5	&	--47.6	&--	\\	
 \hline

\end{tabular}
\end{table*}

\begin{table*}\addtocounter{table}{-1}
  \caption{-- {\emph {continued}}}
  \begin{tabular}{llclcllrlcl} \hline
    \multicolumn{1}{c}{\bf Water} & {\bf OH} & {\bf Sep.} & {\bf Methanol}  & {\bf Sep.} & {\bf Water}& {\bf OH} & {\bf Methanol} & {\bf Continuum} & {\bf Sep.} \\
     \multicolumn{1}{c}{\bf ($l,b$)}  &{\bf ($l,b$)} & &{\bf ($l,b$)} & & {\bf Vpeak} &  {\bf Vpeak} &  {\bf Vpeak}&{\bf ($l,b$)} \\	
      \multicolumn{1}{c}{\bf (degrees)} & {\bf (degrees)} & {\bf (arcsec)}&{\bf (degrees)} &{\bf (arcsec)}& {\bf (\kms)} & {\bf (\kms)} &{\bf (\kms)} & {\bf (degrees)} &{\bf (arcsec)} \\ \hline \\
G\,336.018--0.827	 	&    	G\,336.018--0.827	&	0.6	&	G\,336.018--0.827	&	0.9	&	--54	&	--41.5	&	--53.4	&	G\,336.018--0.828		&	0.5	&\\				
G\,336.359--0.137	 	&    	G\,336.358--0.137	&	3.0	&	G\,336.358--0.137 	&	3.1	&	--67	&	--82	&	--73.6	&	G\.336.360--0.137		&	3.8	&\\
G\,336.433--0.262	 	&	--				&		&    	G\,336.433--0.262	& 	1.9    &	--89	&		&	--93.3	&	--\\
G\,336.830--0.375	 	&	--				&		&    	G\,336.830--0.375	&	1.6    &	--20	&		&	--22.7	&	--\\ 
G\,336.864+0.005	 	&	G\,336.864+0.005	&	1.2	&	G\,336.864+0.005	&	0.7	&	--66	&	--89	&	--76.1	&	--\\
G\,336.983--0.183	 	&	G\,336.984--0.183	&	4.2	&	G\,336.983--0.183	&	3.0	&	45	&	--80.5&	--80.8	&	G\,336.984--0.184      	&	2.5	&\\
G\,336.991--0.024	 	&	--				&		&	--			 	&            	&		&		&		&	G\,336.990--0.025		&	2.3	&\\
G\,336.994--0.027	 	&	G\,336.994--0.027	&	1.0	&	G\,336.994--0.027  	&	0.6	&	--120	&	--123	& --125.8	& --	\\
G\,337.258--0.101	 	&	G\,337.258--0.101	&	0.7	&	G\,337.258--0.101	&	1.2	&	--69	&	--70	&	--69.3	&	--\\
G\,337.404--0.402	 	&	G\,337.405--0.402	&      1.5	&	G\,337.404--0.402	&	0.7	&	--40	&	--38	&	--39.7	&	G\,337.404--0.403		&	2.0	&\\
G\,337.612--0.060	 	&	G\,337.613--0.060	&      0.6	&	G\,337.613--0.060	&	0.9    & 	--51	&	--42	&	--42	&	--\\
G\,337.687+0.137		&	--				&		&	G\,337.686+0.137	&	2.3	&	--74	&		&	--74.9	&	--\\
G\,337.705--0.053	 	&	G\,337.705--0.053	&      0.6	&	G\,337.705--0.053	&	0.9	&	--49	&	--49	&	--54.6	&	G\,337.706--0.054	& 1.1\\
G\,337.916--0.477	 	&	G\,337.916--0.477	&      0.6	&	--				&	      	& 	--33	&	--51	&		&	--\\		
G\,337.920--0.456	 	&	G\,337.920--0.456	&      0.7	&	G\,337.920--0.456	&	0.9	&	--40	&	--39.5	&	--38.8	&	--\\
G\,337.998+0.137	 	&	G\,337.997+0.136	&      1.3	&	G\,337.997+0.136	&	1.1	&	--38	&	--35.5	&	--32.3	&	--\\
G\,338.075+0.012	 	&	G\,338.075+0.012	&      2.1	&	G\,338.075+0.012	&	2.3	&	--50	&	--47	&	--53.0&	G\,338.075+0.012		&	2.9	&\\
G\,338.075+0.010	 	&	--				&		&	G\,338.075+0.009	&	1.3   	&	--48	&		&	--38.2	&	--\\
G\,338.281+0.542	 	&	G\,338.280+0.542	&      1.2	&	G\,338.280+0.542	&	1.1	&	--64	&	--61	&	--56.8	&	--\\
G\,338.433+0.057	 	&	--				&		&	G\,338.432+0.058	&    	4.4    &	--29	&		&	--30.2	&	--	\\
G\,338.461--0.245	 	&	G\,338.461--0.245	&      0.4	&	G\,338.461--0.245   	&	0.8	&	--52	&	--56	&	--50.4	&	--\\
G\,338.472+0.289	 	&	G\,338.472+0.289	&      1.2	&	G\,338.472+0.289	&	1.1	&	--29	&	--32	&	--30.5	&	--\\
G\,338.562+0.217	 	&	--			       	&		&	G\,338.561+0.218	&	2.0    &	--39	&		&	--40.8	&	--\\
G\,338.567+0.110	 	&	--				&		&	G\,338.566+0.110	&	1.4    &	--76	&		&	--75	&	--\\
G\,338.682--0.084	 	&	 G\,338.681--0.084	&      1.4	&      --				&	      	&	--16	&	--22	&		&	G\,338.681--0.085		&	1.5	& 	\\	
G\,338.920+0.550	 	&	 --			       	&		&	G\,338.920+0.550	&	0.8  	&	--68	&		&	--61.4	&	--\\	       	 
G\,338.925+0.556	 	&	 G\,338.925+0.557	&      0.9	&	G\,338.925+0.557	&	1.3	&	--62	&	--61	&	--62.3	&	--\\	
G\,339.582--0.127	 	&	 --				&		&	G\,339.582--0.127	&	0.6    &	--28	&		&	--31.3	&	--\\	
G\,339.622--0.121	 	&	 G\,339.622--0.121	&      0.8	&      G\,339.622--0.121	&	1.3	&	--33	&	--37.3	&	--32.8	&	--\\
G\,339.762+0.055	 	&	 --			       	&		&	G\,339.762+0.054	&      1.6    &	--57	&		&	--51	&	--\\
G\,339.884--1.259	 	&	G\,339.884--1.259b	&	0.7	&	G\,339.884--1.259 	&	0.8   	&	--51	&	--36	&	--38.7	&	--\\
					&	G\,339.884--1.259a	&	1.2	&	--			        	&            	&		&	--29	&		&	\\
G\,340.054--0.243	 	&	G\,340.054--0.244	&      1.4	&	G\,340.054--0.244	&	0.8	&	--54	&	--53.6	&	--59.7	&	--\\
G\,340.785--0.096	 	&	G\,340.785--0.096   	&      2.2   	&	G\,340.785--0.096	&	1.6	&	--120	&	--102	&  --105.1	&	--\\
G\,341.218--0.212	 	&	G\,341.218--0.212 	&	1.2	&	G\,341.218--0.212	&	1.0	&	--39	&	--37.3	&	--37.9	&	--\\
G\,341.276+0.062	 	&	G\,341.276+0.062	&    	0.9	&	G\,341.276+0.062	&	1.1  	&	--64	&	--73	&	--73.8	&	--\\
G\,342.484+0.183	 	&	--				&		&	G\,342.484+0.183	&      1.1    	&	--43	&		&	--41.8	&	--\\
G\,343.127--0.063	 	&	G\,343.127--0.063	&	2.1	&    	--				&            	&	--30	&	--31.5	&		&	--\\	
G\,344.228--0.569	 	&	G\,344.227--0.569	&     1.4	&	G\,344.227--0.569	&	1.0	&	--25	&	--30.5	&	--19.8	&	--\\
G\,344.421+0.046	 	&	--				&		&	G\,344.421+0.045	&	3.4	&	--26	&		&	--71.5	&	--\\
G\,344.582--0.024	 	&	G\,344.582--0.024	&     2.2	&	G\,344.581--0.024	&	2.5	&	--4	&	--2.3	&	1.4	&	G\,344.582--0.024		&	1.7	&\\
G\,345.004--0.224	 	&	G\,345.003--0.224 	&     3.0	&	G\,345.003--0.224   	&      3.1    &		&		&	--26.2	&	G\,345.004--0.225		&	3.6	&\\
					&			      		&		&	G\,345.003--0.223   	&      3.3	&	15	&	--27	&	--22.5	&	\\
G\,345.010+1.793	 	&	G\,345.010+1.793	&     1.5	&	G\,345.010+1.792	&      2.1 	&	--17	&	--22.5	&	--18	&	G\,345.010+1.792		&	4.3	&\\		
G\,345.012+1.797	 	&	--				&		&	G\,345.012+1.797	&	2.2	&	--12	&		&	--12.7	&	--\\			
G\,345.408--0.953	 	&	G\,345.407--0.952	&     4.6	&	G\,345.407--0.952	&	4.9	&	--15	&	--17.6	&	--14.4	&	G\,345.408--0.952		&	3.2	&\\
G\,345.425--0.951	 	&	--				&		&	G\,345.424--0.951  	&	1.6	&	--13	&		&	--13.5	&	--	\\
G\,345.438--0.074	 	&	G\,345.437--0.074	&     1.9	&	--				&		&	--12	&	--24.3	&		&	--\\
G\,345.487+0.314	 	&	--				&		&	G\,345.487+0.314	&	0.6    &	--13	&		&	--22.6	&	--\\
G\,345.493+1.469	 	&	G\,345.494+1.469	&     3.3	&	--				&	      	&	5	&	--12.7	&		&	--\\
G\,345.505+0.348	 	&	G\,345.504+0.348	&     1.8	&	G\,345.505+0.348	&	2.2	&	--4	&	--19.5	&	--17.7	&	--\\
G\,345.699--0.090	 	&	G\,345.698--0.090	&     1.2	&	--				&	      	&	--5	&	--6	&		&	--\\
G\,346.480+0.132	 	&	G\,346.481+0.132 	&     1.5	&	G\,346.481+0.132	&	1.4	&	--10	&	--8	&	--5.5	&	--\\
G\,346.522+0.085	 	&	--				&		&	G\,346.522+0.085	&	0.7    &	4	&		&	5.5	&	--\\
G\,347.628+0.149	 	&	G\,347.628+0.148	&     2.0	&	G\,347.628+0.149 	&	0.9	&	--125	&	--94.3	&	--96.6	&	--\\
G\,347.632+0.210	 	&	--				&		&	G\,347.631+0.211	&	2.9	&	--88	&		&	--91.9	&	G\,347.632+0.210		&	0.9	&\\
G\,348.551--0.979	 	&	G\,348.550--0.979	&     3.7	&	G\,348.550--0.979n	&	1.9	&	--18	& --19.7	& --20.0	&	--\\
					&			      		&		&	G\,348.550--0.979	&      3.4 	&		&		&	--10	&	\\
G\,348.726--1.038	 	&	--				& 		&	G\,348.727--1.037 	&	4.4	&	--10	&		&	--7.6	&--\\
G\,348.885+0.096	 	&	G\,348.884+0.096	&    1.2	&	G\,348.884+0.096	&	1.4	&	--80	& --73.2	&	--76.2	&	--\\
	
\hline

\end{tabular}
\end{table*}

\begin{table*}\addtocounter{table}{-1}
  \caption{-- {\emph {continued}}}
  \begin{tabular}{llclcllrlcl} \hline
     \multicolumn{1}{c}{\bf Water} & {\bf OH} & {\bf Sep.} & {\bf Methanol}  & {\bf Sep.} & {\bf Water}& {\bf OH} & {\bf Methanol} & {\bf Continuum} & {\bf Sep.} \\
     \multicolumn{1}{c}{\bf ($l,b$)}  &{\bf ($l,b$)} & &{\bf ($l,b$)} & & {\bf Vpeak} &  {\bf Vpeak} &  {\bf Vpeak}&{\bf ($l,b$)} \\	
      \multicolumn{1}{c}{\bf (degrees)} & {\bf (degrees)} & {\bf (arcsec)}&{\bf (degrees)} &{\bf (arcsec)}& {\bf (\kms)} & {\bf (\kms)} &{\bf (\kms)} & {\bf (degrees)} &{\bf (arcsec)} \\ \hline \\
G\,348.892--0.180	 	&	G\,348.892--0.180  	&     0.6	&	G\,348.892--0.180	&	1.0	&	7	&	9.5	&	1.4	&--\\
G\,349.067--0.018	 	&	G\,349.067--0.017	&     1.1	&	G\,349.067--0.017	&	1.1	&	13	&	15	&	6.9	&--\\
G\,349.092+0.105	 	&	G\,349.092+0.106	&     0.8	&	G\,349.092+0.106	&	0.9	&	--80	&	--80	& --80.4	&	--\\
					&			      		&		&	G\,349.092+0.105	&      2.0 	&		&		& --76.5	&\\
G\,350.015+0.433	 	&	G\,350.015+0.433	&     0.1	&	G\,350.015+0.433	&	1.0	&	--35	&	--33	& --31.7	&--	\\ 
G\,350.113+0.095	 	&	G\,350.113+0.095	&     1.3	&	--				&	      	&	--64	&	--71	&		&--	\\	G\,350.105+0.084	 	&	--				&		&	G\,350.104+0.084	&	2.0	&	--71	&		& --68.4	&--	\\
					&					&		&	G\,350.105+0.083	&	3.3  	&		&		& --74.0	&	\\
G\,350.299+0.122	 	&	--				&		&	G\,350.299+0.122	&	0.7	&	--68	&		& --62.1	&--	\\
G\,350.330+0.100	 	&	G\,350.329+0.100	&     2.5	&	--				&		&	--62	&	--64	&		&	G\,350.331+0.099		&	3.9	&\\		
G\,350.686--0.491	 	&	G\,350.686--0.491	&     0.4	&	G\,350.686--0.491	&	1.3	&	--14	& --14.5	& --13.8	&	--\\
G\,351.160+0.696	 	&	G\,351.160+0.697	&     2.5	&	G\,351.160+0.697	&	1.3	&	--3	&	--8.5	&	--5.2	&	G\,351.161+0.696		&	3.0	&\\      
G\,351.243+0.671		&	--				&		& 	G\,351.243+0.671	&	3.4	& 	--77	&		&	2.5	&\\
G\,351.246+0.668	 	&	--				&		&	--				&            	&		&		&		&	G\,351.247+0.667		&	3.6	&\\			      
G\,351.417+0.646	 	&	G\,351.417+0.645	&     3.7	&	G\,351.417+0.646	&      1.7	&	--10	&	--9.1	& --11.2	&	--\\
					&			      		&		&	G\,351.417+0.645	&      3.1    &	       	&		& --10.4	&	\\                                               
G\,351.582--0.353	 	&	G\,351.581--0.353	&     1.6	&	G\,351.581--0.353n	&    	2.1	&	--89 	&  --97.6 	&  --91.1 	&	--\\				      		      
					&			      		&		&	G\,351.581--0.353	&	3.4    &       	&	       	&  --94.4 	&	\\
G\,351.775--0.536	 	&	G\,351.775--0.536	&     1.8	&	G\,351.775--0.536	&	1.9	&	--2     	&   --2     	&  1.3     	&--	\\ 	                                                       
G\,352.111+0.176	 	&	--			      	&		&	G\,352.111+0.176	&      2.8	&	 --60  	&         	&  --54.8 	&  --   	\\
G\,352.133--0.944	 	&	--				&		&	G\,352.133--0.944	&    	2.8	&	--11	&        	&  --16    	&--	\\					      	      
G\,352.162+0.199	 	&	G\,352.161+0.200	&     1.0	&	--				&		&	--45 	&  --42.2 	&         	&--	\\	      
G\,352.517--0.155	 	&	G\,352.517--0.155	&     0.2	&	G\,352.517--0.155	&      0.3	&	--49 	&  --50.6 	& --51.2  	&--	\\			      
G\,352.525--0.158		&	--				&		&       G\,352.525--0.158    &      0.3    &       --51  &                &     --53          & -- \\
G\,352.623--1.076	 	&	--			      	&		&	G\,352.624--1.077   	&   	4.7	&	--6     	&	   	&    5.8   	&--	\\		      
G\,352.630--1.067	 	&	G\,352.630--1.067	&     0.5	&	G\,352.630--1.067	&      0.4	&	0	&	0     	&    --2.8 	&--	\\			      
G\,353.273+0.641	 	&	--				&		&	G\,353.273+0.641   	&     0.3	&	--49	&         	&   --5.2  	&--	\\			      
G\,353.411--0.362	 	&	--				&		&	--				&		&          	&        	&         	&	G\,353.411--0.362		&	1.7	&\\	
G\,353.464+0.562	 	&	G\,353.464+0.562	&     0.9	&	G\,353.464+0.562	&	1.9	&	--60  &    --45  	&   --50.7	&--	\\
G\,354.615+0.472	 	&	G\,354.615+0.472	&     1.9	&	G\,354.615+0.472	&	1.8	&	--38	&  --15.4 	&  --24.6 	&--	\\
G\,355.343+0.147	 	&	G\,355.344+0.147	&     2.0	&	G\,355.344+0.147	&	1.7	&	17	&  19      	&  20      	&--	\\	
					&					&		&	G\,355.343+0.148	&	2.7    &    	 	&	     	&  5.7   	&	\\
G\,355.345+0.149		&	--				&		&      G\,355.346+0.149     &      1.4     &    72       &                &      10      & --   \\
G\,357.965--0.164	 	&	--			      	&		&	G\,357.965--0.164    	&     1.5	&	--19   	&         	&   --8.8  	&	--\\
G\,357.967--0.163	 	&	G\,357.968--0.163	&     1.7	&	G\,357.967--0.163	&     0.5	&	--65   	& --6.3    	& --3.2    	&	--\\
G\,358.371--0.468	 	&	--			  	&		&	G\,358.371--0.468 	&	1.0	&	1       	&        	&     1     	&--	\\
G\,358.386--0.483	 	&	G\,358.387--0.482a	&    	2.3	&	G\,358.386--0.483	&	1.5	&     0       	&	--6.3 	& --6.0   	&	G\,358.387--0.483		&	3.4	&\\
					&	G\,358.387--0.482b 	&	3.3	&	--			 	&	      	&              	&    --7.8 	&           	&	\\
G\,359.137+0.032	 	&	G\,359.137+0.032	&     1.3	&	G\,359.138+0.031	&	1.5	&	 --1	&   --1     	&  --3.9   	&--	\\	
G\,359.436--0.102	 	&		--			&     		&	G\,359.436--0.102	&	0.4	&	--59	&    		&  --53.6 	&--	\\
G\,359.436--0.104	 	&	G\,359.436--0.103	&	1.9	&	G\,359.436--0.104	&	0.7	&	--47	&       --52    	&    --52  	&--	\\
G\,359.615--0.243		&	G\,359.615--0.243	&     0.8	&	G\,359.615--0.243	&	0.6	&	64	&   22.5  	&  22.5   	&--	\\
G\,359.969--0.457	 	&	G\,359.970--0.457	&     1.2	&	G\,359.970--0.457	&	1.3	&     11     	&   15.5  	&  23.0   	&--	\\	
G\,0.209--0.002	  	&		--			&     		&	--			 	&		&           	&         	&         	&	G\,0.209--0.002			&	2.4	&\\		 
G\,0.212--0.002	   	&		--			&     		&    	G\,0.212--0.001  	&	3.7	&	56     &         	&   49.2  	&--	\\		  			 
G\,0.316--0.201	   	&		--			&     		&      G\,0.316--0.201        	&	0.7	&	23	&		&   21     	&--	\\				  
					&					&		&	G\,0.315--0.201	 &     2.1    	&		&          	&   18     	&	\\		 				   	
G\,0.376+0.040	        		&	G\,0.376+0.040  	&	1.3   	&	G\,0.376+0.040	 	&	1.6	&	40   	&      36	&   37.1  	&--	\\
G\,0.497+0.188	        		&	G\,0.496+0.188   	&   	2.1   	&	G\,0.496+0.188	 	&	2.1	&	26	&     --5.5 	&   0.8    	&--	\\		 			  
G\,0.547--0.851	        	&	G\,0.546--0.852   	&  	2.7   	&	G\,0.546--0.852 	 &	3.2	&	20	&	13.5	&    13.8 	&--	\\
G\,0.657--0.042		&	G\,0.658--0.042	&	0.4	&       G\,0.657--0.041         &      4.8    &       62    &       52         &            & -- \\ 
G\,0.668--0.035	        	&	G\,0.666--0.035   	&  	4.0   	&	--		 		&		&	59	&	61	&         	&--	\\					   	
G\,2.143+0.009	     		&    	G\,2.143+0.009     	&	0.8   	&	G\,2.143+0.009    	&	0.1	&	37	&	59.8	&    62.7 	&   -- 	\\		
G\,2.536+0.198	      		&	--				&		&	G\,2.536+0.198	 	&	2.5	&     25	&          	&    3.2   	&--	\\		
G\,5.886--0.392		& G\,5.885--0.392		&	6.3	&	G\,5.885--0.392	&	5.5	&11		&	13.9	&	6.7	& --\\
G\,5.901--0.430	      	&	--				&		& 	G\,5.900--0.430 	 &	1.7	&	14	&		&    10    	& -- 	\\	
G\,6.049--1.447	 	& 	G\,6.048--1.447 	&	2.0	&	--				&		&	20	&     11.2 	&         	&	--\\		
G\,6.611--0.082	    	&	--				&		&      G\,6.610--0.082 	 &	2.1	&	6	&		&   0.7     	&--	\\		   	
G\,6.796--0.257	    	&     G\,6.795--0.257 		&	2.1	&	G\,6.795--0.257  	&	2.4	&	1	&	16.1	&   26.6   	&--	\\			   
G\,8.139+0.226	        		&     --				&		& 	G\,8.139+0.226 	&	1.2	&	16	&	    	&   20.0   	&--	\\
	
 \hline
\end{tabular}
\end{table*}

\begin{table*}\addtocounter{table}{-1}
  \caption{-- {\emph {continued}}}
  \begin{tabular}{llclcllrlcl} \hline
     \multicolumn{1}{c}{\bf Water} & {\bf OH} & {\bf Sep.} & {\bf Methanol}  & {\bf Sep.} & {\bf Water}& {\bf OH} & {\bf Methanol} & {\bf Continuum} & {\bf Sep.} \\
     \multicolumn{1}{c}{\bf ($l,b$)}  &{\bf ($l,b$)} & &{\bf ($l,b$)} & & {\bf Vpeak} &  {\bf Vpeak} &  {\bf Vpeak}&{\bf ($l,b$)} \\	
      \multicolumn{1}{c}{\bf (degrees)} & {\bf (degrees)} & {\bf (arcsec)}&{\bf (degrees)} &{\bf (arcsec)}& {\bf (\kms)} & {\bf (\kms)} &{\bf (\kms)} & {\bf (degrees)} &{\bf (arcsec)} \\ \hline \\
G\,8.670--0.356	        	&     G\,8.669--0.356 		&     2.9	&	G\,8.669--0.356	 &	2.7	&	36	&    39.2  	&  39.3    	&	G\,8.670--0.356			&	1.1	&\\		
G\,9.620+0.194	        		&     G\,9.620+0.194 		&	1.6	&	--				&		&	 6	&	22	&    		&--	\\			
G\,9.622+0.195	        		&     G\,9.621+0.196  	&	2.9	&	G\,9.621+0.196	 	&	3.3	&	22	&	 1.4	&    1.3	&--	\\
G\,9.986--0.028	        	&     --				&		&	G\,9.986--0.028 	 &	1.3	&	49     &     		&    47.1   	&--	\\	
G\,10.288--0.125	       	&     --				&		&  	G\,10.287--0.125	&	1.9	&	9	&	     	&    5        	&--	\\
G\,10.323--0.160	        	&     --				&		& 	G\,10.323--0.160	&	1.5	&	--3	&          	&   10      	&--	\\
G\,10.342--0.143	        	&     --				&		& 	G\,10.342--0.142	&	1.8	&	8	&		&   14.8   	&--	\\
G\,10.445--0.018	 	&     G\,10.444--0.018 	&	3.3	&	G\,10.444--0.018	&	3.7	&	70	&    75.5 	&    73.2   	&--	\\
G\,10.473+0.027	        	&     G\,10.473+0.027   	&	0.9	&	G\,10.473+0.027	&	1.9	&	62    	&    51.5 	&    75      	&	G\,10.473+0.027		&	1.8	&\\
G\,10.480+0.034	        	&     G\,10.480+0.033 	&	4.5	&	G\,10.480+0.033	&	4.5	&	64	&    66    	&     65     	&--	\\
G\,10.623--0.383	        	&     G\,10.623--0.383	&	0.6	&	--		 		&		&     2       	&      --2    	&         	&--	\\		
G\,10.959+0.022	        	&     --				&		& 	G\,10.958+0.022	&	1.4	&	25	&		&     24.4	&       G\,10.959+0.022		&	0.7	&\\				 
G\,11.034+0.062		&     G\,11.034+0.062 	&	1.6	&	G\,11.034+0.062	&	1.5	&	18	&	21.7	&     20.6	&   --    \\				 
G\,11.498--1.486	      	&     --				&		&  	G\,11.497--1.485	&	2.9	&	17	&		&     6.7    	&--	\\     				 
G\,11.903--0.142	        	&     G\,11.904--0.141	&	3.5	&	G\,11.904--0.141	&	4.4	&	 36	&	40.5	&     42.8	&--	\\	       

G\,12.203--0.107	      	&     --				&		&   	G\,12.203--0.107	&	0.2	&	35     &          	&    20.5  	&--	\\
G\,12.209--0.102	 	&     --				&		&      G\,12.209--0.102	&	1.0	&     22     	&          	&    19.8  	&	G\,12.209--0.102		&	3.0	&\\	
G\,12.216--0.119	        	&     G\,12.216--0.119 	&	1.5	&	--     			 	&		&	26     &    27.9  	&          	&	--\\
G\,12.681--0.183	        	&    G\,12.680--0.183    	&      1.0	&	G\,12.681--0.182	&	1.1	&	61     &    64.5  	&    57.6   	&	--\\						 				 	
G\,12.889+0.489	        	&    G\,12.889+0.489  	&	3.0   & 	G\,12.889+0.489	&	2.7	&	30	&	33	&	39.3	&	--\\	
G\,12.908--0.260	        	&    G\,12.908--0.260 	&     1.2	&	G\,12.909--0.260	&    1.8	&	37	&    38	&    39.9   	& 	--\\			 	 
G\,17.637+0.156		&   G\,17.638+0.157		&	2.1	&	G\,17.638+0.157	&	2.2	&	27	&    20     	&    20.7    &    	--\\ \hline

\end{tabular}
\end{table*}

\begin{table*}
  \caption{22-GHz continuum sources detected towards water maser sources (i.e. continuum sources within 4.5 arcsec of detected water masers). See Tables 1 and 3 for details of the water maser sources that these continuum sources are associated with. Columns one through to 5 show: the name of the continuum source in Galactic coordinates, the source right ascension, declination, peak flux density (mJy/beam) and integrated flux density (mJy). We present two additional continuum sources that fall just outside our association threshold and we distinguish these sources with a `$^{\#}$' following the source name in column 1.} \label{tab:cont}
  \begin{tabular}{lllcc} \hline 
    \multicolumn{1}{c}{\bf Continuum} & {\bf RA(2000)}  & {\bf Dec(2000)} & {\bf Ipeak} & {\bf Total Flux} \\
     \multicolumn{1}{c}{\bf ($l,b$)} & & &&{\bf Density}\\	
      \multicolumn{1}{c}{\bf (degrees)}  &{\bf (h m s)}&{\bf ($^{o}$ $'$ $``$)}& {\bf (mJy/beam)} & {\bf (mJy)} \\ \hline \\ 
G\,291.611--0.529	& 11 15 02.62 & --61 15 51.4 &  350			& 446\\
G\,291.626--0.531$^{\#}$	& 11 15 09.66 & --61 16 15.7 & 137			& 357\\
G\,299.012+0.128	& 12 17 24.12 & --62 29 05.6 & 25			& 38\\
G\,301.136--0.226	& 12 35 34.96 & --63 02 31.6 & 1017			& 1116\\
G\,311.643--0.380 	& 14 06 38.73& --61 58 21.4 & 174			& 181  \\
G\,313.458+0.193	& 14 19 35.04 & --60 51 52.1 & 205			& 272      \\
G\,316.412--0.308	& 14 43 23.25 & --60 12 59.5 & 160			& 167\\
G\,317.430--0.561	& 14 51 38.03 & --60 00 19.5 & 32			& 35\\
G\,319.399--0.012	& 15 03 17.60 & --58 36 11.2 & 230			& 298\\
G\,320.234--0.283	& 15 09 52.63 & --58 25 32.4 & 282			& 293\\
G\,327.402+0.445   & 15 49 19.35 & --53 45 13.3 & 86			& 94\\
G\,328.236--0.547	& 15 57 58.15 & --53 59 23.4 & 26			& 36\\
G\,328.307+0.431	& 15 54 06.24 & --53 11 38.8 & 3361			& 3647  \\
G\,328.808+0.633	& 15 55 48.33 & --52 43 07.0 & 1208		  	& 1357\\
G\,330.879--0.367	& 16 10 19.95 & --52 06 05.3 & 324			& 476\\
G\,330.954--0.182	& 16 09 52.51 & --51 54 54.1 & 2954			& 3151\\
G\,331.443--0.187	& 16 12 12.83 & --51 35 10.2 & 40			& 44  \\
G\,331.512--0.103	& 16 12 10.10 & --51 28 37.2 & 109			& 113    \\
G\,332.826--0.549	& 16 20 11.12 & --50 53 13.7 & 2510			& 2670   \\
G\,333.030--0.063	& 16 18 56.94 & --50 23 53.5 & 27			& 25    \\
G\,333.466--0.163$^{\#}$	& 16 21 19.71 & --50 09 45.6 & 141		& 254     \\

G\,336.018--0.828	& 16 35 09.39 & --48 46 47.6 & 85.1			& 80.8    \\
G\,336.360--0.137	& 16 33 29.64 & --48 03 38.8 &  189			& 293\\
G\,336.984--0.184	& 16 36 12.60 & --47 37 57.8 & 51			& 56.8   \\
G\,336.990--0.025	& 16 35 32.48 & --47 31 14.6 & 91			& 102\\
G\,337.404--0.403	&16 38 50.59  & --47 28 02.8  & 117			& 121\\
G\,337.706--0.054	& 16 38 29.83 & --47 00 35.7 & 244			& 262\\
G\,338.075+0.012	& 16 39 39.15 & --46 41 26.2 & 818			& 1144       \\
G\,338.681--0.085	& 16 42 24.19 & --46 18 00.4 & 74			& 74   \\
G\,344.582--0.024	& 17 02 58.03 & --41 41 52.7 & 19			& 21    \\
G\,345.004--0.225	& 17 05 11.36 & --41 29 06.5 & 353			& 355  \\
G\,345.010+1.792	& 16 56 47.85 & --40 14 25.8 & 362			& 367    \\
G\,345.408--0.952	& 17 09 35.62 & --41 35 54.6 & 493			& 608\\
G\,347.632+0.210	& 17 11 36.22 & --39 07 06.2 & 46			& 48\\
G\,350.331+0.099	& 17 20 02.09 & --36 59 12.8 & 55			& 65\\
G\,351.161+0.696	& 17 19 57.65 & --35 57 51.8 & 238			& 271\\
G\,351.247+0.667	& 17 20 19.29 & --35 54 39.4& 1280		&      1567\\
G\,353.411--0.362	& 17 30 26.66 & --34 41 46.0 & 384			& 788        \\
G\,358.387--0.483	& 17 43 37.96 & --30 33 49.2 & 109			& 113     \\
G\,0.209--0.002	& 17 46 07.57 & --28 45 30.5 & 75			& 95\\
G\,8.670--0.356	& 18 06 19.15 & --21 37 32.1 & 677			& 689\\
G\,10.473+0.027   	& 18 08 38.41 &  --19 51 47.9 & 152			& 151\\
G\,10.959+0.022 	& 18 09 39.43 &  --19 26 27.0 & 150			&153\\
G\,12.209--0.102	& 18 12 39.85 &  --18 24 20.0 & 119			& 140\\ \hline

\end{tabular}
\end{table*}

\section{individual sources}\label{sect:ind}

Here we draw attention to information on the sources that can not be 
adequately conveyed in the tables and spectra.  We discuss 
some entries in the Table 1 where close companions may be either discrete 
extra maser sites, or merely multiple features in an unusually extended 
site.  Our interpretation of the likely systemic velocity, based on 
association with methanol or OH maser emission, is given in some cases 
where water shows high velocity features that dominate the spectrum.  
Extreme variability is sometimes evident from Table 1, and a few examples 
of this variability are demonstrated by spectra shown from both 2003 and 
2004.  The absence of an entry at one epoch is occasionally due to 
confusion from nearby features, as remarked in 
these notes.  There are many instances where high velocity emission is 
present (as indicated in Table 1), but is rather weak and barely 
visible on the spectrum, so  we draw attention to it here.  

\subparagraph{284.350--0.418.} This water maser is coincident with 
main-line OH 
maser 284.351--0.418, with accompanying emission at 6035-MHz 
\citep{Cas97}, but clearly offset from nearby methanol maser 
284.352--0.419 by 6 arcsec. Spectra from epoch 2003 as well as 2004 are 
shown as an example of the typical variability seen between our 
observing epochs.

\subparagraph{285.260-0.067 and 285.263-0.059.} The latter is a very 
strong water maser 
associated with an OH maser and having similar velocity of its 
strongest emission.  In the case of OH masers this systemic velocity is a 
reliable indication of the systemic velocity.  The first maser, offset by 
1 arcmin, may be loosely associated, but it is difficult to recognise 
possible emission near the systemic velocity because of confusion with the 
stronger companion.  Its spectrum varied markedly from 2003 to 2004, and 
highly blue-shifted emission dominates the 2004 spectrum.  

\subparagraph{290.374+1.661 and 290.384+1.663.} The first of these sources 
is 
coincident with both OH and methanol maser emission. The positions 
measured in 2003 and 2004 are coincident, but detected features do not 
overlap in velocity ranges.   
There has been a substantial decrease in the peak flux density from 
3.5~Jy to 0.26~Jy from the first to the second epoch. 

The second source 290.384+1.663, is offset from the OH and methanol target 
and appears to be solitary, with no associated OH or methanol maser 
emission.

\subparagraph{291.270--0.719, 291.274--0.709 and 291.284--0.716.} The 2003 
data for these sources are presented in \citet{Cas04}, along with 
extensive discussion on associated sources.  Note that the OH 
target 291.274--0.709 was a supplementary addition to the list of C98.  The 
variability of the water masers between the two epochs is 
moderate, with minimal changes in the velocity ranges of the detected 
emission but many changes in the relative flux densities of individual 
features. 

291.270--0.719 is associated with methanol maser emission and has weak 
emission near the systemic velocity but much stronger emission 
blue-shifted by almost 80 \kmsns.   291.274--0.709, shows emission only 
near 
the systemic velocity and is coincident with both OH and methanol maser 
emission. The strongest source, 291.284--0.716, is associated with neither 
OH or methanol maser emission and shows no detectable water maser emission 
at the systemic velocity but has strong blue-shifted emission.  
\citet{CP08} regard this source and 291.270-0.719 as members of a 
distinct class of water masers that are dominated by blue-shifted 
outflows.

\subparagraph{291.578--0.434, 291.579--0.431 and 291.581--0.435.} The main 
strong 
source 291.579--0.431 is a persistent maser detected in both 2003 and 
2004, and also 1981 \citep{C+89}, with intensity varying by a factor of 4.  The other 
sources, detected at a single epoch, are even more variable.  All are 
associated with NGC 3603 \citep{Cas04}.

\subparagraph{291.610--0.529, 291.627--0.529 and 291.629--0.541.} The 
three water masers of this cluster were all detected both in 2003 and 2004 
and are associated with NGC 3603 \citep{Cas04}.  Only G291.610--0.529 is 
coincident with an OH maser and none of the sources is associated with 
any methanol maser emission. 

\subparagraph{297.660-0.974.}  The strongest water maser peaks, at 29 
\kms\ in 2003 and 26 \kms\ in 2004, agree well with the associated OH 
maser peak at 27.6 \kmsns, the probable systemic velocity for the region.  
There is a weak high velocity water maser feature of 0.6~Jy at a velocity 
of --78 \kmsns, slightly more than 100 \kms\ from the systemic velocity.

\subparagraph{299.012+0.125 and 299.013+0.128.}  299.013+0.128, with 
strongest peak at +19 \kms\ in both 2003 and 2004, was first detected by 
Caswell et al. (1989) and is in good agreement with the position and 
velocity of methanol and OH masers. In 2004 we detected an additional weak 
source, 299.012+0.125, offset by $\sim$9 arcsec.  We regard the latter as 
most likely a distinct source, probably in the same cluster but lacking 
emission at its systemic velocity.  

\subparagraph{300.968+1.143 and 300.971+1.143.} 300.968+1.143 was first 
observed by \citet{C+89}, with high velocity emission extending to --85 
\kmsns, 
remaining similar in our observations in both 2003 and 2004. This source 
is solitary, offset from the target OH maser by 18 arcsec.  Observations 
in 2004 uncovered an additional source, 300.971+1.143, with a flux 
density of 3~Jy. As can be seen in Table~\ref{tab:masers}, in 2003 we 
placed an upper limit on the flux density of this source of 3~Jy, quite 
crude because at this velocity there was confusion by strong emission 
from 300.968+1.143 at this epoch.

\subparagraph{301.136--0.225, 301.136--0.226a, 301.136--0.226b and 
301.137--0.225.}  
Although we list four distinct water maser positions, they probably 
represent a single maser site, with the last three locations all within 3 
arcsec of OH and methanol maser emission,  Confusion between features in 
2003 prevented separate position measurement of any but the strongest 
feature.  

\subparagraph{305.191--0.006 and 305.198+0.007.}  305.198+0.007 is 
probably the same source as 305.20+0.01 in \citet{C+89}. Both of these 
water masers are offset from 
the target OH maser, 305.200+0.019 (with methanol maser, 305.199+0.005).

\subparagraph{305.208+0.207}  Two strong peaks have positions separated by 
about 1 
arcsec, with the weaker peak slightly closer to the associated OH and 
methanol emission.   

\subparagraph{306.318--0.331.}  This is new solitary water maser source 
found offset 
from target OH and methanol maser positions by $\sim$15 arcsec.

\subparagraph{308.754+0.549.}  The OH target is an addition to the C98 
list.  Details of methanol, OH and water are given by \citet{Cas04}.  

\subparagraph{308.918+0.124.} The position of this water maser falls 
within our 3 
arcsec association threshold of OH maser G\,308.918+0.123 but lies 3.6 
arcsec from methanol maser G\,308.918+0.123. The OH and methanol masers 
are almost certainly coincident, with a measured separation of 0.6 arcsec, 
and we treat all three species as coincident.  

\subparagraph{309.921+0.479.} \citet{C+89} reported the detection of a 
4.5-Jy water maser 
at --70 \kms\ towards this OH and methanol site.   We detect no emission 
($<$ 0.3 Jy) in either 2003 or 2004.  

\subparagraph{310.144+0.760 and G\,310.146+0.760.} The first of this close 
pair of 
sources is located at the site of both OH and methanol maser emission and 
the second source is located at the site of an isolated 1720-MHz OH maser 
\citep{Cas04c} and is strongly variable.

\subparagraph{311.94--0.14.} This is a methanol site from Caswell (2009) with position 14 07 49.72, -61
23 08.3 (uncertainty 0.4 arcsec) from which there was no water detection
in 2003 (this position was not observed in 2004).
However a water maser was reported by Caswell et al. (1989) with peak flux
density 38 Jy from the position 14 07 49.9 -61 23 20, nominally offset
from the methanol by 12 arcsec but with rms position uncertainty of
about 10 arcsec and thus possibly coincident with the methanol.  An OH
maser is listed by C98 at 14 07 48.7 -61 23 22, nominally offset from the
methanol by 16 arcsec, but again, possibly coincident (to within the OH
position uncertainty of more than 15 arcsec).  Like the water maser, the
OH has varied, and later
observations to attempt an improved position determination failed to
detect it. We have omitted this source from the statistics since for our
position coincidence threshold of 3 arcsec. it may be
an OH site accompanied by water, by  methanol, by both or by neither,
depending on the precise positions yet to be determined.

\subparagraph{312.106+0.278 and 312.109+0.262.}  The second source is 
located towards 
the targeted methanol maser and consists of a single feature, with 
velocity similar to the methanol. The first source is a slightly 
stronger single feature, offset from the second by nearly 1 arcmin, and 
clearly offset in velocity by 6 \kms.  

\subparagraph{312.596+0.045 and 312.599+0.046.} These sources are a close 
pair, 
separated by $\sim$10 arcsec. The second of these sources was also 
detected by \citet{C+89} and is coincident with both OH and methanol 
masers. The first source, 312.596+0.045, is coincident with a methanol 
maser site.

\subparagraph{313.457+0.193 and 313.470+0.191.} These sources have an 
angular separation of $\sim$45 arcsec.  313.470+0.191 was also detected in  
\citet{C+89} and is associated with both OH and methanol masers, with peak 
water maser velocity comparable to the systemic velocity of the region as 
traced by the coincident methanol maser with mid-range velocity of --8 
\kmsns.  313.457+0.193 has an emission peak near --1 \kms\ at both epochs, 
suggesting its systemic velocity (and distance) is similar to that of its 
companion.  We regard the slightly stronger emission at 45 \kms\ seen 
only in 2004 as a strongly varying high velocity feature.

\subparagraph{316.360--0.361 and 316.361--0.363.}  Near these water masers 
lie an 
isolated methanol site (316.381--0.379) and a `methanol with OH' site 
(316.359--0.362).  Water maser 316.360--0.361 is 
offset from the methanol in the latter pair by 3.2 arcsec, and slightly 
further from the OH. Thus the water maser is formally rejected as part of 
an intimate association, but this remains somewhat uncertain.  
The second water maser source is further offset from these sources than 
the first and is thus very clearly a distinct, isolated site. 

\subparagraph{318.044--1.404}  The OH and methanol maser counterparts 
show that the systemic velocity is indeed near +45 \kms\ and indicate 
a distant location 
outside the solar circle.  The large latitude offset from the 
Galactic plane is consistent with the Galactic warp known to be present in 
this outer region of the Galaxy.  

\subparagraph{318.948--0.196a and 318.948--0.196b.} This pair of sources is 
essentially 
one extended source.  
Using the 2004 data, we were able to distinguish two main sites (with an angular separation of just over 2 arcsec) which conveniently reveals the 
association with nearby OH and methanol masers that otherwise would have 
fallen outside our 3 arcsec coincidence threshold. 

\subparagraph{320.221--0.281, 320.232-0.284, 320.233-0.284, 320.255--0.305 
and 
320.285--0.308.}  We suggest that all 5 sites are at similar distance in a 
cluster with systemic velocity near --65 \kms.  320.232-0.284 has 
associated methanol and OH maser emission and 320.233-0.284, offset by 
nearly 5 arcsec, seems to be a separate site.  
320.255-0.305 shows strong emission only near --126 \kmsns, highly 
blue-shifted from our suggested systemic velocity near --65 \kmsns, and 
apparently in a small class of water masers where blue-shifted emission 
dominates (Caswell \& Phillips 2008).  
Interestingly, like 291.284--0.716, 320.255--0.305 is not associated
with either OH or methanol maser emission.
Emission from this region was 
reported, with lower position precision, by Caswell et al. (1989).

\subparagraph{321.028--0.484 and 321.033--0.483.} Both water masers are 
offset from the 
target OH and methanol maser 321.030--0.485 by more than 7 arcsec; the 
second water maser is associated with methanol maser 321.033--0.483.

\subparagraph{321.148--0.529} The spectrum of this maser is shown at both 
epochs as 
an example of extreme variability with no common features seen in spectra observed less than 1 
year apart.  

\subparagraph{323.459-0.079.}  A water maser was discovered (but without a 
precise position) by \citet{C+89} towards this OH and methanol site, but 
was below our detection threshold of 0.2 Jy in both 2003 and 2004.

\subparagraph{324.716+0.342.} This source was detected by \citet{C+89} at 
a peak flux density 
of 138~Jy, coincident with both OH and methanol maser emission. The water 
maser shows marked variability with peaks of 10 and 26~Jy during 2003 and 
2004 respectively. 

\subparagraph{326.662+0.521, 326.665+0.553 and 326.670+0.554.} This group 
of three 
sources is spread over 2 arcmin. 326.670+0.554 coincides with an OH 
maser;  the water maser peak flared in 2004 relative to 2003.  This 
maser is probably the same as one reported, with peak of 780 Jy, but with 
poor position, by Batchelor et al. (1980).  In 2004 no targeted observation 
was made of the first source and although recognised at the beam edge 
in another observation, it was confused by a strong flare from 
326.670+0.554.  Measurement for the second source (326.665+0.553) is reported only from 
the 2004 measurements since the 2003 observations were confused by 
326.662+0.521, preventing a useful upper limit estimate.  

\subparagraph{326.780-0.241} This is a new water maser coincident with an 
OH maser listed by Caswell (1998) with approximate coordinates 
326.77-0.26;  subsequent (previously unpublished) ATCA measurements of the 
OH show it to be coincident with the water maser.  

\subparagraph{326.859-0.676.}  This weak new water maser coincides 
spatially with
methanol maser 326.859--0.667 (offset by 3.4 arcsec), whose systemic velocity is --58.0 \kms.  It seems likely to 
be an association in which only a heavily blue-shifted water maser 
feature, at --103 \kmsns,  is seen.  

\subparagraph{327.291-0.578.}  This strong water maser with peak of 
several hundred 
Janskys is associated with OH and methanol masers.  Spectra are shown for 2003 
and 2004 revealing great variability of the water maser, and can be 
compared with a spectrum shown by Batchelor et al. (1980) when its peak 
was over 1000Jy.  
 
\subparagraph{327.402+0.445.}  In contrast to the previous source, this 
water maser has 
greatly increased intensity compared to measurements by Batchelor et al. 
(1980).  The water maser is associated with a strong methanol maser.  An 
OH maser is close to the methanol.  We therefore treat the OH as an 
association also, although its separation from the water is just outside 
our formal association criterion.  

\subparagraph{328.306+0.432.} This water maser is associated with 22-GHz 
radio
continuum but appears to be truly offset from the target OH maser 
328.307+0.430 (which has no accompanying methanol) by more than 5 arcsec. 

\subparagraph{328.808+0.633.} Detected towards OH maser site 
328.809+0.633.
Also associated are many transitions of methanol and OH, including; 6.6-, 12.2-, 19.9-, 85.5- and 107-GHz methanol masers (Caswell 2009; Caswell et al. 1995; Ellingsen et al. 2004; Ellingsen et al. 2003; Val'tts et al. 1999) as well as 1720-, 4765-, 6030-, 6035-MHz and 13.4-GHz transitions of OH (Dodson \& Ellingsen 2002; Caswell 2003, 2004b, 2004c). This source is also associated with strong 22-GHz radio continuum.

\subparagraph{329.021--0.186, 329.029--0.199, 329.030-0.205 and 
329.031--0.198.}    
This cluster of four sources is spread over about 1 arcmin. 
The first source is solitary (no OH or methanol), the second (which varied 
below the detection limit in 2004) has OH, and the third and fourth are 
associated with both OH and methanol.  Ellingsen (2006) showed that these OH, methanol and previously known water masers are associated with a filamentary infrared dark cloud as well as Class I methanol masers.

\subparagraph{329.342+0.130.} This water maser is clearly offset by 
almost 1 arcmin from the target OH maser 329.339+0.148 and no other water 
emission was detectable in the field.  The water maser velocity is similar 
to that of its OH neighbour and they presumably reside in the same star 
formation cluster, discussed in detail by Caswell (2001, 2004).  Note that 
the OH target is an addition to the C98 list.  

\subparagraph{329.404--0.459, 329.405--0.459 and 329.407--0.459.} 
The second source is at an OH and methanol maser site, and its velocity is 
similar, presumably representative of the systemic velocity.  The first 
source is offset only 3 arcsec from the second site, whereas its 
velocity is close to that of the previously discussed, more distant, 
cluster around 329.339+0.148.  None the less, we suggest it most likely 
represents high velocity emission related to 329.405--0.459.  
Water maser 329.407--0.459, associated with a
methanol maser, has shown extreme variability with peak flux density of 
80~Jy in 2003 but not detected above our detection limit
of 0.2~Jy in 2004.

\subparagraph{329.421--0.167, 329.424--0.164 and 329.426--0.161.}  
The only other maser species nearby, both spatially and in velocity, is a 
1720-MHz maser 329.426--0.158 \citep{Cas04c}.  We conclude that all four 
masers lie in the same star forming cluster but are not closely 
associated.  

\subparagraph{330.879--0.367} The  water maser has been known for many 
years (Batchelor 
et al. 1980) and our precise position confirms that it coincides with one 
of the strongest known OH masers, accompanied by very weak methanol 
emission (Caswell 2009).  

\subparagraph{330.954--0.182.} The  water maser has remained  very strong 
for many 
years (see Batchelor et al. (1980)) and has prominent high 
velocity emission (a feature at -191 \kms\ has a peak flux density of 
0.3 Jy but is too weak to be seen on the spectra displayed here).  
OH and methanol emission are present nearby but spread over several 
arcsec,  and the most detailed maps (Caswell et al. 2010) show the water 
to be associated with OH emission only, with the methanol emission clearly 
offset to another site offset to the south-west by more than 3 arcsec.  
We measure the flux density at 22-GHz of an associated strong \UCHII 
region as 3.2~Jy.

\subparagraph{331.418+0.252.} This 0.6-Jy water maser was detected almost 
50 arcsec from the targeted methanol maser 331.425+0.264. The water maser
emission is observed towards an \UCHII region that we detect at 22-GHz.

\subparagraph{331.512--0.103.} Batchelor et al. (1980) observed this source with a 
peak flux
density of 4300~Jy. Observations of this source in 2003 and 2004 showed a
decrease in the flux density to 700~Jy and 534~Jy in 2003 and 2004 
respectively.  The water maser is coincident with an OH maser (but no 
methanol) and an \UCHII region that we detect.  The strongest water 
emission is near the systemic velocity, and almost symmetric about this 
there are multiple high velocity features extending for 70 \kms.  

\subparagraph{332.826--0.549.} This is a strong new water maser coinciding 
with 
methanol and a 6035-MHz OH maser but offset 7 arcsec from a 1665-MHz OH 
maser.  

\subparagraph{333.219--0.062 and G\,333.234--0.060.} 333.234--0.060 was 
detected in
both 2003 and 2004 and has been previously observed by Batchelor et al. (1980).  At all 
epochs the peak has been more than 100 Jy at a velocity near -88 \kmsns, 
which is the mid-range velocity of an associated OH maser, and the likely 
systemic velocity.  In our 2004 water observations, an extreme high 
velocity feature was observed with a peak flux density of 0.3~Jy at +81 
\kms, more than 160 \kms\ from the systemic velocity.
333.219--0.062 is offset from 333.234--0.060 by almost an arcmin and is 
solitary. This source was observed with a peak flux density of 0.5~Jy in 
2004 and was not detected in 2003 above 0.3~Jy.  In the 
absence of an association with OH or methanol, its systemic velocity is 
unknown.  

\subparagraph{333.387+0.032.} This weak water maser is located at the site 
of both OH 
and methanol maser emission.  It was detected with a peak flux density of 
0.4~Jy in 2003 and had decreased to 0.14~Jy in 2004. Spectra 
from both epochs are presented in Fig.~\ref{fig:spectra}.

\subparagraph{333.608--0.215.} This was one of the earliest observed water 
masers,
discovered by \citet{John72} and was later observed by Batchelor et al. (1980) in 1976 with a
peak flux density of 100~Jy. \citet{Breen} carried out interferometric
observations of this source in 2006, deriving a precise position for the
source that is within 0.4 arcsec of the independent position quoted in
Table~\ref{tab:masers}.  Both the Breen spectrum of 2006 and that 
found in the present observations of 2003 differ markedly from the 2004 
spectrum shown here, revealing strong variability of high velocity 
emission.  This source is associated with an OH maser and 
is offset by 15 arcsec from a bright \UCHII region which was erroneously
reported by \citet{Breen} as having an integrated flux density of 631~mJy
at 22~GHz. The present observations find that the \UCHII region has an
integrated flux density of more than 16~Jy at 22-GHz.

\subparagraph{333.930--0.134.} This very weak water maser was observed at 
the
2004 epoch only. While the peak of the detected emission is a mere
0.18~Jy, the source position is in remarkably good correspondence with the
targeted methanol maser, the separation being less than 1.5 arcsec.

\subparagraph{335.059--0.428, 335.060--0.427 and 335.070--0.423.} The 
first two of 
these sources are separated by 3.3 arcsec and therefore may be
essentially a single source spread over a few arcsec.  The second 
source, 335.060--0.427, with measurements in both 2003 and 2004, shows 
best positional agreement with an associated OH and methanol maser site.   
335.059--0.428 was not recognisable as a distinct source in the 2003 data 
owing to confusion from the stronger companion 335.060--0.427. 

335.070--0.423 is offset from the previous two sources by 43 
arcsec and is solitary.

\subparagraph{335.585--0.285, 335.586--0.290 and 335.588--0.264.} These 
three sources
are spread over almost 80 arcsec. The first two sources were detected at
both epochs and are both coincident with both methanol and OH masers.
The third source, 335.588--0.264, was detected with a peak flux density 
of 16~Jy in 2003 but not detectable above 0.2~Jy in 2004.  It is isolated 
from other maser species and is devoid of detectable 22-GHz radio 
continuum emission. 

\subparagraph{335.787+0.177, 335.789+0.174 and 335.789+0.183.} 
335.789+0.174 was detected by Batchelor et al. (1980) as a 25~Jy source towards an OH maser 
with the same Galactic coordinates. Our observations detect a water maser 
of 3~Jy in 2003 and detect no emission above 0.2~Jy in 2004.  
335.787+0.177, detected in both 2003 and 2004, is a solitary maser 
and has no detectable radio continuum at 22-GHz. The third source, 
335.789+0.183, is also isolated from other maser species as well as 22-GHz 
radio continuum emission. It was 
detected with a peak flux density of 4.2~Jy at a velocity of
--91 \kms\ in 2003 and was not detectable above 0.2~Jy in 2004. If this 
velocity represents the systemic velocity, then its distance is likely 
to differ greatly from its apparent companions.  Alternatively, it may 
be a companion at similar distance but showing no significant emission 
near the systemic velocity, and only blue-shifted emission.   Such high 
velocity features are notoriously variable.

\subparagraph{336.864+0.005, 336.864--0.002 and 336.870--0.003.} These 
three
sources are located within 40 arcsec of each other and were all detected
in both the 2003 and 2004 observations. The first source is associated
with both OH and methanol maser emission and the other two sources are
solitary.

\subparagraph{336.983--0.183.}  Figure~\ref{fig:spectra} shows both the 
2003 and 2004 spectrum for this weak source. The only significant 
feature in 2003 is a peak at --76 \kms.  In 2004, emission near this 
velocity is weaker, and a high velocity feature near +45 \kms\ is 
marginally stronger.  The source is associated with a methanol maser as 
well as an \UCHII region that we list in Table 
7. The methanol maser is strong, with a well-measured position and velocity near --81 \kms.  Nearby 
is a weak OH maser just outside our criterion for an association with 
the water position, but slightly closer to the methanol.  Furthermore, 
the position of 6035-MHz OH emission (Caswell 2001) is acceptably within 
our coincidence criterion, so 
we add this as an OH maser association with both methanol and water.

\subparagraph{336.991--0.024, 336.994--0.027 and 336.995--0.024.} These 
three
sources appear clustered within $\sim$15 arcsec. 
336.994--0.027 is the strongest of the three (160~Jy), detected in 
2003, 2004 and by Batchelor et al. (1980),
and is associated with both OH and methanol maser emission, with systemic 
velocity near -120 \kmsns. 
The 2 weaker water sites have quite different radial velocities, 
near -50 \kmsns, and might be at a distance quite different from the 
strongest one.  336.991--0.024 was
detected in both 2003 and 2004, with a peak flux density of 4 and 1~Jy at
the respective epochs, and is associated with an \UCHII region that we
detect.  336.995--0.024 was detected only in 2004 and had a peak flux 
density of 1.1~Jy.

\subparagraph{337.994+0.133 and 337.998+0.137.} Batchelor et al. (1980) detected 
337.998+0.137 with a
peak flux density of 200~Jy and we detected a decreased flux density of 30
and 27~Jy in 2003 and 2004 respectively. This source is coincident with
both OH and methanol maser emission.  337.994+0.133 is a solitary 
maser, offset from the previous source by 19 arcsec.

\subparagraph{338.069+0.011, 338.075+0.012, 338.075+0.010 and 
338.077+0.019}  
Water maser emission from 338.069+0.011 is the strongest in this cluster 
at both epochs and has no other maser counterpart.  338.075+0.012 detected 
only in 2003, and the weakest of the group, coincides with OH and methanol 
maser emission.  338.075+0.010 is associated with a methanol maser with 
systemic velocity near -38 \kms;  the water maser emission in 2003 was 
strongest at a highly blue-shifted velocity, but by 2004 this had faded 
below detectability, leaving only features closely straddling the systemic 
velocity.  338.077+0.019 was detected in both 2003 and 2004, the spectrum 
remaining unchanged; it has no apparent association with other masers.  

\subparagraph{338.920+0.550 and 338.925+0.556}  The stronger site, 
338.925+0.556 coincides with OH and methanol masers.  The weaker site, 
338.920+0.550, is associated with a methanol maser with 
systemic velocity near -60 \kmsns, similar to the other site, and the 
water spectrum in 2003 was dominated by highly blue-shifted emission.  

\subparagraph{343.126--0.065 and 343.127--0.063.} These two 
sources are separated by 9.3 arcsec. The second source is strong and 
associated with an 
OH maser, but not methanol.  The first source is much weaker but clearly 
distinct and is solitary.

\subparagraph{345.004-0.224}  In 2003 the only water emission was near 
the systemic velocity, as defined by the associated methanol and OH 
masers.  In 2004, high velocity features dominated.  Spectra from both 
epochs are shown to demonstrate this interesting variability.  

\subparagraph{345.010+1.793, 345.010+1.802 and 345.012+1.797}  
In this small cluster, the first water maser coincides with an OH and 
methanol site, the second has no other maser counterpart, and the third, 
the strongest, has a methanol counterpart.  

\subparagraph{345.397-0.950, 345.402-0.948, 345.405-0.947, 345.406-0.942, 
345.408-0.953, 345.412-0.955 and 345,425-0.951}  One of these sites, 
345.425-0.951, is coincident with a methanol maser site.  We also accept a 
coincidence between 345.408-0.953 and an OH and methanol maser site, 
despite an offset of 4.6 arcsec, formally just outside our our criterion 
for a single epoch water measurement; there is evidence (from 
comparison of features common to 2003 and 2004, and a feature common to 
2004 and FC89) that the 2004 observation of this field yields positions 
slightly too far south and at too large an RA, a correction that would 
improve the coincidence.   The systemic velocity of the two associated, 
methanol sites is near -15 \kmsns.  In view of the large offset of the 
cluster from the Galactic plane in Galactic latitude, all water sites are 
likely to be clustered at a similar distance, irrespective of the water 
maser velocity.  Their separation from each other is sufficiently large to 
suggest that each site has its own exciting star, and thus it seems likely 
to be a remarkable physical cluster of massive stars. Radhakrishnan et 
al.  (1972) suggest that the distance to the complex is half the distance 
to the Galactic Centre.  

\subparagraph{345.487+0.314.}  As seen in the notes 
tabulating associations, there is a coincident methanol maser and no 
nearby ground-state OH maser.  There is, however, a 6035-MHz excited state OH 
maser offset just over 3 arcsec to the north (Caswell 2001).  

\subparagraph{345.493+1.469, 345.494+1.470 and 345.495+1.473}  The first 
site coincides with an OH maser that has no accompanying methanol 
maser.  Another nearby OH maser site shows no methanol or water 
emission.  All four sites are likely to lie in a nearby cluster, as 
evident from the large Galactic latitude, and the systemic velocity 
probably lies between -15 and 0 \kms, based on the OH velocities.  

\subparagraph{347.623+0.148 and 347.628+0.149}  The first water maser, 
detected in both 2003 and 2004, has no coincident maser of OH or methanol.  
The second water maser, at similar velocity, was detected only in 2003 and 
coincides with OH and methanol, with systemic velocity near -95 \kms.

\subparagraph{348.533--0.974, 348.534--0.983 and 348.551--0.979}  The 
third water site is close to the original OH target but is more 
precisely coincident with the methanol site 348.550-0.979n (Caswell 
2009).  The original `OH with methanol' target, 348.550-0.979, is 
regarded by Caswell (2009) as a nearby but distinct site and, on this 
interpretation, it is a site without detected water maser emission.  
The water masers 348.533--0.974 and 
348.534--0.983 are new, chance, detections in the vicinity.  

\subparagraph{348.726--1.038}  This water maser is offset from the 
target OH (with methanol) by more than 4 arcsec and is not formally an 
association.  However, the large spread over several arcsec in the 
positions of individual water maser spots reveals a larger than usual 
maser site and the possibility of an association will require further 
investigation.  

\subparagraph{350.105+0.084 350.112+0.089 and 350.113+0.095}  The first of these is 
associated with a methanol site, and the second with an
OH site.  In a cluster of 6 water maser sites showing peak emission at 
velocities between -72 and -44 \kmsns, they are the only 2 
accompanied by maser emission of another species.  Spatially in this 
same cluster, the water maser 350.112+0.089 displays a quite different 
velocity range, from -175 to -106 \kms\ suggesting that it might be at a 
quite different location, perhaps in the near side of the 3-kpc arm, whose 
characteristic velocity at this longitude extends to approximately -110 
\kms\ (Green et al. 2009);  see also the note for 351.582--0.353.  

\subparagraph{350.299+0.122.} This is the weakest single epoch 
water maser that
we list, with a peak flux density of 0.17~Jy. The position of the
water maser is only 0.7 arcsec from the targeted methanol maser and its
emission peak velocity of --68 \kms\ is within 6 \kms\ of the
associated methanol maser peak emission. Furthermore, there are no nearby
water maser sources that could confuse the region and be detected as a 
sidelobe here;  there is therefore little doubt that this is a genuine 
weak water maser.

\subparagraph{351.240+0.668, 351.243+0.671 and 351.246+0.668} 
The first 2 sites were listed by Caswell \& Phillips (2008), with 
special discussion of a blue-shifted outflow that dominates 
the emission from 351.243+0.671, and remarks that the water maser spots 
in the outflow are distributed over several arcsec.  
All three sites appear to be distinct, with large separations of more than 
15 arcsec, yet close enough to suggest that they all lie in the same 
star-forming cluster, with systemic velocity +2.5 \kms as defined by the 
methanol maser counterpart to 351.243+0.671 (Caswell \& 
Phillips 2008), and all lying in the large NGC 6334 complex, at a commonly 
accepted distance of 1.7 kpc

\subparagraph{351.417+0.645}  This maser coincides with an HII region 
NGC 6334F, a very strong methanol maser, and an OH maser (Caswell 1997) 
and its water maser emission was mapped by Forster \& Caswell (1999), 
showing a scatter of spot positions over several arcsec, including an 
apparent jet-like feature.

\subparagraph{351.582--0.353}  This strong water maser is coincident with 
an OH and methanol maser site which is located in the near side of the 
expanding 3-kpc arm (Green et al. 2009;  Caswell et al. 2010).  

\subparagraph{352.623-1.076 and 352.630-1.067}  The second water maser 
coincides with an OH and methanol site.  It flared from a peak of 35 Jy in 
2003 to 700 Jy in 2004.  It appears to be the same maser that was first 
reported by Sakellis et al. (1984) with peak flux density 346 Jy. The 
first source is solitary with no other maser counterpart and offset by 
more than 40 arcsec.     

\subparagraph{353.273+0.641} This site is the prime example of a class 
of water maser dominated by a blue-shifted outflow (Caswell \& Phillips 
2008);  its systemic velocity is estimated from a coincident 
methanol maser.  

\subparagraph{353.408--0.350 to 353.414--0.363 inclusive}  This is 
a cluster of 6 solitary water maser sites.  The target OH and methanol maser site 
353.410--0.360 (Caswell 1997) lies in this cluster but none of the water 
maser sites coincide with it.   

\subparagraph{354,703+0.297, 354.712+0.293 and 354,722+0.302}  The 
velocity of all three solitary water sites is near +100 \kmsns, similar to that of 
the nearby OH with methanol site 354.724+0.300.  The velocity of the 
latter has been interpreted as evidence of a location in the 
the Galactic bar (Caswell 1997), which we suggest is an appropriate 
interpretation for all 4 sites.  

\subparagraph{357.965--0164 and 357.967--0.163}   The two water sites  
are separated by 9 arcsec.  The first site has a weak methanol 
counterpart but no OH;  the second has stronger methanol and also OH.  
The methanol and OH emission is confined to a small range between --9 
and +3 \kms, and thus the systemic velocity for both sites probably lies 
in this range.
Water emission at the first site was strong in 2003 but weak in 
2004, and confined to within about 15 \kms\ of the systemic velocity. Water emission at the second site is seen not only near the 
systemic velocity but also at many blue-shifted and red-shifted 
high velocities; in 2004 the high velocity features were stronger than 
emission near the systemic velocity.

\subparagraph{359.436--0.102 to 359.443--0.104 inclusive}   The 
six water maser sites in this cluster all have similar velocities.  
Confusion prevented a useful upper limit estimate for emission from 2 of 
them not detected in 2003.  One site has an OH with methanol counterpart 
and another, a methanol counterpart.  The methanol sites appear to be 
located in the near side of the 3 kpc arm (Green et al 2009) and this 
interpretation can be applied to all six water sites in the cluster.   

\subparagraph{0.655--0.045 to 0.677--0.028}  These five sites lie within 
the Sgr B2 complex and are clearly distinguishable with our spatial 
resolution. Two of them have associated OH maser emission but none have associated methanol masers.

\subparagraph{5.886--0.392}  This site has OH and water masers spread over 
more than 5 arcsec, associated with a compact HII region that is likely to 
be approaching the end of its masing phase (Caswell 2001) and with various 
distance estimates, of which 2 kpc is currently favoured (Stark et al. 
2007).  It was observed by Forster \& Caswell (1989; 1999) and our 2003 
water maser observations were made to assess the changes over a decade.  
We cite the position of the strongest feature during our observations and 
note that there are many components at offset positions.  Our water 
reference position and the OH reference position from Caswell (1998) are 6 
arcsec apart but we list it as an association with OH on the basis of the 
intermingled features mapped by Forster \& Caswell (1999).  

\subparagraph{9.620+0.194 and 9.622+0.196}  These coincide respectively 
with an OH maser site with no reported methanol, and the strongest known 
methanol site which also has a coincident OH maser.   
A third methanol maser site in this cluster, 9.619+0.193, shows no 
detectable water maser emission.  

\subparagraph{10.473+0.027 and 10.480+0.034}  The first source was 
detected in both 2003 and 2004, and the spectrum shown is from 2004 
when a strong flare occurred.  The spectrum shown for 10.480+0.034 is 
from 2003 when 10.473+0.027 was not flaring, and confusion from 
10.473+0.027 was much less.   

\subparagraph{10.623--0.383.}   Note a small correction to the C98 OH 
position, for which the RA should be 18$^{h}$10$^{m}$28.67$^{s}$ (not 
28.61).  Our water measurement in 2003 is in good agreement with the 
revised OH position and the measurement of FC89.  

\subparagraph{11.903-0.141.}  We detect a weak water 
maser at this site which is probably the same source that was first 
reported by Caswell et al. (1983) with peak flux density 7.9 Jy (but with 
position uncertainty exceeding 10 arcsec). 
Note a small transcription error in the target OH 
position from FC89, FC99 and C98, for which the RA should be 
18$^{h}$12$^{m}$11.46$^{s}$ (not 11.56).  Our water maser position is in 
satisfactory agreement with the corrected OH position.   The corrected OH 
position also agrees better 
with the 6035-MHz maser position (C97), and with its associated \UCHII 
region (C97) for which improved measurements by Forster \& Caswell (2000) 
at 8.7 GHz give a flux density of 33.9 mJy. 

\subparagraph{15.016--0.679 to 15.034-0.667 inclusive} These all reside in 
the well-known nearby star-forming complex M17.  
All seven masers were recognisable and spatially distinct in the 2004 
observations but only 3 were distinct in the 2003 observations, partly due 
to the stronger confusing emission from 15.028--0.673 at the earlier epoch. Spectra for the two strongest sources from the 2003 epoch are shown in Fig. 1 and are in addition to the 2004 spectra.

\section[]{Discussion}

\subsection{Water maser variability}\label{sect:var}

Water masers have been noted on many occasions for their often extreme
variability over relatively short timescales, and some studies have 
extended over several decades \citep[e.g.][]{Felli07}. As our data are 
confined to only two epochs, and limited to coarse spectral resolution, we 
do not attempt a detailed study of the variability of our sources.  
However, we are able to highlight some 
interesting examples of variability, and the large 
size of our sample allows us to derive several interesting statistics.  

For the 207 sources observed and detected at both epochs, we see 
variability ranging from sources showing no measurable variability to 
occasional extreme levels, and many intensities varying by factors of 
more than two.
High velocity features can be particularly variable, with many spectral
features of sources not being common to both epochs.

As noted in Section 3, Fig.~\ref{fig:spectra} includes the spectra of 
eight sources from both
epochs (284.350--0.418, 321.148--0.529, 327.291--0.578, 333.387+0.032, 
336.983--0.183, 345.004--0.224, 15.026--0.654 and 15.028--0.673).  They illustrate changes seen over 
the 10-month time scale  and highlight the fact that, for four of the 
examples, the feature with the strongest peak at the two epochs is 
at a different velocity.  

\begin{figure}
\vspace{-1cm}
	\psfig{figure=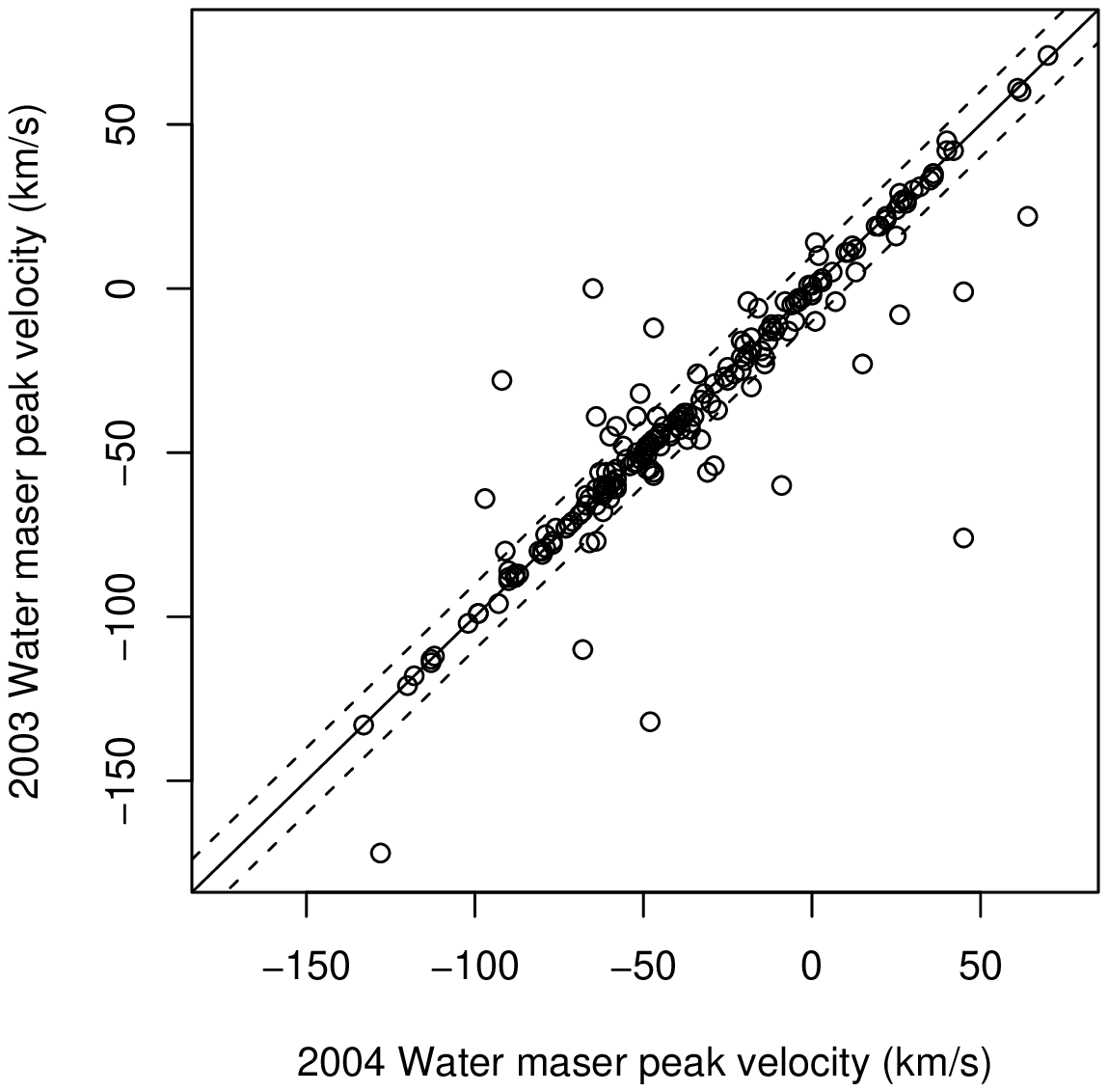,width=9cm,angle=270}
	 \caption{2004 peak water maser velocity versus 2003 peak water maser velocity. Overlaid is a solid line with a slope of 1 and two dashed lines showing a deviation of 10 \kms\ either side of the solid line.}
	 \label{fig:water_water}
\end{figure}

A qualitative impression from the full set of spectra at both epochs 
(including our unpublished material for the 2003 epoch) is 
that the spectra of many sources show little resemblance at the 
two epochs.  
Quantitatively, we use the data of Table 1 to derive the plot of 
Fig. ~\ref{fig:water_water} which 
compares the velocity of the water maser peak emission in 2003 and 2004.  
An interesting statistic for the sources measured at two epochs 
shows that the strongest peak is at a significantly different velocity 
(offset more than 2 \kms) for 38 per cent (78 of 207) of the sites.  
However, from Fig.~\ref{fig:water_water}  we see that a much smaller fraction of velocity 
differences exceed 10 \kms.  
These large velocity differences generally represent the truly high velocity features which, in a few sources, can dominate the spectrum at some epochs since they have highly variable intensities. For example, the source 357.967-0.163 has the 2003 peak intensity at 0 \kms (near systemic) but the 2004 peak at -65 \kms (a high velocity feature);  and the source 336.983-0.183 has the 2003 peak intensity at -76 \kms (near systemic) but the 2004 peak at +45 \kms (a high velocity feature).


We have observations available in both 2003 and 2004 for 253 water 
masers; 46 of them (17 per cent) were detected at only one epoch.
Only 16 (one-third) of these sources that varied below our detection 
limit at one of the epochs were stronger than 2~Jy when detectable, with 
the strongest source being 80~Jy. Half of the sources detected only at 
one epoch show a single feature and the majority of the
others exhibited 3 or fewer features. In comparison, the vast majority of water maser sources that are associated with OH or methanol masers exhibit five or more spectral features. We have investigated
the associations with other maser species for these 46 sources to search 
for other possible properties in common. Of the 46 sources, 12 are associated with both OH and methanol masers, 2 are
associated with OH masers only, 3 are associated with methanol masers
only and 28 are solitary (no association with another maser species). A chi-squared test was carried out on the percentage of sources in
each of these groups compared to what would be expected if these sources
were distributed in the same way as our entire sample. We find that a
statistically significant higher percentage of the sources detected at
only one epoch are solitary (p-value 0.02), compared to the distribution
of associations in our full sample. The numbers of sources that varied
below the detection limit at one epoch and are associated with
combinations of OH and/or methanol masers are as expected from the
distribution of the entire sample.

\citet{Claussen96} suggested that water masers 
associated with low mass stars
were in general both weaker and more variable than those associated with
high mass stars. As we find that the sources only detectable at one epoch
are biased towards solitary sources and are in general relatively weak, it
is possible that a number of these water masers are in fact associated
with lower mass stars perhaps residing in the same stellar clusters as the
high mass star formation regions towards which the observations were
targeted (see also end of Section 5.6 and Section 5.7).

The potential of water masers for mapping the distribution of massive SFRs 
throughout the Galaxy has been demonstrated for a few sources by 
astrometry sufficiently precise to achieve parallax measurements and 
precise distances (e.g. Sato et al. 2008).  Water masers appear to provide 
the largest population to make these investigations, but with several 
caveats.  Firstly, a site must have individual maser spots persisting for 
more than a year.  Despite the extreme variability shown by all spots at 
some sites, and some spots at most sites, our data reassuringly 
demonstrate that there still remain an enormous number of suitable sites.  
A second reservation concerns the ability to associate a systemic 
velocity with the precise distance, enabling mapping of the Galactic 
velocity field.  As we shall see in Sections 5.3 and 5.4, the estimate 
of the systemic velocity for an isolated water maser is uncertain, but an 
excellent estimate can be obtained from the OH or methanol masers 
accompanying many water masers.  Note that the spatial correspondence 
between maser 
species is usually sufficient to yield very high confidence associations, 
whereas associations with more diffusely distributed thermal emission in 
molecular clouds are less reliable.  So, associated OH or methanol masers 
are the key to establishing the systemic velocity of a water maser.   We 
note that for sites with OH but no methanol, 
parallax determinations are beyond present capabilities except through the 
use of associated water masers.   And even for some methanol sites, it may 
turn out that an associated water maser provides the best parallax 
measurement.  
Thus the important role of water masers in these Galactic 
studies is assured.

\subsection{Spatial distributions of maser spots}

The VLA study of water masers by FC89 and FC99 examined the distributions
of maser spots, both in velocity and spatially, for the masers that were
quite strong and/or displayed many spectral features.  It was shown that,
where many maser spots were present, they lay either within a diameter
rarely exceeding 30 mpc, or in a few clusters of this size separated by
distances at least several times larger.  The quite large beamsize used in 
the present observations precludes detailed study of the
spot distributions, but allows recognition of clusters with several
distinct members, of which there are many. We do however, find a
substantial number of water maser sources that show distinguishable
angular separations between clusters of maser spots emitting
near the systemic velocity, and those emitting at high velocities.  The 
separations are generally of the order of 2 or 3 arcsec, and are 
plausibly attributed to associated outflowing material. Occasionally,
separations between systemic and high velocity components exceed 4 arcsec, 
and even in these cases it  seems credible that these are associated 
outflows.

\subsection{Detection statistics and relationship to 
ground-state OH and methanol masers}

The position measurements and new detections of water masers reported in
Table~\ref{tab:masers} mostly arose from a search at all the
positions of southern SFR maser sites with ground-state OH main-line (1665
and 1667-MHz) masers that had not previously been searched.  The 
target list corresponded to table 1 of C98, plus a few 
modifications which we briefly summarise here. 
Small position corrections were needed for 10.623-0.383 and 11.904-0.141;  
and an improved (previously unpublished) position has been determined for 
326.780-0.241, listed by C98 at the approximate position 326.77-0.26.  
The list was augmented by 291.274--0.709, 308.754+0.549 and 329.339+0.148 
(see notes for these sources in Section 4 and Caswell (2001)).  The OH source 311.94+0.14 still has no precise position measurement and,
although searched for water, has been omitted from the statistics,
as discussed under the note on 311.947+0.142.

  Because the observations were targeted toward OH masers detected in a
blind search, the detection statistics can be meaningfully computed. The
new search has established sensitive upper limits for water masers toward
42  main-line OH maser sources, and a net 
detection rate for water masers
towards OH masers of 79 per cent.  Additional observations were made
towards a selection of 104 methanol masers with no reported OH
counterpart \citep[chiefly from][]{C09}.

As noted in Section 3, of the 379 detected water maser sources, 128 are associated
with both OH and methanol masers, 33 are associated with OH masers only, 70 are associated with methanol masers only and 148 are 
solitary (i.e. not associated with either OH or methanol maser emission). The water maser detection rate towards sources exhibiting both OH and methanol maser emission is 77 per cent (128 of 166). In contrast, the water maser detection rate towards OH maser sites (with no methanol) is 89 per cent (33 of 37). In order to determine if water masers were
preferentially detected towards OH masers without associated
methanol masers we carried out a chi-squared test. The resultant p-value
of 0.6 means that the higher detection rate of water masers towards OH maser sources without associated methanol is not statistically significant.


Because the sample of methanol masers targeted in 2004 was not 
homogeneous, it is difficult to draw strong conclusions concerning 
methanol/water associations from the detection
statistics. However, the fact that we find 70 associations of water
towards methanol masers without OH (from a total sample of 104), and 33 associations of
water towards OH masers with no methanol (from a total sample 
of 37), along with a detection rate of 77 per cent towards sources exhibiting both OH and methanol masers, indicates that the overlap between the lifetimes of water,  OH and methanol masers is large. A possible interpretation is that during the evolution of the star the methanol masers are not only the first maser species to appear but also the first species to turn off whereas the OH and water both persist for longer.

\begin{table}
  \caption{The average and median water maser flux densities for all the sources we detect. }
  \begin{tabular}{lll}\hline
    \multicolumn{1}{l}{\bf Water} & {\bf Average flux} & {\bf Median flux}	 \\
     \multicolumn{1}{c}{\bf classification}  &{\bf density (Jy)}  &{\bf density (Jy)}  \\ \hline
  all sources		& 	57.1	&	5	\\   
  with OH			&	96.1	&	15	\\
  with methanol		&	68.3	&	9	\\
  with continuum	&	74.9	&	18	\\ 
  only OH			&	138.2& 	25\\
  only methanol		&	26.1	&  3.5\\
   solitary			&	18.9	&	2.8	\\ \hline
    \label{tab:fluxcomp}
  \end{tabular}
\end{table}

Table~\ref{tab:fluxcomp} presents both the average and median flux
densities of the water masers that we detect, broken up into categories
according to their association with OH and methanol masers as well as
22-GHz radio continuum. Where a source was detected in both 2003 and 2004
we have used the two recorded values of flux density as separate
entries. In column 1 the types of association are listed. Here `all
sources' incorporates all water maser peak flux densities detected at
either epoch;  `with' OH, methanol or continuum refers
to water masers that are associated with the afore-mentioned source but 
not
limited to associations with only these sources;  `only' OH or methanol
incorporates only those water maser sources exclusively associated with
either OH or methanol masers, but places no restrictions on their association with 22-GHz radio continuum; and `solitary' refers to water sources that are not associated
with either OH or methanol masers. We find that there is trend of increasing
water maser flux density from solitary sources to sources associated with
methanol masers to sources associated with OH masers and 22-GHz radio
continuum. This may indicate that the water masers increase in flux
density as the sources evolve, similarly to that found by \citet{Breen09}
for 6.6-GHz methanol masers. We note, however, that the average and median flux density of the solitary water masers is likely to be at least partly due to some of these sources being associated with low-mass stars.

The recent completion of the Southern Hemisphere component of the Methanol
Multibeam (MMB) survey (Green et al. 2009; Caswell et al. 2010) for 
6.6-GHz methanol masers will soon provide an even more extensive catalogue 
of SFRs than the OH catalogue of \citet{C98}
to search for associated water maser emission. The MMB survey is the most
sensitive survey yet undertaken for young high-mass stars in the Galaxy
and is complete within 2 degrees of the Galactic plane. Water maser
observations towards this unbiased catalogue of methanol masers will
enable meaningful detection statistics for methanol and water maser
associations to be derived and allow comparison with our statistics
derived from the comparison of OH and water maser associations. Due to
this, coupled with the fact that our methanol targeted sample is not 
homogeneous, we do not attempt to draw further conclusions from this 
methanol sample.  We anticipate that comparisons
between the sources detected in the MMB survey and infrared data, in 
conjunction with follow-up observations for 22-GHz water masers, 
12.2-GHz methanol masers and other non-masing molecular species will 
uncover unique insights into the differing physical conditions 
responsible for the presence/absence of the
different methanol maser transitions.

\citet{Beuther02} conducted a high resolution study of the water and
methanol masers in 29 massive star formation regions and found that 
10 methanol masers coincide spatially with water masers to within 
their uncertainty of about 1.5 arcsec.  They remarked that no
spatial correlations exists between the two maser species, which is 
clearly not an inference from their results and refers presumably to the 
common expectation that detailed correlations of maser spots is unlikely.  

Our finding is that occurrence of a water maser at nearly 80 per cent of 
the OH maser targets is comparable to the occurrence of methanol at OH 
targets.  This might seem surprising in view of the difference in 
favoured pumping schemes, where both OH and methanol depend on far IR 
radiation, whereas the favoured pumping scheme for water masers is 
collisional (Elitzur, Hollenbach \& McKee 1989).  
However, it is consistent with the expectation that in most locations 
where methanol and OH masers occur there are coincident, or nearby, 
shocked regions with high densities suitable for the excitation of water 
masers.  Pursuing this further, the apparently larger number of water 
masers than other maser species is consistent with the extremely common occurrence of such shocked 
regions, not only in the envelope of a high-mass star, but also in outflows and in lower mass stars with less IR flux.  

Interestingly, pumping of the formaldehyde 
masers (which are uncommon, but also trace SFRs)
is unclear, since the Boland \& de Jong (1981) radiative pump seems 
deficient and may need shocks and collisions, according to Hoffman et 
al. (2003).

\subsection{Velocity distributions of maser features}\label{sect:vel_dist}

\begin{figure}
\vspace{-1cm}
	\psfig{figure=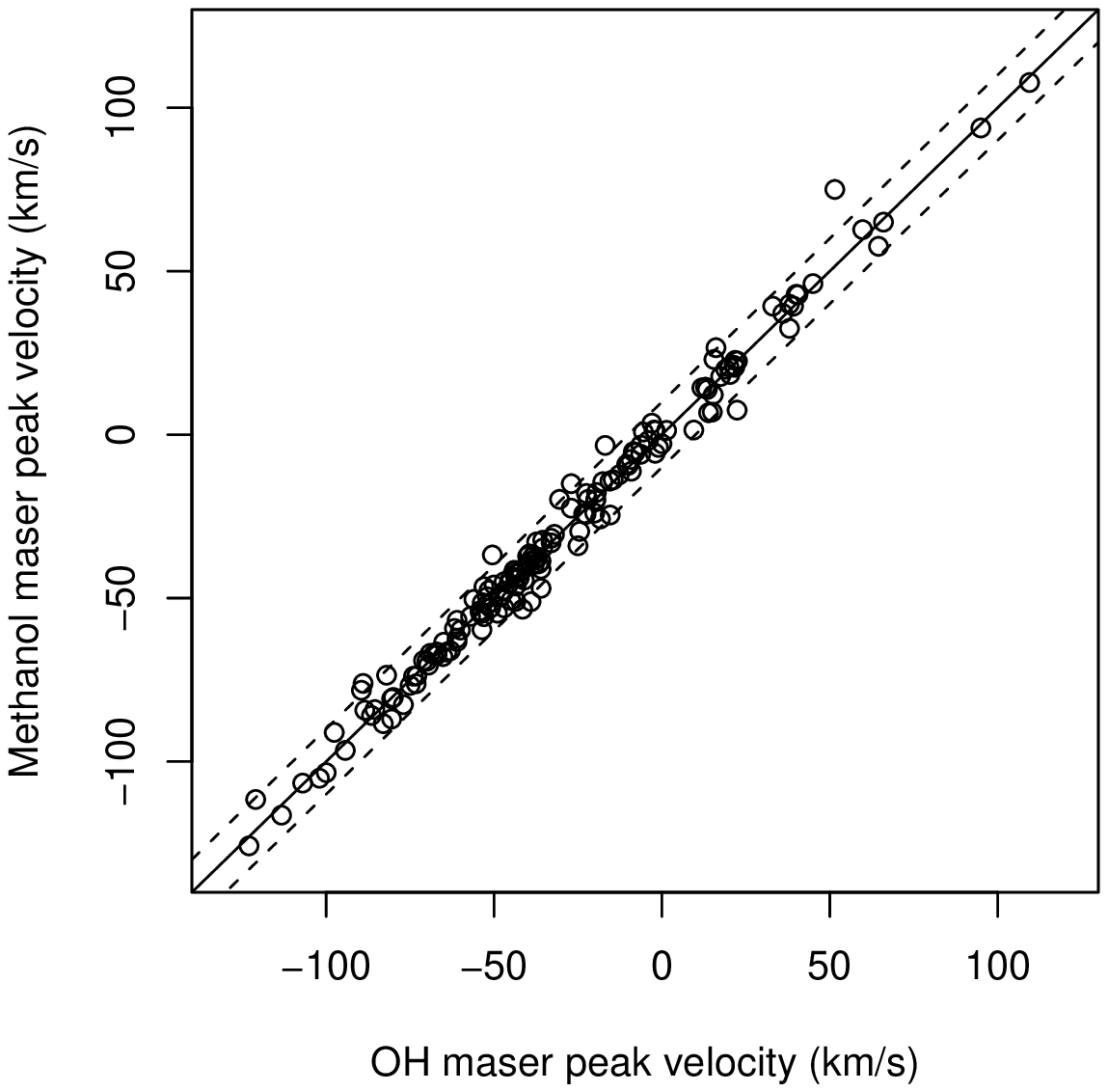,width=9cm,angle=270}
	 \caption{OH maser peak velocity versus methanol maser peak velocity. Overlaid is a solid line with a slope of 1 and two dashed lines showing a  deviation of 10 \kms\ either side of the solid line.}
	 \label{fig:oh_meth}
\end{figure}

The velocity distributions for the water masers have intrinsic interest 
but are most fruitfully studied in comparison with OH and methanol 
counterparts where available.   The velocity range of water maser emission 
at many sites
is larger than for OH or methanol, with the velocity range of water masers
measured in 2003 having an average of 27 \kms\ and a median of 15 \kms\
and those measured in 2004 showing an average velocity range of 30 \kms\
and a median of 15 \kmsns. In contrast, methanol masers rarely show emission 
that exceeds a velocity range of 16 \kms\ \citep{C09}, and C98 found 
the median velocity range of 100 OH masers with flux densities greater 
than 2.7~Jy to be 9 \kmsns.

Sometimes the water emission is remarkably symmetric about the 
systemic velocity but, more often, is asymmetric. The strongest 
water maser 
emission is generally confined to the velocity ranges of the associated OH 
and methanol masers.
The velocity of methanol maser emission is regarded as a
reliable indication of the systemic velocity for the regions that these
masers are tracing \cite[e.g.][]{C09,P09}, allowing kinematic distances
for the sources to be computed. OH masers are slightly less reliable 
tracers of systemic velocities because small changes to the apparent 
radial velocity are caused by the Zeeman effect. Water masers however, are
generally regarded as unreliable tracers of systemic velocities, as they
commonly trace high velocity outflows and have large velocity ranges.

We first compare the peak velocities of methanol and OH masers in Fig.~\ref{fig:oh_meth}, 
(using the population of 165 sources studied in this paper, including sources both with and without associated water maser emission).  
This clearly demonstrates the close similarity in the velocity of their 
peaks. The average difference between the peak velocities of the OH and methanol masers is 3.4 \kmsns, with a median difference of 2.2 \kmsns.

\begin{figure}
\vspace{-1cm}
	\psfig{figure=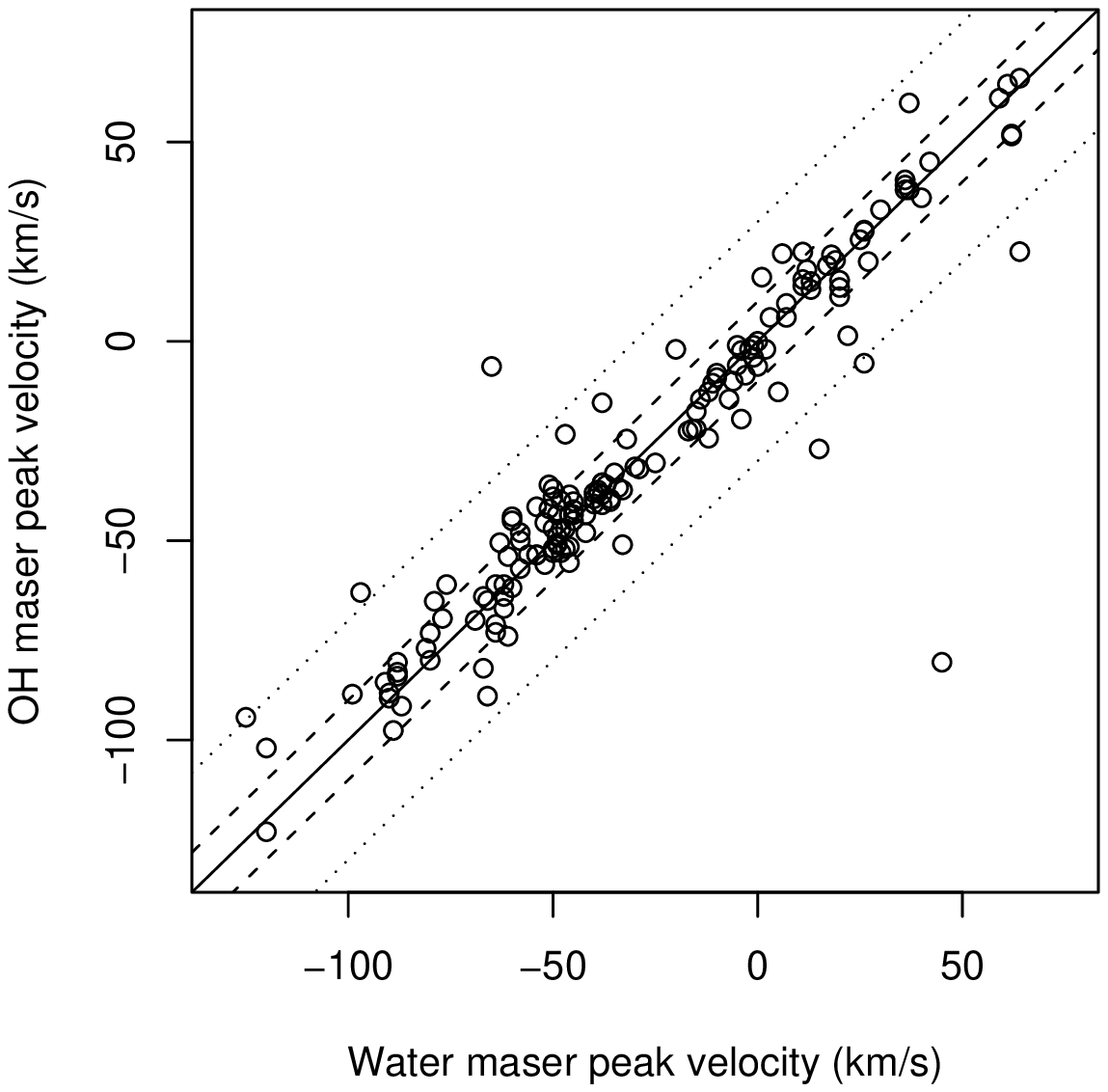,width=9cm,angle=270}
	 \caption{Water maser peak velocity versus OH maser peak velocity. Overlaid is a solid line with a slope of 1 and two dashed lines showing a deviation of 10 \kms\ either side of the solid line. An additional pair of lines (dotted) shows a deviation of 30~\kmsns.}
	 \label{fig:oh_water}
\end{figure}

\begin{figure}
\vspace{-1cm}
	\psfig{figure=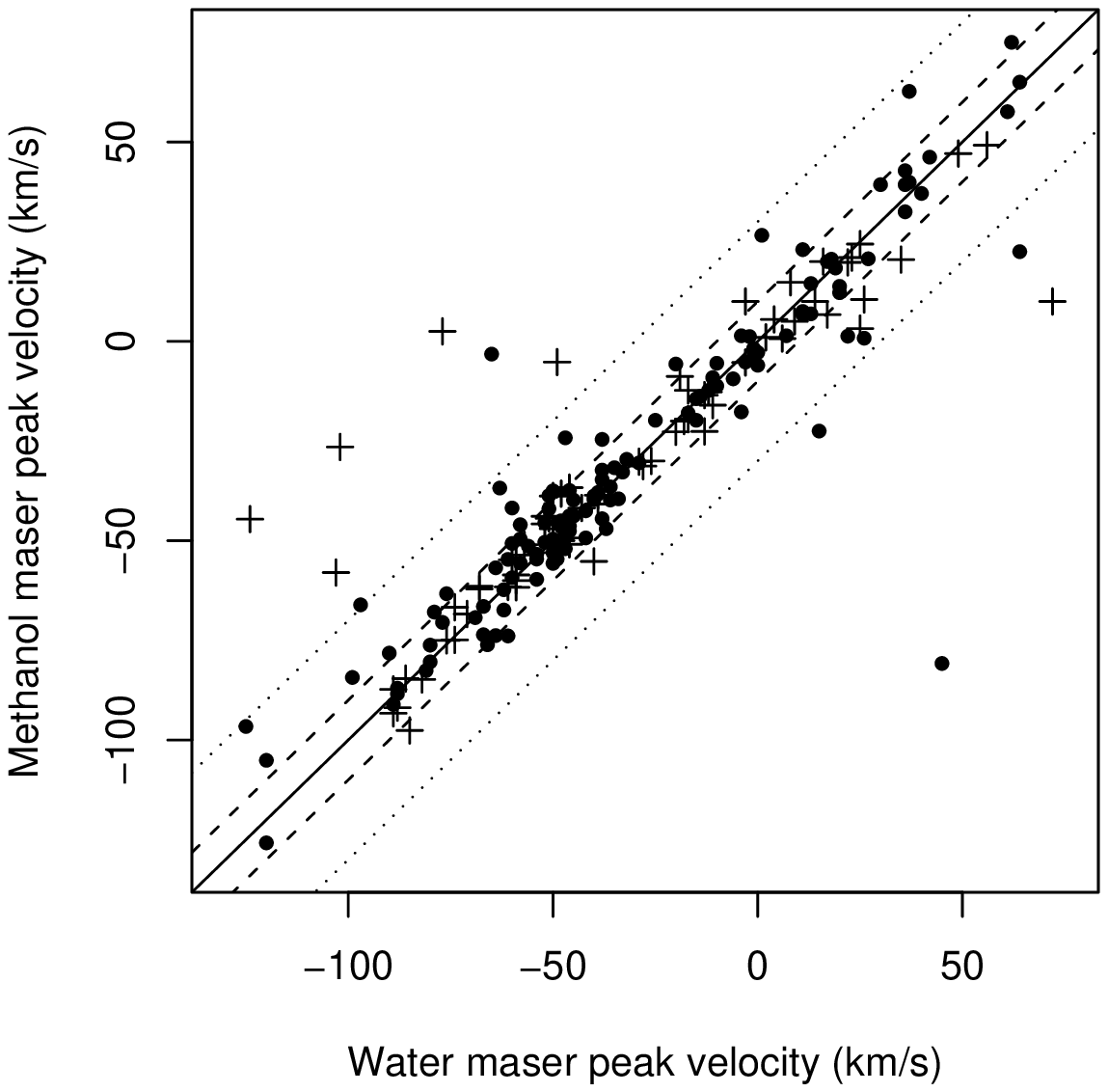,width=9cm,angle=270}
	\caption{Water maser peak velocity versus methanol maser peak velocity. Overlaid is a solid line with a slope of 1 and two dashed lines showing a deviation of 10 \kms\ either side of the solid line. An additional pair of lines (dotted) shows a deviation of 30~\kms. We distinguish sources with only methanol (cross) from those with OH as well as methanol (dot).}
	\label{fig:meth_water}
\end{figure}

We now compare the velocity of the peak emission of our large sample of
water masers with the peak of associated OH and methanol masers
(Figs~\ref{fig:oh_water} and ~\ref{fig:meth_water}, respectively). The 2004
peak velocities were used where possible and 2003 peak velocities were
used otherwise.   
In the case of 160 OH-water maser associations, we find that the average 
difference in the peak velocities is 7.8 \kms\ and the median difference is 
4 \kmsns. 
In the case of 197 methanol-water maser associations we find that the 
average difference in the peak velocities is 8.8 \kms\ and the median 
difference is 4.2 \kmsns. 
Thus for the majority of sources there is quite good correspondence 
between the velocity of the peaks of water masers and
those of the associated OH or methanol maser peak emission, but there are 
some striking outliers. 

Comparison of Fig.~\ref{fig:oh_meth} with Figs~\ref{fig:oh_water} and~\ref{fig:meth_water} highlights the closer correspondence between the peak velocities of the OH and methanol masers than either of these species compared to the associated water maser peak velocity. 93 percent of the OH and methanol maser peak velocities are in agreement to within 10 \kmsns, whereas for the water-methanol and water-OH peak velocities this number falls to 78 and 79 per cent.

We now look in more detail at the outlying sources in Figs~\ref{fig:oh_water} and~\ref{fig:meth_water}. The isolated source at the lower right of the plots is 336.983--0.183 and, as remarked in the note of Section~\ref{sect:ind}, the high-velocity water feature is accompanied by emission at the systemic velocity (methanol and OH peak velocity) which is of comparable intensity (or stronger at the 2003 epoch). Disregarding this source, Fig.~\ref{fig:oh_water} shows only a small number of dominant high-velocity features (offset more that 30 \kms) with no preference for red or blue shifts. In Fig.~\ref{fig:meth_water}, we distinguish the sources with only methanol from those with OH as well as methanol. We note that there is a striking group of six highly blue-shifted features of which five have no OH emission. These are some of the distinct population of dominant blue-shifted outflows discussed by Caswell \& Phillips (2008).


\subsection{Clustering of maser sites and association with other 
masers of OH}

\begin{figure}
	\psfig{figure=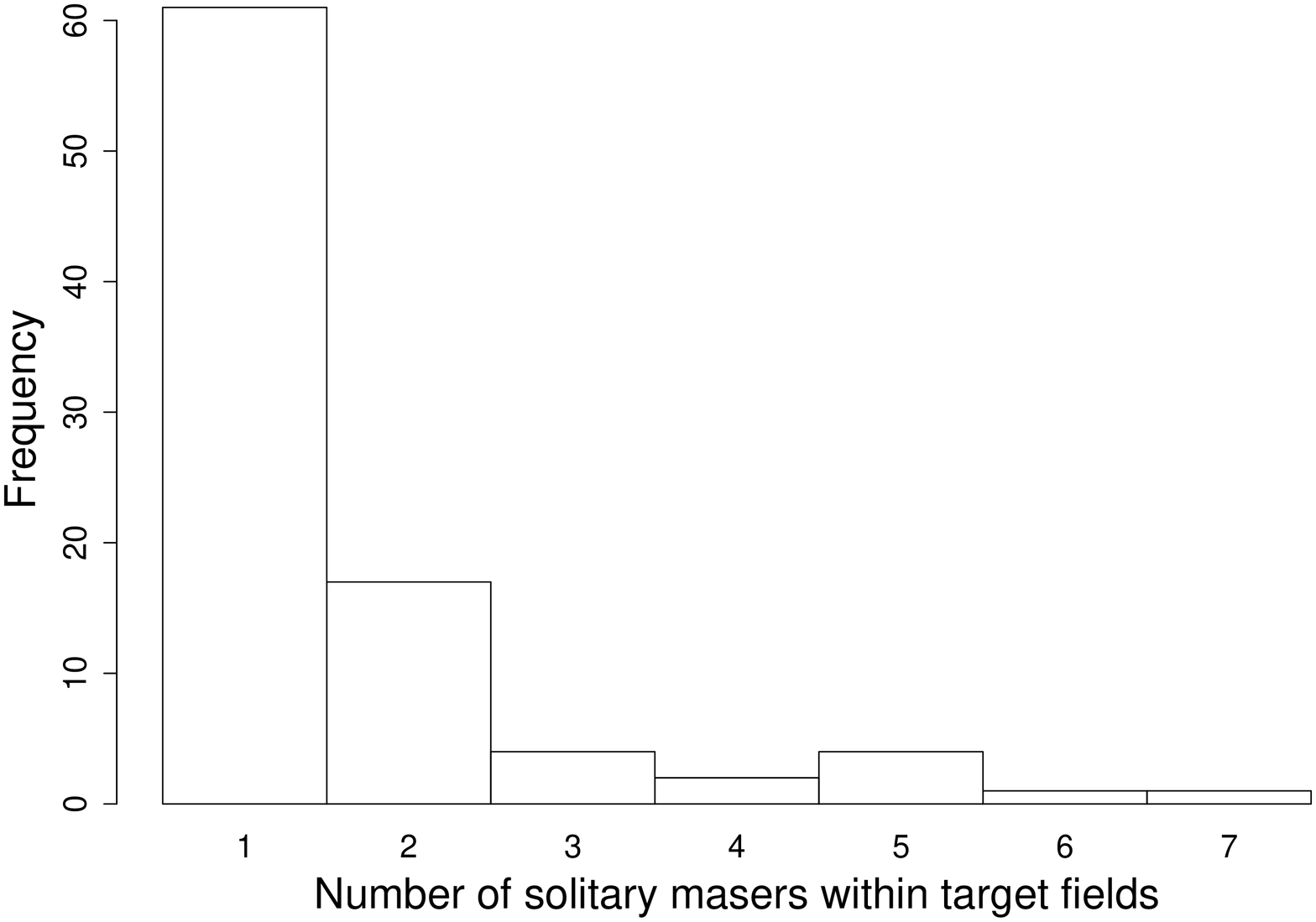,width=9cm,angle=270}
	\caption{Histogram of the number of solitary water masers detected in single ATCA fields.}
	\label{fig:sol_hist}
\end{figure}

In addition to the water maser sources that we find to be intimately
associated with the OH and methanol masers that were targeted, we
frequently detect water masers separated from the target OH and methanol
masers by $\sim$10 arcsec or more. Furthermore, we find the occurrence of
multiple water maser sites within the HPBW of the ATCA primary beam to be common,
with the number of sources often exceeding two and reaching as high as
seven. Fig.~\ref{fig:sol_hist} shows a histogram of the number of water masers in the targeted ATCA fields that are solitary (i.e. in addition to the water masers detected
at the targets of OH and methanol). There are 90 cases where there is one or more
solitary maser within the HPBW of a target OH or methanol maser. This
means that observations targeted towards OH and methanol masers have a 29
per cent chance of detecting at least one unrelated water maser within
$\sim$2 arcmin of the target source.

Clusters comprising combinations of water, OH and methanol masers spread over $\sim$20 arsec are common. Comparative studies of sites within these clusters, where we can infer that 
near-contemporaneous formation of several massive stars has occurred, hold
promise for unravelling the preferred environments and stellar evolutionary
stages for different maser species;  future studies of water masers and
associated IR sources will play a major role in this investigation.

Six sites of 1720-MHz maser emission, believed to be of the SFR variety 
but with no other maser species, were listed by \citet{Cas04c}, and these 
have also 
been searched for water masers.  A new water maser was detected towards 
one of them, 310.146+0.760, as well as towards its cluster companion the 
OH 1665-MHz and methanol maser 310.144+0.760, offset nearly 10 arcsec.  
The 1720-MHz maser 329.426-0.158 was previously the only maser detected in 
a putative SFR (with an HII region nearby both spatially and in velocity).  
Although no water was found at the 1720-MHz site, two 
new water maser sites were discovered, with offsets of only 20 arcsec, 
and therefore credibly within this SFR and therefore now increasing its
known maser population to a cluster of three masers.  

Two sites that currently are known as masers only at the 6035-MHz 
excited-state of OH were also searched for water.  No water 
detection was made towards either 311.596-0.398 or 345.487+0.314.  

\clearpage

\clearpage

\subsection{Association with continuum \UCHII regions}\label{sect:cont}

The present observations, although focused on spectral line emission, also 
allowed a search at each maser site for an 
associated \UCHII region, to a limit of $\sim$~30 mJy if the 
region is not too confused. Although this is two orders of magnitude less
sensitive than can be achieved with a targeted wide bandwidth survey  
(Forster \& Caswell 2000), it is at higher frequency, and provides in many 
cases a useful first estimate. 

The sensitivity to radio continuum is not uniform for all sources, and is
especially poor for observations with maser emission over an extensive
velocity range, leaving little of the bandpass free from line emission.  
Due to this, and our short integration times we are sensitive only to
relatively strong \UCHII regions.

We have not used the continuum data collected at the 2003 epoch since the 
{\em uv}-coverage (with an EW array) was significantly poorer than  
achieved with the H168 array in 2004.
The 29 water maser sources observed only during the 2003 observations have
therefore been removed from the subsequent statistics. 
Table~\ref{tab:contcomp} presents the number of water maser sources 
observed in 2004, broken up
into the categories of solitary, associated with an OH maser, associated
with both OH and methanol masers, and associated with a methanol maser
(numbers shown in column 2). Column 3 shows the number of water maser
sources in each category that are also associated with 22-GHz radio
continuum from \UCHII regions that we detect, while column 4 shows the
percentage of sources with detectable 22-GHz radio continuum in each
category. We find associations with \UCHII regions, indicative of an
embedded massive early type star, for 42 of the water maser sources that
we detect. 

Due to the targeted nature of this search, with the OH sample being
complete but the methanol sample incomplete, percentages are presented 
in Table 5 to reveal more clearly the correlations. The percentages of 
water maser sources with associated \UCHII
region in Table~\ref{tab:contcomp} show that \UCHII regions are
preferentially detected toward water maser sources with associated OH
masers.

\begin{table}
  \caption{Comparison between water maser associations and the presence of associated 22-GHz radio continuum. Column 1 describes the water maser associations, column 2 shows the number of water maser sources observed in 2004 that fall under the given association. Column 3 gives the number of sources within each category that are associated with 22-GHz radio continuum and column 4 shows this number as a percentage of the water maser sources in each category.}
  \begin{tabular}{llll}\hline
    \multicolumn{1}{l}{\bf Water} & {\bf \# of sources} & {\bf \# of sources}	& {\bf \% with}  \\
     \multicolumn{1}{c}{\bf association}  &{\bf total (2004)}  &{\bf with cont} & {\bf cont} \\ \hline
 OH \& methanol	&	112	&	24	&	21.4\\
  OH				&	28	&	6	&	21.4\\
  methanol		&	67	&	5	&	7.5	\\ 
   solitary 			&	143	&	7	&	4.8\\
  \hline
  \label{tab:contcomp}
  \end{tabular}
\end{table}

We find that our overall detection rate for \UCHII regions towards water maser sources with associated OH masers (with or without associated methanol) is 21.4 per cent while our detection rate towards water maser sources without associated OH masers (solitary or with methanol) is 5.0 per cent. \citet{FC00} conducted a sensitive search at 8.7-GHz for \UCHII regions towards OH and water masers which showed that 52 per cent of the OH masers they targeted had an associated \UCHII region. While our observations are almost two orders of magnitude less sensitive than was achieved by \citet{FC00}, they are at a higher frequency allowing us to potentially detect emission from hyper-compact (HC) \HII  regions which are typically optically thick at centimetre wavelengths. Comparison with the detection rate of \citet{FC00} suggests that a more sensitive search at 8.7-GHz for \UCHII regions towards our OH maser associated water maser sources would more than double our detections from 30 sources to $\sim$71.

Our results support arguments (\citet{C01,Beuther02,Breen09}) that
methanol maser emission is often seen prior to any OH maser emission,
but is sensitive to the onset of emission from \UCHII regions, and
less able to survive the later stages of the evolution of the \UCHII
region. Carrying on these arguments to include water masers we find that
the water masers are also present at the early stages of formation, like
the methanol masers, prior to the onset of OH maser emission. These
statistics for solitary water sites naively suggest that these water masers precede the onset of strong \UCHII regions. However, as mentioned in Section 5.1, it is possible that a significant population of the solitary water masers are associated with low mass stars and this provides an alternative explanation.

\subsection{Comparison with GLIMPSE objects}

\subsubsection{Association with GLIMPSE point sources}

We have compared the positions of the 379 water maser sources with the 
positions of sources in the GLIMPSE point source catalogue. We find that 343 of our water maser sources are within the Galactic longitude and latitude ranges observed by GLIMPSE and that 165 of these are within 3 arcsec of a GLIMPSE point source (48 per cent). This number increases to 211 if sources from the GLIMPSE archive are included (62 per cent). The fraction of the water maser sources associated with point sources contained in either the GLIMPSE point source catalogue or the supplementary archive catalogue, is similar to that found by \citet{Ellingsen06} when comparing the positions of 56 methanol masers with GLIMPSE sources (68 percent).

We have further investigated the associations between water maser and GLIMPSE  sources by comparing the GLIMPSE source association rates of water masers in their association categories (i.e. associated with both OH and methanol masers, associated with OH masers, associated with methanol masers etc.). Association rates are as follows:

\begin{itemize}
\item 76 of the 165 GLIMPSE detections have both OH and methanol (i.e. 46\% of the GLIMPSE sources are associated with OH, methanol and water and 60\% of the OH methanol and water sources have an associated GLIMPSE source). 
\item 91 of 165 GLIMPSE detections have OH detections (i.e. 55\% of the GLIMPSE sources have an associated OH maser and 57\% of OH maser sources have an associated GLIMPSE source). 
\item 109 of 165 GLIMPSE detections have methanol detections (i.e. 66\% of the GLIMPSE sources have an associated methanol maser and 56\% of methanol maser sources have an associated GLIMPSE source).
\item 18 of 165 GLIMPSE detections have associated radio continuum (i.e. 11\% of the GLIMPSE sources have an associated HII region and 43\% of HII regions have an associated GLIMPSE source).
\item 41 of 165 GLIMPSE detections only have an associated water maser (i.e. 25\% of the GLIMPSE sources are only associated with a water maser and 29\% of the water only sources have an associated GLIMPSE source).

\end{itemize}

The GLIMPSE point source association rates are similar in all categories
except for solitary water masers and water masers associated with radio
continuum where the association rates are significantly lower. In the case
of the radio continuum, it is likely that a large number of sources
exhibiting strong radio continuum would no longer be  point sources at
mid-infrared frequencies (because sources exhibiting strong radio continuum are likely to be more evolved) and this would therefore 
account for the lower
association rate. The lower association rate between GLIMPSE point sources
and solitary water masers could be explained by a tendency for
these water sources to be associated with more extended objects, or, alternatively, that the
solitary water masers are commonly associated with lower luminosity
sources.  

\begin{figure}
	\psfig{figure=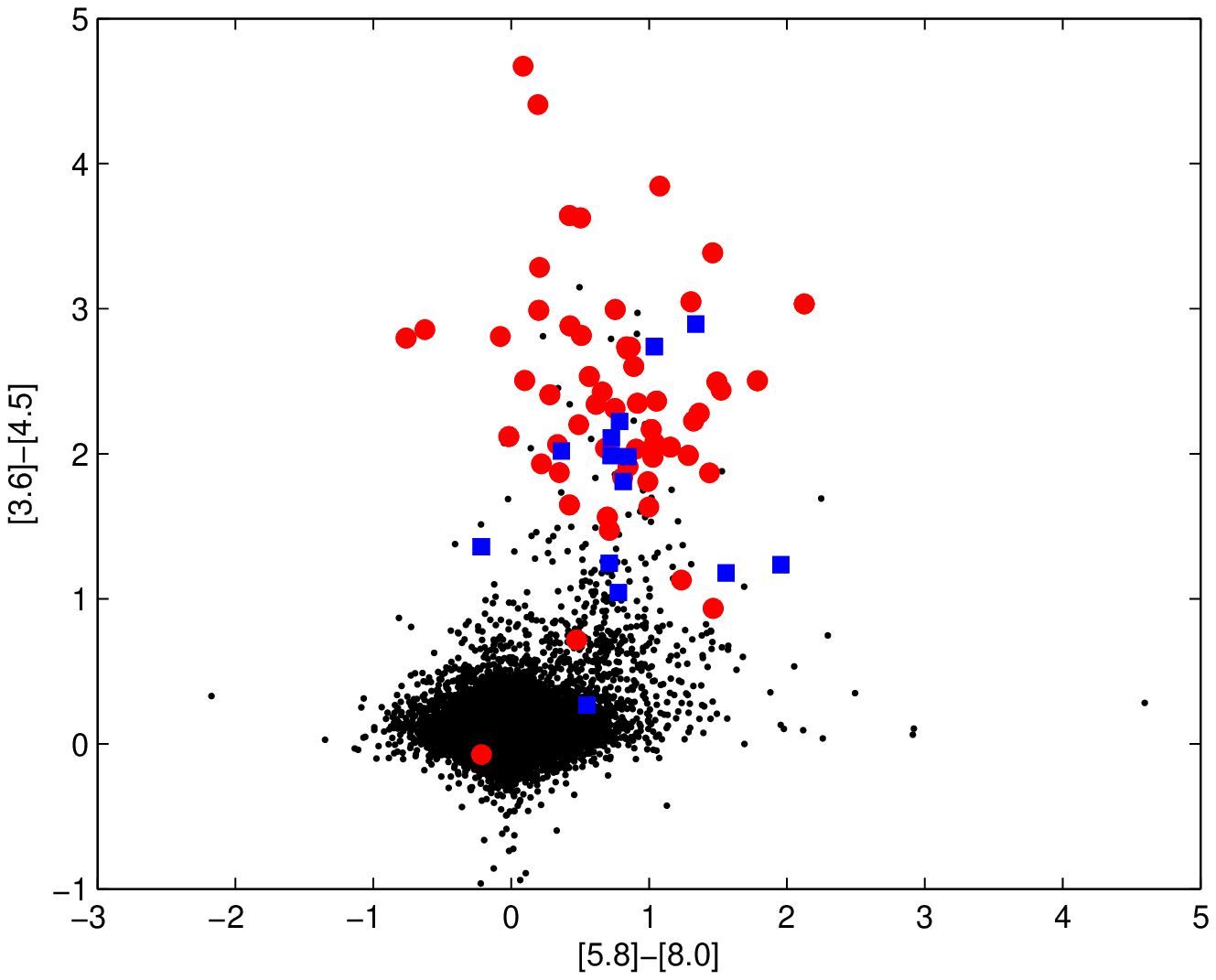,width=10cm,angle=0}
	\caption{Colour-colour plot of GLIMPSE point source data. Water maser sources with associated OH and/or methanol maser emission are represented by red circles and solitary water maser sources are represented by the blue squares. The black dots represent all of the GLIMPSE point sources within 30 arcmin of {\em l} = 326$\fdg$5, {\em b} = 0\fdg0.}
	\label{fig:glimpse}
\end{figure}

Figure~\ref{fig:glimpse} shows a plot of the [3.6]--[4.5] $\mu$m versus [5.8]--[8.0] $\mu$m colours of the GLIMPSE point sources associated with the water masers. Flux density measurements for all four of the IRAC bands had to be available for the inclusion in this plot thus limiting the plotted sample to 14 solitary water masers (i.e. with no methanol or OH maser counterpart) and 58 water masers with either a methanol or an OH counterpart. We find, similarly to previous comparisons \citep[e.g.][]{Ellingsen06,Breen09}, that the GLIMPSE sources associated with the masers are located above the majority of the comparison sources in the colour-colour plot. This figure also reveals an apparent difference in the ranges of the [3.6]--[4.5]~$\mu$m colours for solitary water masers compared with water masers that are associated with methanol, OH or continuum (or a combination of these). 

We have carried out a t-test (testing the hypothesis that there is no difference between the means) on both the [3.6]--[4.5] and [5.8]--[8.0]~$\mu$m colours of those GLIMPSE sources associated with solitary water masers compared to those associated with water masers as well as OH, methanol or radio continuum. In the case of the [5.8]--[8.0]~$\mu$m colours we find that there is no statistically significant difference between the values associated with the two groups of water maser sources. For the [3.6]--[4.5]~$\mu$m colours we find that there is a statistically significant difference (p-value 0.007) between the GLIMPSE sources associated with solitary water masers and those water masers with associated methanol, OH or radio continuum sources.

As can be seen in Fig.~\ref{fig:glimpse}, the [3.6]--[4.5]~$\mu$m colour tends towards smaller values in the case of the solitary water masers. Since there is no difference in the [5.8]--[8.0]~$\mu$m colours between the two groups of sources this means the solitary water maser associated GLIMPSE sources have a much less steep spectrum at wavelengths $<$ 5~$\mu$m than at wavelengths greater than this. This indicates that these sources may be colder in general. Another explanation may be that the GLIMPSE sources associated with the solitary water masers have a relative excess of 4.5~$\mu$m flux density, similar to extended green objects (EGOs) \citep{Cyg08}. 

In Section~\ref{sect:var} we suggested that some fraction of the solitary
water masers are likely to be associated with low-mass stars rather than
the high-mass star formation regions where these observations were
targeted. According to \citet{Cyg08}, GLIMPSE is too shallow to detect
emission from outflows associated with low-mass stars (except perhaps for
the closest low-mass star formation regions). Furthermore, if GLIMPSE
detected infrared emission associated with low-mass stars it certainly
would not detect it as a point source because the space density would be
much too high. We therefore conclude that none of the sources included in
Fig.~\ref{fig:glimpse} are associated with low-mass stars and therefore
can not be responsible for the difference. However, it is possible that
the lower association rate for solitary water maser sources with GLIMPSE
point sources is partially because some fraction of the solitary water 
maser sources are associated with low-mass stars.

\clearpage

\subsubsection{Association with extended green objects (EGOs)}

We have compared the locations of the EGOs presented in \citet{Cyg08} with
our 379 water masers. In order to avoid large numbers of chance associations between EGOs and water masers, we consider an EGO to be
associated with a nearby water maser when the angular separation is less
than 10 arcsec.  \citet{Cyg08} compared the locations of 6.6-GHz methanol
masers with the images of their EGOs and show that this separation captures most of the associations while minimizing the chance coincidences that would result from a larger threshold. We find that 63 of the
water masers are coincident with an EGO identified by \citet{Cyg08}.

\begin{table}
  \caption{Comparison between water maser associations in our full sample with water maser associations for sources associated with EGOs. Column 1 describes the water maser associations, column 2 shows the percentage of water maser sources in each category (from the full sample) and column 3 gives the percentage of sources in each category that are also associated with an EGO \citep{Cyg08}.}
  \begin{tabular}{lll}\hline
    \multicolumn{1}{l}{\bf Water} & {\bf \% of full} & {\bf \% of sources}	\\
     \multicolumn{1}{c}{\bf association}  &{\bf sample}  &{\bf  with EGOs}\\ \hline
  OH \& methanol	&	33.8	&	55.6	\\
  OH				&	8.7	&	7.9	\\
  methanol		&	18.5	&	20.6 \\ 
    solitary 			&	39.0	&	15.9	\\

  Total			&	100	& 	100 \\ \hline
  \label{tab:egos}
  \end{tabular}
\end{table}

Table~\ref{tab:egos} shows in the second column the percentage of the full sample of
379 water masers that fall within the four categories: associated
with both OH and methanol masers, associated with only OH masers,
associated with only methanol masers, and solitary; and in the third column, the number of water sources in
each category that are also associated with EGOs as a percentage of the
total number EGO-associated sources. This table shows that those water maser
sources coincident with EGOs and associated with only methanol or OH masers are distributed in 
similar fashion to our
complete sample of water masers, with little difference between the
percentage of water sources presented in columns two and three. However, in the
case of the solitary water sources, the association rate with EGOs is much lower
than would be expected (similar to solitary water masers associated with
GLIMPSE point sources). The absence of EGO associations with a large
number of the water maser only sources may suggest that solitary water
masers are associated with lower luminosity sources. Alternatively,
considering that the water maser only sources tend to be associated with
GLIMPSE point sources with dominant 4.5~$\mu$m emission, a characteristic
shared by EGOs, the water maser only sources may represent a class of
younger sources, the majority of which have not yet produced an extended
outflow. Perhaps this indicates that these solitary water masers are
associated with outflow related sources: compact green objects,
pre-cursors to EGOs.

We find that there is a higher association rate with EGOs for those 
water maser sources accompanied by both methanol and OH masers.  This
indicates that EGOs persist into the stage of star formation that is
evolved enough to have produced an OH maser but not so evolved that the
production of an associated strong \UCHII region has caused the
methanol maser emission to cease. Eighty-nine of the water maser sources
we detect that are associated with both OH and methanol masers are within
the regions covered by GLIMPSE and have been inspected for the presence of
EGOs \citep{Cyg08}. We find that 35 of these 89 sources are associated
with an EGO, a rate of 39 per cent. This lends further credence to the
idea the EGOs are not exclusively tracing the earliest stages of massive
star formation but persist well into the stage where OH masers are
present. Furthermore, as there is a large number of these objects, they
must have a significant lifetime.

\section{Conclusions}

From a large sample of water masers measured with precise positions at two 
epochs, we conclude that spectra are highly variable but positions are 
generally persistent.

The occurrence of a water maser at nearly 80 per cent of the OH maser 
targets is comparable to that of methanol at OH sites.  This is despite 
the difference in favoured pumping schemes, where both OH and methanol 
depend on far IR radiation, whereas the favoured pumping scheme for 
water masers is collisional.

Our study of water masers at methanol maser sites is preliminary, 
but the common presence of water at methanol sites is confirmed.  We 
argue that there is indeed an important role for water masers in mapping 
the Galaxy and its velocity field.  The present contribution of a large 
number of water masers with accurate positions in the southern Galaxy has 
been an important step in advancing such a project, and reveals 
the value of conducting even larger  
future surveys with complete Galactic plane coverage.

\section*{Acknowledgments}

The Australia Telescope Compact Array is part of the Australia Telescope which is funded by the Commonwealth of Australia for operation as a National Facility managed by CSIRO. This research has made use of: NASA's Astrophysics
Data System Abstract Service; the NASA/
IPAC Infrared Science Archive (which is operated by the Jet Propulsion
Laboratory, California Institute of Technology, under contract with
the National Aeronautics and Space Administration); and data products from the GLIMPSE
survey, which is a legacy science program of the {\em Spitzer Space
  Telescope}, funded by the National Aeronautics and Space
Administration.

\end{document}